\documentclass[%
 reprint,
superscriptaddress,
 amsmath,amssymb,
 aps,
]{revtex4-2}

\usepackage{graphicx}
\usepackage{dcolumn}
\usepackage{bm}

\usepackage[caption=false]{subfig}

\usepackage[hidelinks,unicode=true]{hyperref}

\usepackage{xr}

\usepackage{xcolor}

\begin{document}


\title{Selective social interactions and speed-induced leadership in schooling fish}

\author{Andreu Puy}
\email{andreu.puy@upc.edu}
\affiliation{Departament de Física, Universitat Politècnica de Catalunya, Campus Nord B4, 08034 Barcelona, Spain}

\author{Elisabet Gimeno}
\affiliation{Departament de Física, Universitat Politècnica de Catalunya, Campus Nord B4, 08034 Barcelona, Spain}
\affiliation{Departament de Física de la Matèria Condensada, Universitat de Barcelona, Martí i Franquès 1, 08028 Barcelona, Spain}

\author{Jordi Torrents}
\affiliation{Departament de Física, Universitat Politècnica de Catalunya, Campus Nord B4, 08034 Barcelona, Spain}
\affiliation{Departament de Física de la Matèria Condensada, Universitat de Barcelona, Martí i Franquès 1, 08028 Barcelona, Spain}

\author{Palina Bartashevich}
\affiliation{Institute for Theoretical Biology, Humboldt-Universität zu Berlin, 10115 Berlin, Germany}
\affiliation{Excellence Cluster Science of Intelligence, Technische Universität Berlin, 10587 Berlin, Germany}

\author{M. Carmen Miguel}
\affiliation{Departament de Física de la Matèria Condensada, Universitat de Barcelona, Martí i Franquès 1, 08028 Barcelona, Spain}
\affiliation{Institute of Complex Systems (UBICS), Universitat de Barcelona, Barcelona, Spain}

\author{Romualdo Pastor-Satorras}
\affiliation{Departament de Física, Universitat Politècnica de Catalunya, Campus Nord B4, 08034 Barcelona, Spain}

\author{Pawel Romanczuk}
\affiliation{Institute for Theoretical Biology, Humboldt-Universität zu Berlin, 10115 Berlin, Germany}
\affiliation{Excellence Cluster Science of Intelligence, Technische Universität Berlin, 10587 Berlin, Germany}
\affiliation{Bernstein Center for Computational Neuroscience, Berlin, 10099 Berlin, Germany}


\date{27 March, 2024}

\begin{abstract} 
Animals moving together in groups are believed to interact among each other with effective social forces, such as attraction, repulsion and alignment. Such forces can be inferred using `force maps’, i.e. by analysing the dependency of the acceleration of a focal individual on relevant variables. Here we introduce a force map technique suitable for the analysis of the alignment forces experienced by individuals. After validating it using an agent-based model, we apply the force map to experimental data of schooling fish. We observe signatures of an effective alignment force with faster neighbours, and an unexpected anti-alignment with slower neighbours. Instead of an explicit anti-alignment behaviour, we suggest that the observed pattern is the result of a selective attention mechanism, where fish pay less attention to slower neighbours. This mechanism implies the existence of temporal leadership interactions based on relative speeds between neighbours. We present support for this hypothesis both from agent-based modelling, as well as from exploring leader-follower relationships in the experimental data.
\end{abstract}

\maketitle


\section*{Significance}
Grouping animals interact socially to move together and make collective decisions. These interactions are typically assumed to be attraction
to distant individuals, repulsion from too close neighbors,
and alignment among neighbors to coordinate movement directions.  
Combining analysis of experimental data of schooling fish with
agent-based modeling, we find that social interactions are
modulated by relative speeds between neighbors, where individuals tend to align  only with faster neighbors and ignore slower ones. These 
selective social interactions can be linked to dynamically changing leader-follower relationships defined by relative speeds of neighbors. Our results provide a deeper understanding of emergent leadership based on variable and adaptive speeds of individuals, and a different perspective on mechanisms of movement coordination in natural and artificial systems.

\section*{Introduction}
Moving animal groups can exhibit a complex and highly coordinated
collective behaviour that results in the formation of spatial patterns changing
dramatically in time, as observed in fish schools or bird
flocks~\cite{sumpterCollectiveAnimalBehavior2010, vicsekCollectiveMotion2012}.
A possible framework for interpreting this complex behavior relies on the
concept of effective forces acting on individuals. However, contrary to physical
forces, these forces mainly originate from individual decision-making processes
and, in particular, do not necessarily fulfil the action-reaction Newtonian
principle \cite{romanczukCollectiveMotionDue2009}. Depending on their origin, we
can classify them into individual forces and social forces.  Individual forces
are independent of the social environment. They can have a physical explanation
related to the mechanics of locomotion, such as propulsion and friction, or be a
consequence of non-physical processes linked to individual movement decision,
including intrinsically stochastic behavior and response to environmental cues,
such as the presence of tank walls, food sources or predators.
Social forces, on the other hand, emerge from the interactions between the
individuals of the group and are usually described in terms of a combination of
attraction, repulsion and alignment
forces~\cite{sumpterCollectiveAnimalBehavior2010, vicsekCollectiveMotion2012,
  herbert-readUnderstandingHowAnimal2016}. More specifically, attraction allows
individuals to aggregate in groups, repulsion prevents collisions and alignment
induces individuals to head in the same direction as their neighbours.

Models of collective motion based on self-propelled particles interacting with neighbours via these three kinds of social interactions are able to reproduce typical features observed in moving animals such as polarized movement, where individuals move globally in the same direction; swarming, where individuals aggregate in a group but move in random directions; and milling, where individuals rotate around a core~\cite{reynoldsFlocksHerdsSchools1998, couzinCollectiveMemorySpatial2002}. Although explicit alignment forces are a straight forward way to induce global orientational state, they are not a necessary condition~\cite{romanczukCollectiveMotionDue2009, strombomCollectiveMotionLocal2011, ferranteElasticityBasedMechanismCollective2013, barberisLargeScalePatternsMinimal2016,bastienModelCollectiveBehavior2020,strombomAttractionVsAlignment2022}.

From an experimental point of view, studies have provided different results regarding the existence of explicit alignment interactions. One possible approach is to postulate a model and optimize its parameters from the experimental data. Lukeman et al~\cite{lukemanInferringIndividualRules2010} assumed a zonal model, where the different attraction, repulsion and alignment  forces are implemented in concentric layers around the individual, and found that the strength of alignment was comparable to that of attraction. In a similar spirit but without choosing a specific model, Theraulaz and collaborators~\cite{caloviDisentanglingModelingInteractions2018, escobedoDatadrivenMethodReconstructing2020} proposed symmetrical constraints to identify the contributions of the different interactions and obtained simple analytical functions, assuming that the effect of different variables can be separated as a product of functions. They reported alignment interactions that depend on the relative headings and  positions between neighbours.
An alternative method is to train a neuronal network with experimental data and infer interactions from it.
Heras et al.~\cite{herasDeepAttentionNetworks2019} found evidence of alignment for large neighbour speeds and an anti-alignment region behind the individual that was associated to possible interactions with the walls.

The most common procedure to measure  forces, however, is the use of `force maps', also known as `averaging method', where forces are inferred from the dependency of the acceleration with some relevant variables and averaging over all others.  While force maps have been used with great success to infer attraction and repulsion forces in moving animal groups~\cite{katzInferringStructureDynamics2011, herbert-readInferringRulesInteraction2011, pettitInteractionRulesUnderlying2013, herbert-readHowPredationShapes2017, zienkiewiczDatadrivenModellingSocial2018, escobedoDatadrivenMethodReconstructing2020}, they are not free from criticism~\cite{escobedoDatadrivenMethodReconstructing2020, herasDeepAttentionNetworks2019, mudaliarExaminationAveragingMethod2020}: since force maps can be visualized mainly up to two-dimensions, one either has to visualize cuts of higher dimensional objects that may lack statistics or average over some variables; it is not necessarily straightforward to obtain simple analytical expressions to reproduce the observed force maps; different interactions in a force map can appear mixed, which can make them difficult to identify and disentangle; and different interaction rules may result in similar force maps.
The existence of effective alignment forces has been investigated through the analysis of changes in relative heading of neighbouring individuals~\cite{mudaliarExaminationForceMaps2023}. Whereas earlier works failed to identify alignment interactions~\cite{herbert-readInferringRulesInteraction2011, katzInferringStructureDynamics2011}, later studies found some evidence for alignment between neighbours~\cite{pettitInteractionRulesUnderlying2013, zienkiewiczDatadrivenModellingSocial2018, escobedoDatadrivenMethodReconstructing2020}.

A common view assumed in most numerical models and experimental analysis is that coherent collective motion emerges spontaneously from symmetrical interactions among identical self-propelled group members. This contrasts with studies of leadership in which the strength of the interactions between pairs is asymmetric and displays a net information flow, sometimes with time delays~\cite{zafeirisWhyWeLive2018}. From the group perspective, one way to represent leader-follower relationships is via complex directed interaction networks~\cite{nagyHierarchicalGroupDynamics2010, rosenthalRevealingHiddenNetworks2015}. Furthermore, leadership structures depend strongly on the characteristics of the species but also on many individual factors such as information, physiological state, dominance and navigational competence.
In previous works on birds~\cite{nagyHierarchicalGroupDynamics2010}, bats~\cite{orangeTransferEntropyAnalysis2015}, and fish~\cite{bumannFrontIndividualsLead1993, katzInferringStructureDynamics2011, herbert-readInferringRulesInteraction2011, rosenthalRevealingHiddenNetworks2015, crosatoInformativeMisinformativeInteractions2018, gimenoLeadershipCollectiveMotion2018, romero-ferreroStudyTransferInformation2023}, information flow was identified to naturally occur from influential individuals at the front of the group towards following individuals located further back.
Leadership was also observed to depend on average individual speeds of previous solo flights for homing pigeons~\cite{pettitSpeedDeterminesLeadership2015} and for schooling fish~\cite{jollesConsistentIndividualDifferences2017, gimenoLeadershipCollectiveMotion2018, jollesGrouplevelPatternsEmerge2020}. Furthermore, deep learning tools have provided evidence that individuals are more influenced by neighbours moving at high speeds~\cite{herasDeepAttentionNetworks2019, romero-ferreroStudyTransferInformation2023}.

Here we review the concept of force maps and introduce a more natural alignment force map in Cartesian coordinates that depends on relative velocities of the focal individual with respect to its neighbours. Using a simple model of collective motion with social and individual forces, we demonstrate the utility of the force maps to capture the attraction, repulsion and alignment forces. Equipped with the proposed alignment force map, we apply it to experimental data of schooling fish and obtain two distinct behaviours: alignment with a neighbour that moves faster than the individual and an unexpected region of apparent anti-alignment when a neighbour moves at slower speeds than the individual. In order to explore the origin of this anti-alignment signal, we investigate an agent-based model with explicit alignment and anti-alignment forces, and show that it fails to reproduce realistic behavioural dynamics. Instead, we are able to reproduce the qualitative behaviour of the experimental data with a selective interactions model, where individuals only interact with neighbours at faster speeds and ignore neighbours with slower speeds.
We further examine the selective-interaction hypothesis by studying the effect of leadership at the level of force maps and demonstrate that a follower displays a pure alignment force and a leader effectively a pure anti-alignment force.
Finally, we validate the hypothesis that individuals with slower and faster neighbours are leaders and followers respectively, studying time-delayed correlations between the velocities of a focal fish and its neighbour.

\section*{Results}
\subsection*{Force maps}
The force map method allows inferring forces from a system. It assumes the system has several force terms depending on sets of independent variables, and that the average in time of each force term is zero. Consider we want to infer a force term $\vec{f}(x)$ as a function of only the variable $x$. We can achieve this from Newton's second law by studying the acceleration depending on $x$ while averaging over all other variables in the system, $\vec{a}(x)$~\cite{katzInferringStructureDynamics2011}. Similar procedures have also been employed in deriving effective equations of motion for motile cells~\cite{bodekerQuantitativeAnalysisRandom2010}. Working in units of mass $m = 1$, Newton's second law will become $\vec{f} (x) \simeq \vec{a} (x)$, as all other forces will cancel out in the average. In practice, this approach is implemented by binning over the whole dataset the relevant variable $x$  as $\{x_k\}_{k=1, \ldots, M}$, where $k$ is the index of the bin, and averaging the values of the acceleration on that bin,  $\vec{f}(x_k) \simeq \left< \vec{a} (x)\right>_{x \in x_k}$.

\begin{figure*}[t!p]
\centering
\subfloat[\label{fig:force_maps:atracRepForce}]{%
  \includegraphics[width=0.4\textwidth]{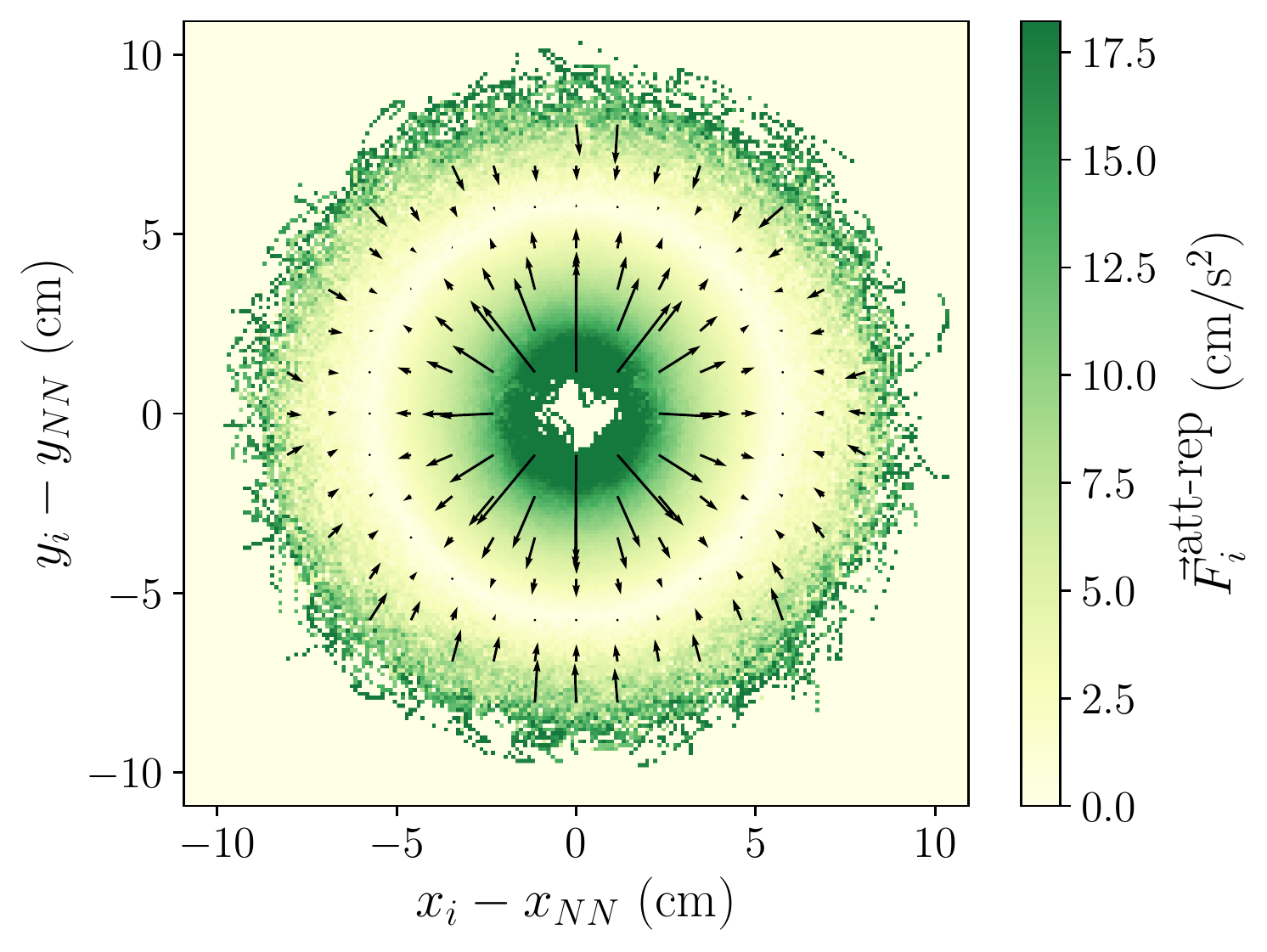}%
}
\hspace{0.01\textwidth}
\subfloat[\label{fig:force_maps:atracRep_basicModel}]{%
  \includegraphics[width=0.4\textwidth]{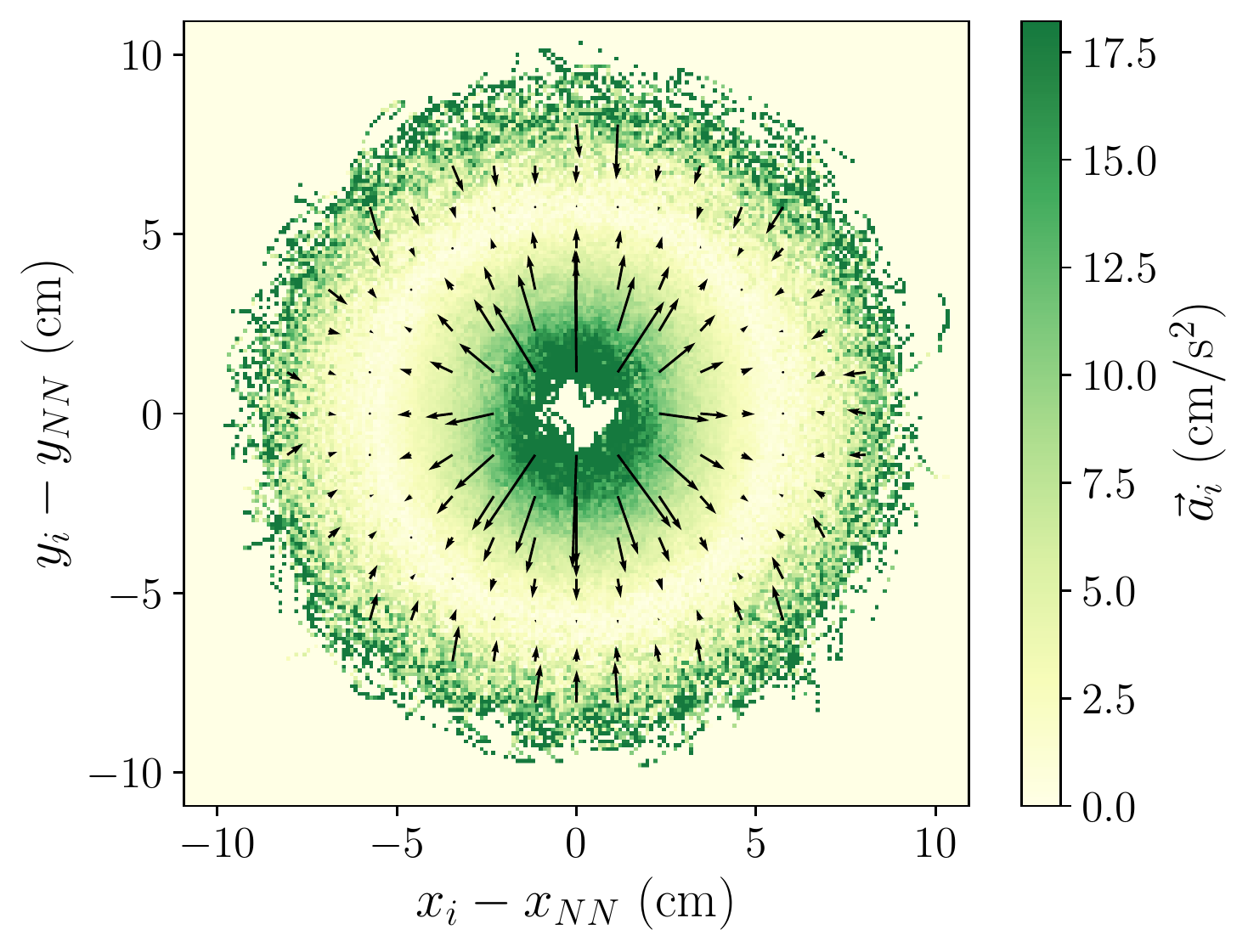}%
}

\subfloat[\label{fig:force_maps:alignForce}]{%
  \includegraphics[width=0.4\textwidth]{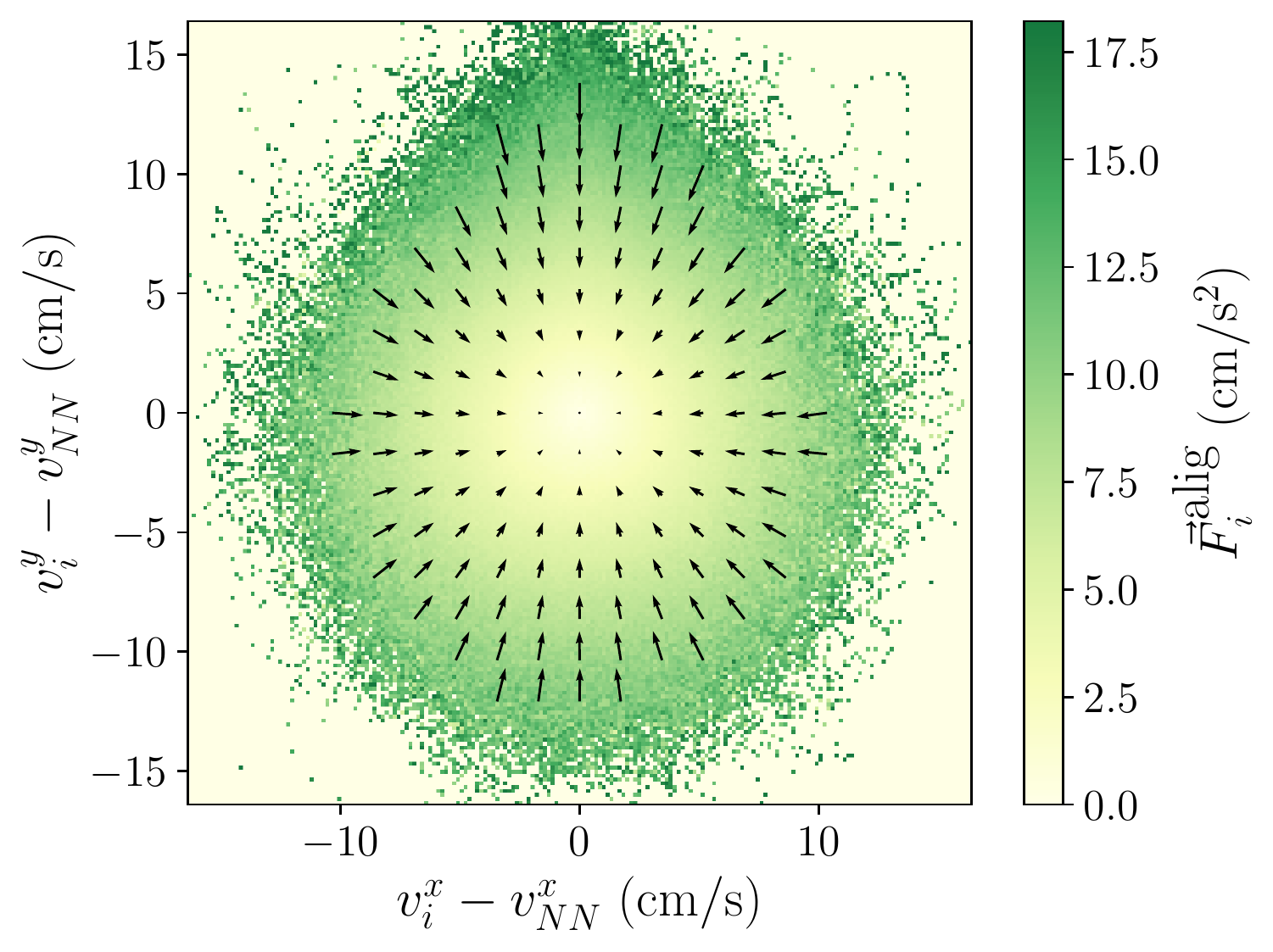}%
}
\hspace{0.01\textwidth}
\subfloat[\label{fig:force_maps:align_basicModel}]{%
  \includegraphics[width=0.4\textwidth]{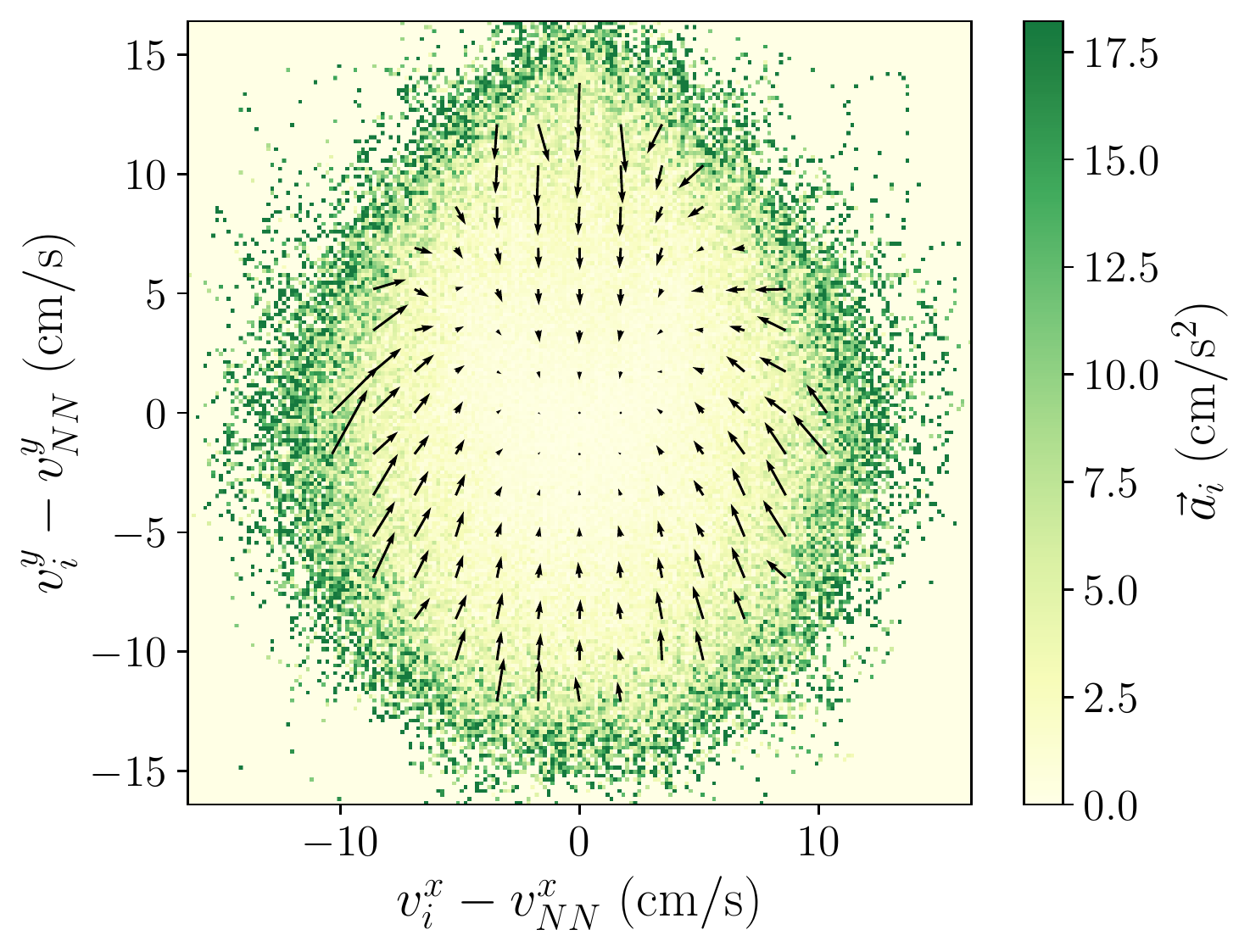}%
}
\caption{\textbf{Attraction-repulsion and alignment force maps for the standard model.} The force maps are able to recover qualitatively the behaviour of the corresponding forces. Average (a) attraction-repulsion force and (b) acceleration (attraction-repulsion force map) acting on an individual $i$ depending on its relative position with the nearest neighbour $NN$, $\vec{x}_i - \vec{x}_{NN}$. We observe for small distances repulsion, for intermediate distances no net force and for large distances attraction. Average (c) alignment force and (d) acceleration (alignment force map) acting on an individual $i$ depending on its relative velocity with the nearest neighbour $NN$, $\vec{v}_i - \vec{v}_{NN}$. In all plots, the $y$-axis is along the direction of motion of the focal individual, the colormap displays the modulus of the force and the arrows its components. Forces are expressed in units of mass $m=1$. We employ values for all focal individuals and time steps in the simulation.} \label{fig:force_maps}
\end{figure*}

\subsubsection*{Attraction-repulsion force map}
This force map was introduced both by Katz et al.~\cite{katzInferringStructureDynamics2011} and Herbert-Read et al.~\cite{herbert-readInferringRulesInteraction2011}. It obtains the attraction and repulsion forces from the average acceleration $\vec{a}_i$ of an individual $i$ as a function of its position relative to the nearest neighbour $NN$, $\vec{x}_i - \vec{x}_{NN}$, assuming this neighbour dominates the interaction. We can demonstrate it with a simple model of schooling fish, which we refer as \emph{standard model}. In this model individuals interact with some social forces, described by attraction-repulsion and alignment terms, and respond to some individual forces, represented by a friction-propulsion term to drive the motion and noise (for more details see \hyperref[sec:methods:standardModel]{Methods}). Employing simulations of the model, we display in Fig.~\ref{fig:force_maps:atracRepForce} the average attraction-repulsion force acting on a focal individual depending on its position relative to the the nearest neighbour. We use values for all focal individuals and time steps (counts are shown in Supplementary Fig.~\ref{supp:fig:basicModel_attRep}a). The components of the force are displayed with arrows and the modulus with the colormap.  The $y$-coordinate is oriented along the direction of motion of the focal individual. We observe for near distances arrows point outwards, indicating the focal fish experiences a repulsive force away from the neighbour. For intermediate distances there is an area with no net force (the equilibrium distance). Finally, for far away distances arrows point inwards, signaling an attractive force towards the neighbour. Now in Fig.~\ref{fig:force_maps:atracRep_basicModel} we display the attraction-repulsion force map given by the average acceleration for an individual depending on its relative position with the nearest neighbour. Indeed, we observe it recovers satisfactorily the attraction-repulsion force. This procedure works adequately because other forces do not depend directly on the relative position of the individual with the nearest neighbour and tend to cancel out in average (Supplementary Fig.~\ref{supp:fig:basicModel_attRep}b and c).

We note that previous works on the attraction-repulsion force map employed the
reversed coordinates $\vec{x}_{NN} - \vec{x}_i$, but they typically separated
the acceleration components in different plots and did not display the force
with arrows (except in~\cite{zienkiewiczDatadrivenModellingSocial2018}). Because
here instead we display the force as a vector, we find more natural to represent
it analogously to a force field from physics, where the force at a point
indicates the force a particle would feel at that point. Furthermore, we also
want to remark that, in absence of other forces, a trajectory observed in the
force map is more complex than just following the direction of the arrows, as
there will be rotations caused by the $y$-axis pointing towards the direction of
motion of the focal individual.

\subsubsection*{Alignment force map}

Previous works used force maps for the alignment force that depend on angular differences as well as on relative positions, which make the alignment force more difficult to disentangle from the attraction and repulsion forces. Here instead we introduce a more natural force map for alignment in Cartesian coordinates obtained from calculating the average acceleration $\vec{a}_i$ of an individual $i$ as a function of its relative velocity with the nearest neighbour $NN$, $\vec{v}_i - \vec{v}_{NN}$. This approach is directly inspired from the expression of the alignment force between neighbours in our standard model (see Eq.~\ref{eq:aligForce} in \hyperref[sec:methods:standardModel]{Methods}). It naturally accounts for the alignment of the full velocity vectors of neighboring individuals, thus not only their direction of movement but also of their speeds.

In Fig.~\ref{fig:force_maps:alignForce} we show the average alignment force acting on a focal individual depending on its relative velocity with the nearest neighbour for the standard model. We employ values for all focal individuals and time steps (counts are shown in Supplementary Fig.~\ref{supp:fig:basicModel_align}a). The components of the force are displayed with arrows and the modulus with the colormap.  The $y$-coordinate is oriented along the direction of motion of the focal individual. We observe the force vectors point inwards and increase with the velocity differences. This results in the force term decreasing the relative velocity and aligning the velocity of the focal individual with that of its neighbour. In Fig.~\ref{fig:force_maps:align_basicModel} we show the alignment force map given by the average acceleration in this representation. This force map also recovers successfully the qualitative behaviour of the alignment force, with arrows pointing inwards (this is clearly observed looking at the projection of the acceleration in the radial direction of the force map in Supplementary Fig.~\ref{supp:fig:radialForceMap:standardModel}). Again, this works because other forces of the model do not depend directly on the relative velocity of the individual with the nearest neighbour and have small contributions in this representation (Supplementary Fig.~\ref{supp:fig:basicModel_align}b and c).

\subsection*{Experimental force maps}

\begin{figure*}[t!p]
\centering
\subfloat[\label{fig:force_maps_exp:exp_atracRepModel}]{%
  \includegraphics[width=0.42\textwidth]{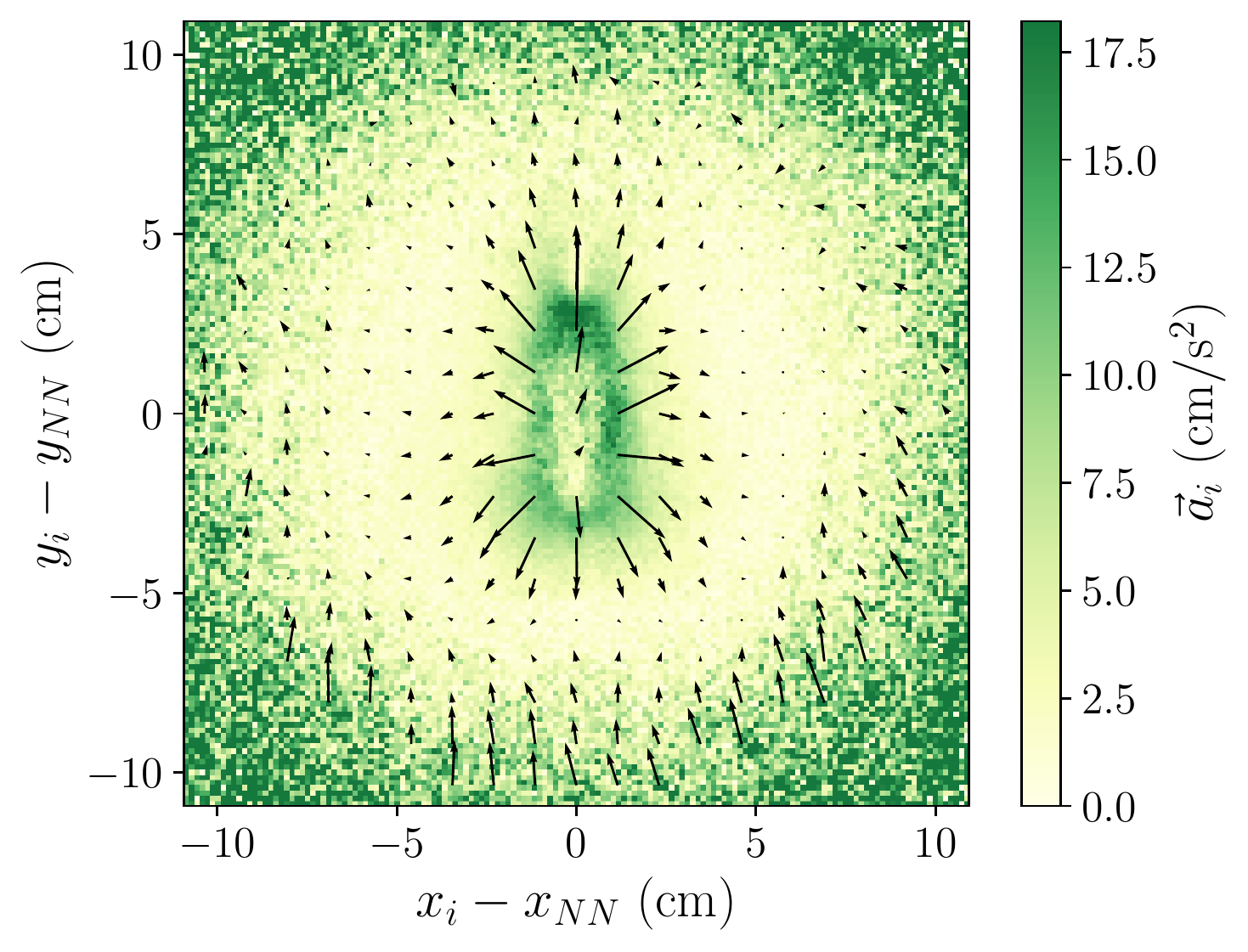}%
}
\hspace{0.01\textwidth}
\subfloat[\label{fig:force_maps_exp:aligExp}]{%
  \includegraphics[width=0.4\textwidth]{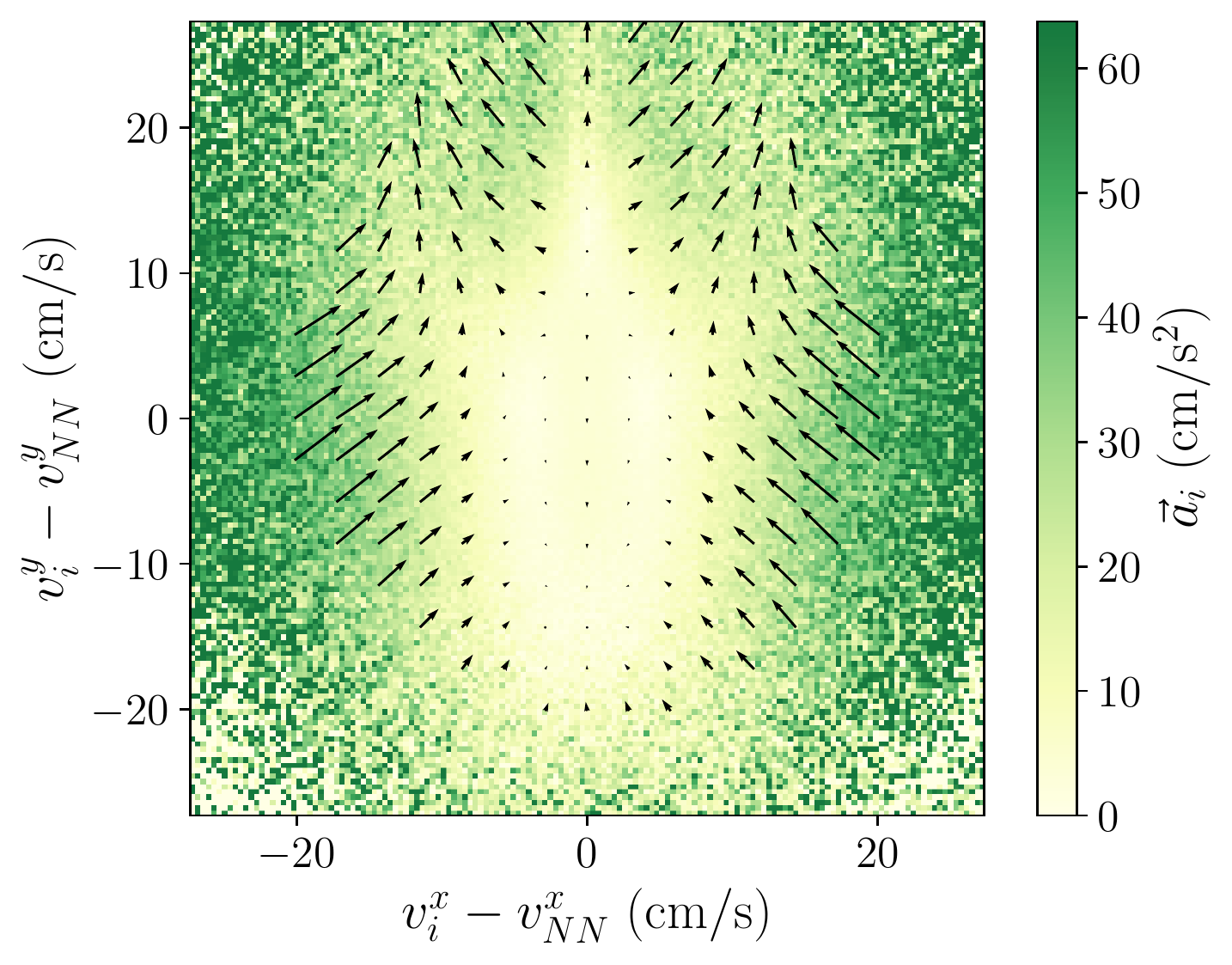}%
}
\caption{\textbf{Force maps for the experimental data:} (a) attraction-repulsion and (b) alignment force map. While the attraction-repulsion force map is similar to the standard model, the alignment force map displays a striking different behaviour: in the lower-half region arrows point inwards indicating alignment but in the upper-half region arrows point outwards indicating anti-alignment.
} \label{fig:force_maps_exp}
\end{figure*}

Equipped with the force maps technique outlined above, we are now able to
explore experimental data of schooling fish. With this purpose, we analyze the
motion of $N=39$ individuals of the species black neon tetra
\emph{Hyphessobrycon herbertaxelrodi}, a highly social fish that tends
  to form polarized, compact and planar schools, freely swimming in an
  approximately two-dimensional experimental tank (see
\hyperref[sec:methods:experimentalData]{Methods} for experimental details).

We start with the attraction-repulsion force map in Fig.~\ref{fig:force_maps_exp:exp_atracRepModel} (the counts of the force map are shown in Supplementary Fig.~\ref{supp:fig:force_maps_counts_exp:atracRep}). We see this force map has a similar qualitative behaviour to the attraction-repulsion force for the standard model (Fig.~\ref{fig:force_maps:atracRepForce}), with repulsion for short distances, no net force for intermediate distances and attraction at large distances.

Now we turn our attention to the alignment force map obtained from experimental data, displayed in Fig.~\ref{fig:force_maps_exp:aligExp} (the counts are shown in Supplementary Fig.~\ref{supp:fig:force_maps_counts_exp:alig}). Contrary to the alignment force from the standard model (Fig.~\ref{fig:force_maps:alignForce}), we can identify two qualitatively different behaviours depending on the sign of the difference of velocities between an individual and its nearest neighbour in the direction of motion of the individual ($y$-axis). If the neighbour is moving at faster speeds than the individual (lower-half part), we observe alignment interactions with the acceleration pointing inwards. However, when the neighbour moves at slower speeds than the individual (upper-half part), the acceleration points outwards and indicates a surprising anti-alignment interaction, with fish turning away from neighbours. This observation is made more apparent with the projection of the acceleration in the radial direction of the force map, in Supplementary Fig.~\ref{supp:fig:radialForceMap:experimentalData}.

We find that the qualitative behaviour of the alignment and anti-alignment regions in the alignment force map is robust against a variety of factors: different number of fish (Supplementary Fig.~\ref{supp:fig:alig_different_N}); different distances between neighbours (Supplementary Fig.~\ref{supp:fig:alig_Nn_dij}), which tests for possible effects originating from the attraction or repulsion regimes in the attraction-repulsion force; different distances to the tank walls (Supplementary Fig.~\ref{supp:fig:alig_dWall}), which discards possible interactions with the walls; for the case the fish is approaching or moving away from the neighbour (Supplementary Fig.~\ref{supp:fig:alig_approachingLeaving}); for different heading orientations between the fish and its neighbour (Supplementary Fig.~\ref{supp:fig:alig_headingOrientations}); and for different velocities of the focal individual (Supplementary Fig.~\ref{supp:fig:alig_modVel}). Only when differentiating the case where the neighbour moves at faster or slower speeds (Supplementary Fig.~\ref{supp:fig:alig_modVelNn_i}), we observe exclusively alignment or anti-alignment interactions in the expected regions of the map.

\subsection*{Explicit anti-alignment model}
We find the standard model of schooling fish reproduces properly the attraction-repulsion force map in the experimental data of schooling fish, but is unable to account for the alignment force map. To correct this, we now consider a model variation implementing explicitly the alignment force observed experimentally. We do so by defining an alignment force given by an interpolation of the values in the experimental alignment force map, which is previously smoothed with a Gaussian filter of window $\sigma = 2$. All other forces and parameters are kept the same as in the previous section.

\begin{figure*}[t!p]
\centering
\subfloat[\label{fig:alignmentForceMap_variations:expAlignModel_alignmentForceMap}]{%
  \includegraphics[width=0.4\textwidth]{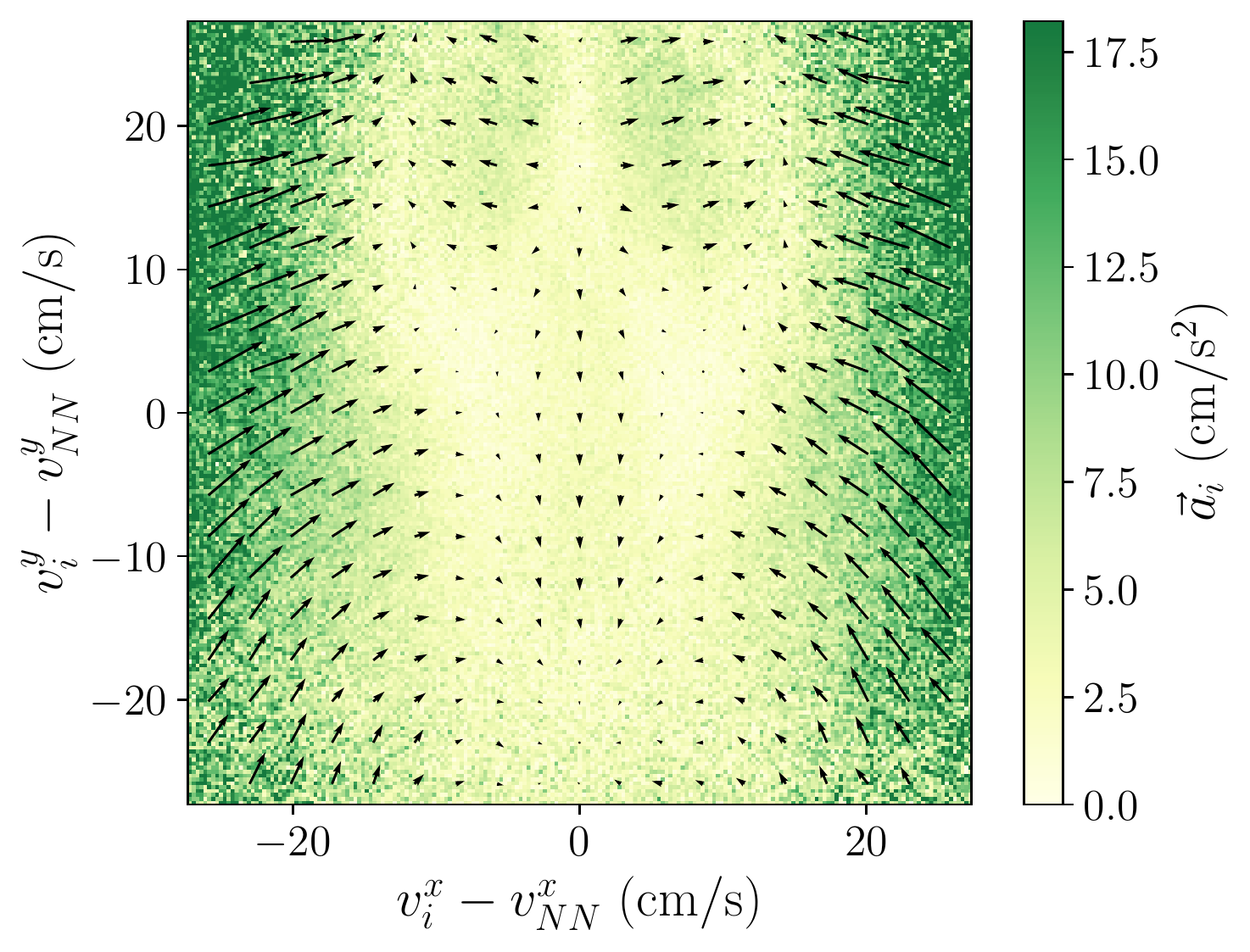}%
}
\hspace{0.01\textwidth}
\subfloat[\label{fig:alignmentForceMap_variations:selecModel_alignmentForceMap}]{%
  \includegraphics[width=0.4\textwidth]{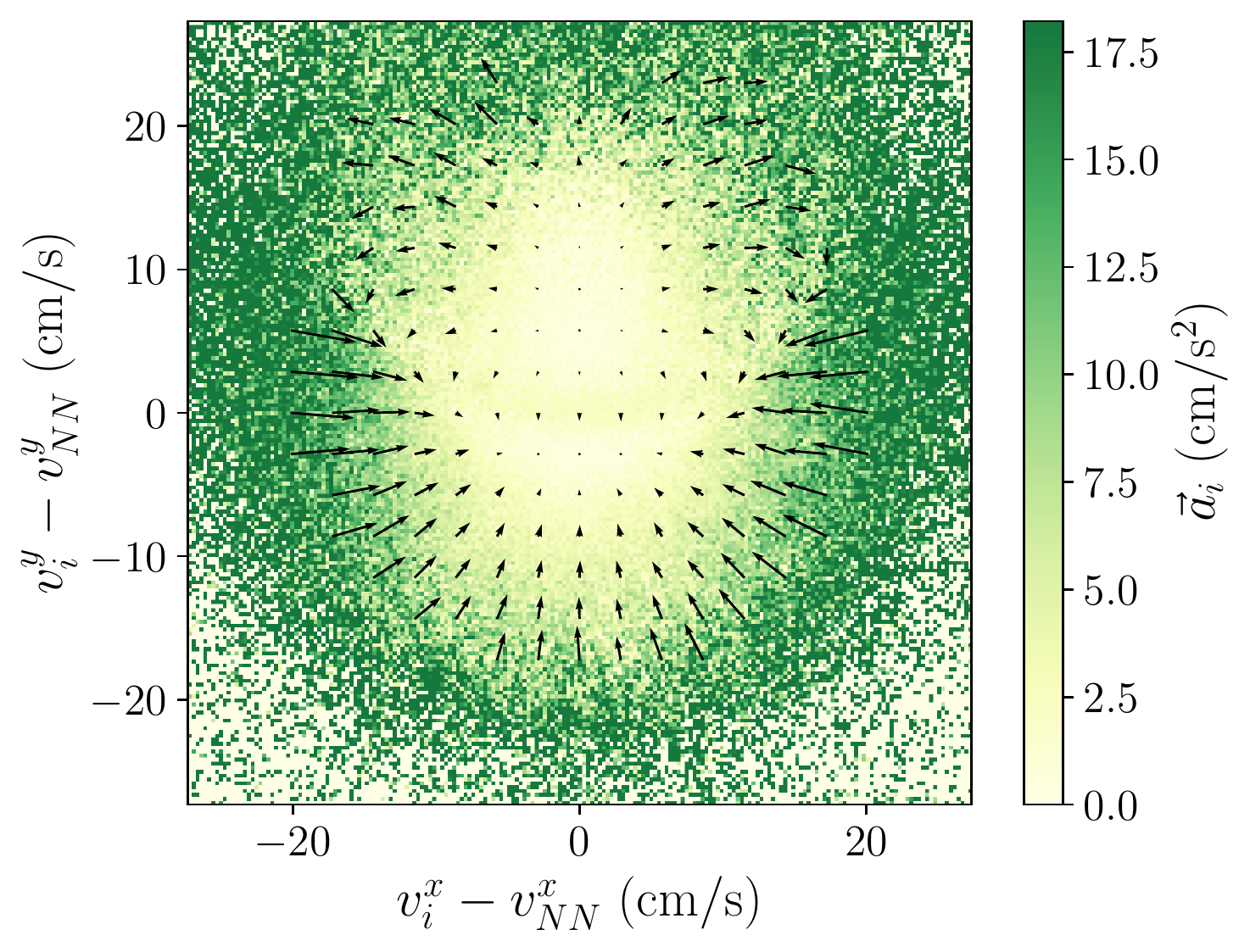}%
}
\caption{\textbf{Model variations with alignment and anti-alignment regions in the alignment force map.} Alignment force maps for (a) the explicit anti-alignment model and (b) the selective interactions model.} \label{fig:alignmentForceMap_variations}
\end{figure*}

Despite the model reproduces qualitatively the expected alignment force map (Fig.~\ref{fig:alignmentForceMap_variations:expAlignModel_alignmentForceMap} and Supplementary Figs.~\ref{supp:fig:expAligmentModel_align} and \ref{supp:fig:radialForceMap:explicitAntiAlignment}), the dynamics is unrealistic. When an individual is moving faster than its neighbours, it is located in the anti-alignment region of the force map and the explicit anti-alignment force causes the individual to accelerate in the forward direction, which in the absence of other forces will result in an even larger speed. This behaviour is observed in the simulations of the model (see Supplementary Video S3), where some individuals acquire large speeds and move towards the front of the group. However, this is not found in the experimental data and there is no biological motivation for fast fish to waste energy by increasing their speed further.

\subsection*{Selective interactions model}
The non-physical and non-biological features of the explicit anti-alignment model suggest that the anti-alignment region observed in the alignment force map of the experimental data may not originate from an explicit anti-alignment rule, but it could instead emerge from some other interaction that appears effectively as an anti-alignment force.

To explore this idea, we first study the effect of a persistent random force in the alignment force map. We propose a variation of the standard model, in which individuals have some probability rate $r$ to switch off their social interactions and feel instead a persistent force in a fixed random direction with a given amplitude $A$ for some time $\tau$. Biologically, this can be motivated as an individual fish may decide to follow private information based on environmental cues and temporarily ignore social cues. Depending on the relative strength of the persistent random force compared to the alignment force, we observe two different outcomes. When the alignment force dominates the persistent random force, we obtain an alignment force map that is still full alignment (Supplementary Fig.~\ref{supp:fig:persistentRandomForce_alig:acc}).
However, the persistent random force contributes to the total force with anti-alignment (Supplementary Fig.~\ref{supp:fig:persistentRandomForce_alig:randomForce}). We can understand this as in a group with overall high directional alignment, a persistent ``private'' force acting temporarily in a fixed random direction will tend to make the focal fish turn away from their neighbour's direction and appear effectively as an anti-alignment interaction. Naively, one could expect that in the regime when the persistent random force dominates the alignment force, the resulting alignment force map would be anti-alignment. However, that is not the case as there is not anymore ordered collective motion and the force map is essentially noise.

To connect this idea with the observed experimental alignment force map, we propose another variation of the standard model, a selective interactions model where individuals pay attention primarily to neighbours exhibiting strong movement cues. We assume that individuals interact socially with neighbours that move at a speed larger than their own, and switch off social interactions with neighbours that move at a smaller speed (see Supplementary Video S4). The weights in the social interactions are calculated as in the standard model (see \hyperref[sec:methods:standardModel]{Methods}) but considering only faster neighbours. We display the alignment force map in Fig.~\ref{fig:alignmentForceMap_variations:selecModel_alignmentForceMap} (see Supplementary Figs.~\ref{supp:fig:selInteractionsModel_align} and \ref{supp:fig:radialForceMap:selectiveInteractions}). We see this model reproduces the features observed in the experimental data,
with alignment for faster neighbours and anti-alignment for slower neighbours. The emergence of anti-alignment for slower nearest neighbours in this model can be traced back to the attraction-repulsion force
(see Supplementary Fig.~\ref{supp:fig:selInteractionsModel_align}). Although this force is switched off for slower neighbours, the focal individual will continue to interact with other non-nearest faster neighbours. Thus, when we construct the alignment force map for the slower nearest neighbours, the actual dynamics of the focal agent is driven by attraction to farther away neighbours, which may be located at random relative positions.
This force then can act in any direction and, therefore, appears effectively as a persistent random force. As we found previously, this results in individuals on average turning away from their nearest neighbour and displaying apparently an anti-alignment interaction.

\subsection*{Comparison of model observables}

\begin{figure*}[t!p]
\centering
\subfloat[\label{fig:observables:polarization}]{%
  \includegraphics[width=0.35\textwidth]{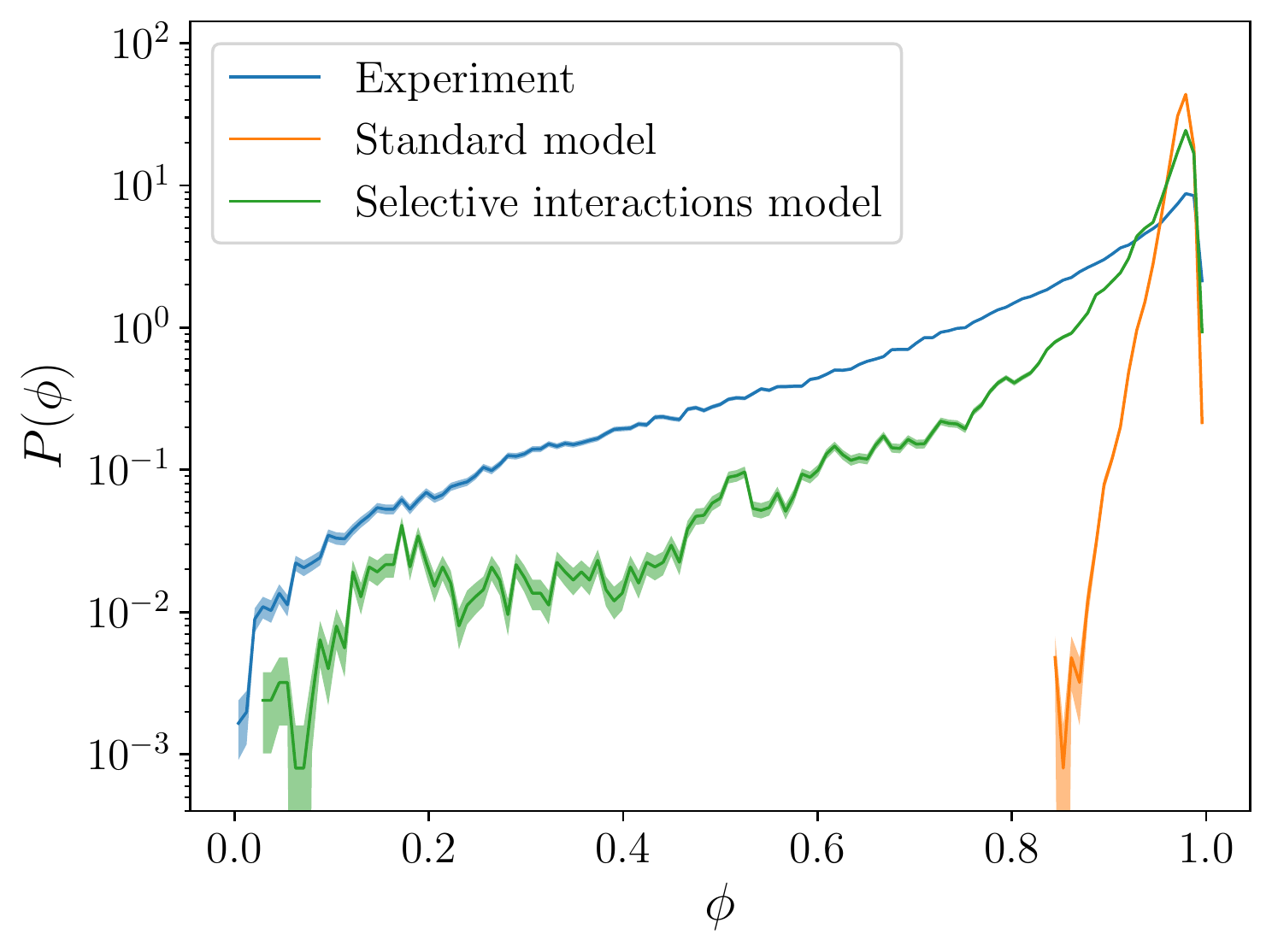}%
}
\hspace{0.01\textwidth}
\subfloat[\label{fig:observables:dNN}]{%
  \includegraphics[width=0.35\textwidth]{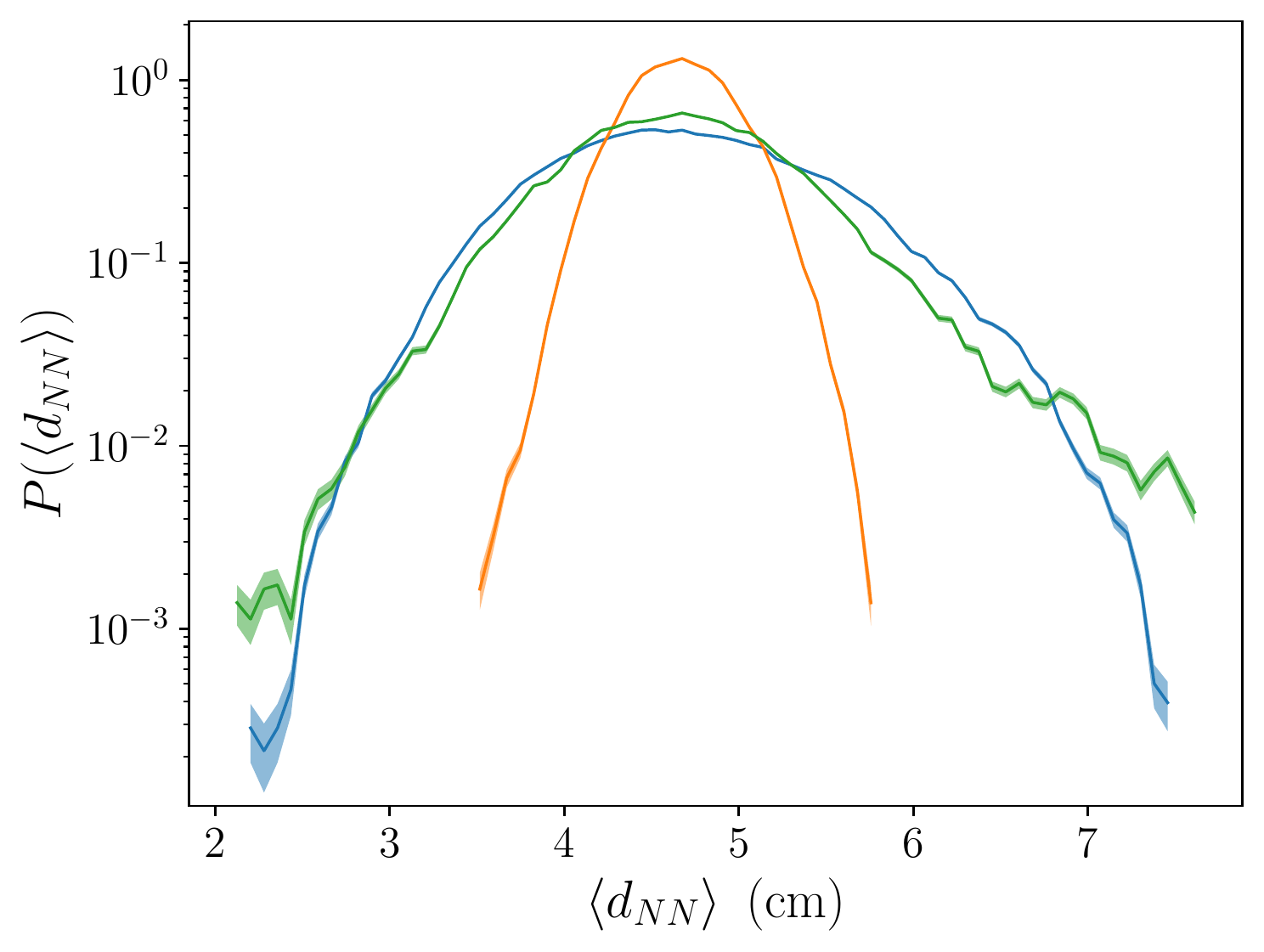}%
}

\subfloat[\label{fig:observables:areaCH}]{%
  \includegraphics[width=0.35\textwidth]{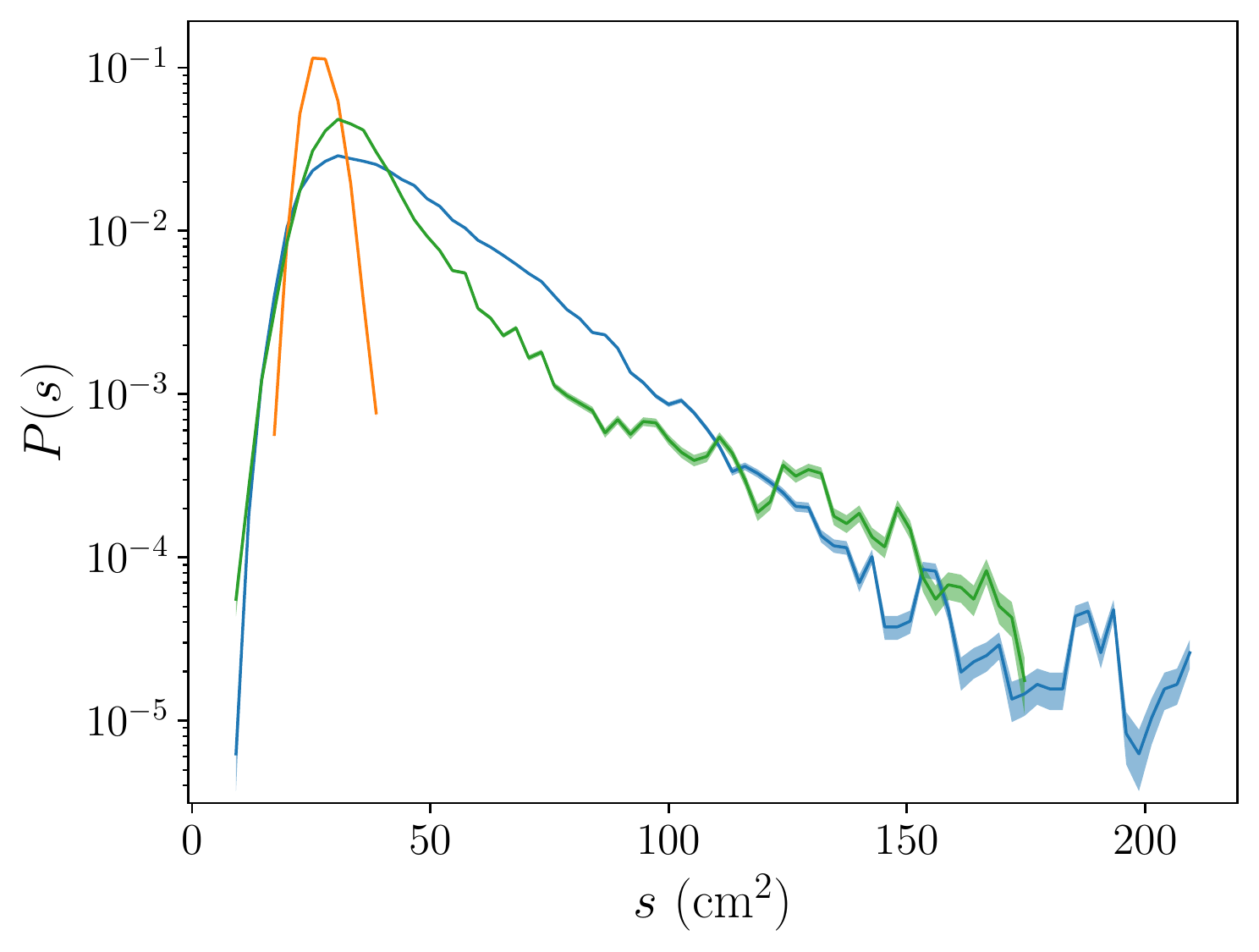}%
}
\hspace{0.01\textwidth}
\subfloat[\label{fig:observables:contactVoronoi}]{%
  \includegraphics[width=0.35\textwidth]{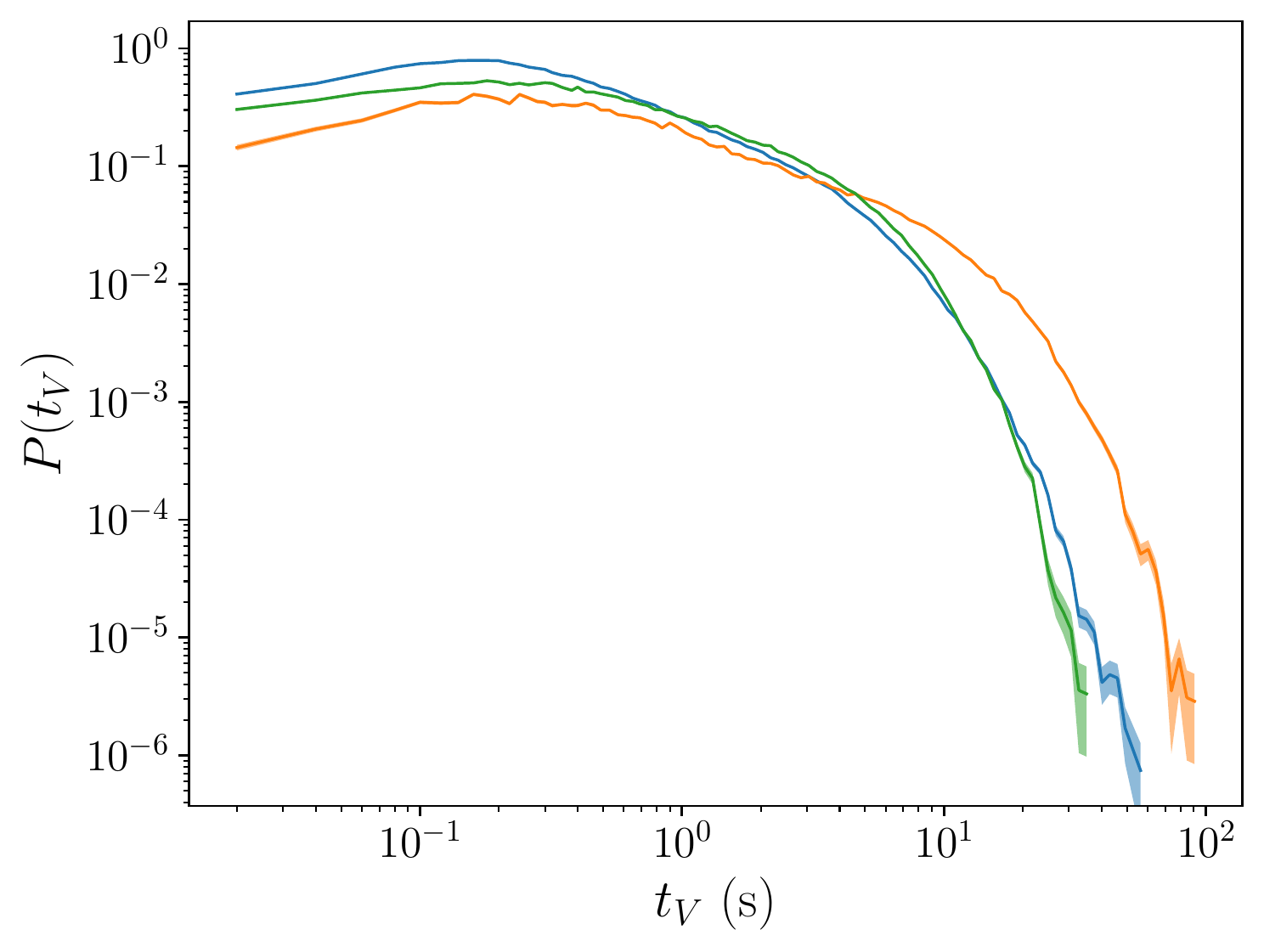}%
}
\caption{\textbf{The selective interactions model is more similar to the experimental data than the standard model for a set of observables.} Probability density functions (PDF) of (a) polarization $\phi$, (b) instantaneous average nearest neighbour distance $\left < d_{NN} \right >$, (c) area of the convex hull normalized by the number of individuals $s$ and (d) contact duration time between Voronoi neighbours $t_V$. The error bands are calculated from the standard deviation of a Bernoulli distribution, with the probability given by the fraction of counts in each bin.} \label{fig:observables}
\end{figure*}

Our analysis based on the alignment force map suggests the selective
interactions model as a more accurate depiction of the experimental data than
the standard model. Here, we extend our investigation by comparing the models
and the experimental data with a set of observables (Fig.~\ref{fig:observables}
and Supplementary Fig.~\ref{supp:fig:pdfs_comparison}). The
polarization $\phi$, defined as
\begin{equation*}
\phi \equiv \left | \frac{1}{N} \sum_i \frac{ \vec{v}  }{  v } \right|,
\end{equation*}
is a measure of the group order that tends to $1$ if all individuals move in the
same direction, and decreases when they move in increasingly different
directions. The nearest neighbour distance averaged for all individuals
$\left < d_{NN} \right >$ is a measure of the cohesiveness of the group. The
convex hull of the group is defined as the smallest convex polygon that contains
all individuals. Its area normalized by the number of individuals $s$
is a measure of the extension occupied by the group. The contact duration time
between Voronoi neighbours $t_V$ is a measure of the persistence of
the local neighbours structure. In addition we also consider
  individual speeds $v_i$ and turning rates $\omega_i$.

The selective interactions model outperforms the standard model in its ability to reproduce the various experimental observables: it is on average less ordered by displaying a broad distribution of order parameter values (with more low polarization values in Fig.~\ref{fig:observables:polarization}); it exhibits a larger variability in the nearest neighbour distances and in individual speeds (wider PDFs in Fig.~\ref{fig:observables:dNN} and Supplementary Fig.~\ref{supp:fig:pdfs_comparison:speeds}, respectively), more extended groups (with higher values for the convex hull areas in Fig.~\ref{fig:observables:areaCH}) and larger rates of neighbours shuffling (with shorter contact duration times of Voronoi neighbours in Fig.~\ref{fig:observables:contactVoronoi}). On the other hand, individual turning rates are similar for the models and the experimental data (Supplementary Fig.~\ref{supp:fig:pdfs_comparison:omega}), despite the unbounded space of the models result in less large turning rates. We also observe in Supplementary Fig.~\ref{supp:fig:pdfs_comparison:speed_temporal_evo} that both models display oscillations in individual speeds that resemble the oscillations in the experimental data, which are a consequence of the burst-and-coast mechanism of fish locomotion~\cite{weihsEnergeticAdvantagesBurst1974,
  harpazDiscreteModesSocial2017, liBurstandcoastSwimmersOptimize2021}. 
  
The similarities in a set of observables between the selective interactions model and the experimental data provide additional support for the selective interactions hypothesis. However, given the model's parsimonious approach, we do not anticipate it to account for many specific details of the system.

\subsection*{Leadership and alignment}
The above results suggest that the features of the experimental alignment map
may be explained by selective social interactions with faster individuals. In
this section, we want to validate this hypothesis, connecting it to the possible
presence of leader-follower relationships.

\begin{figure*}[t!p]
\centering
\subfloat[\label{fig:delayedAlignmentForceMap:diagram}]{%
  \includegraphics[width=0.35\textwidth]{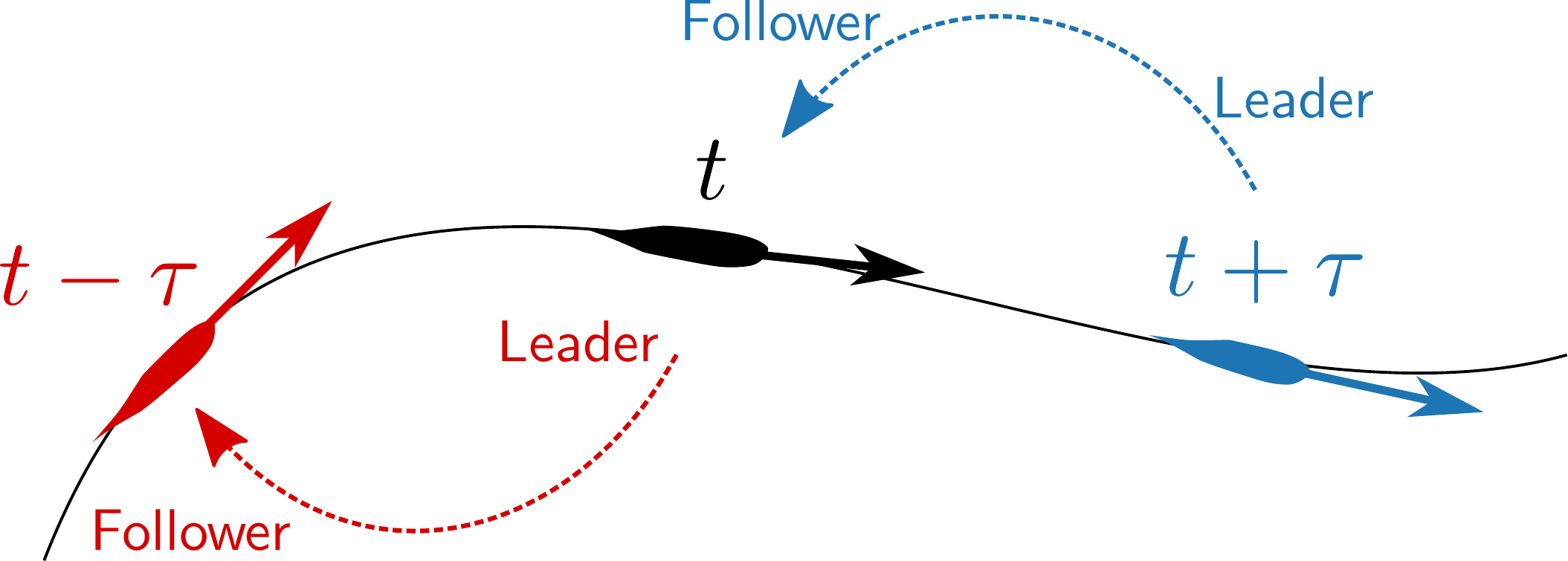}%
}

\subfloat[\label{fig:delayedAlignmentForceMap:pos}]{%
  \includegraphics[width=0.4\textwidth]{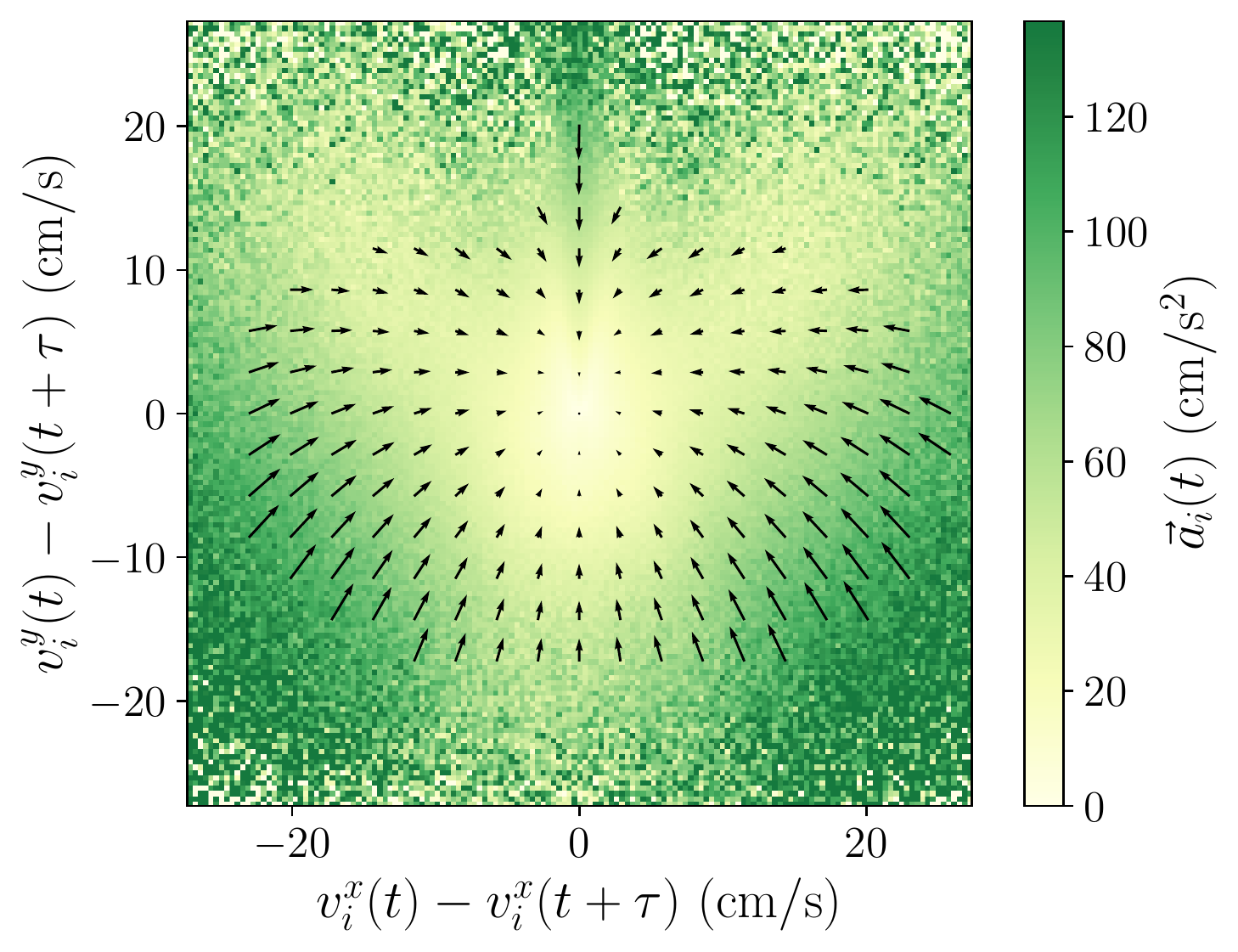}%
}
\hspace{0.01\textwidth}
\subfloat[\label{fig:delayedAlignmentForceMap:neg}]{%
  \includegraphics[width=0.4\textwidth]{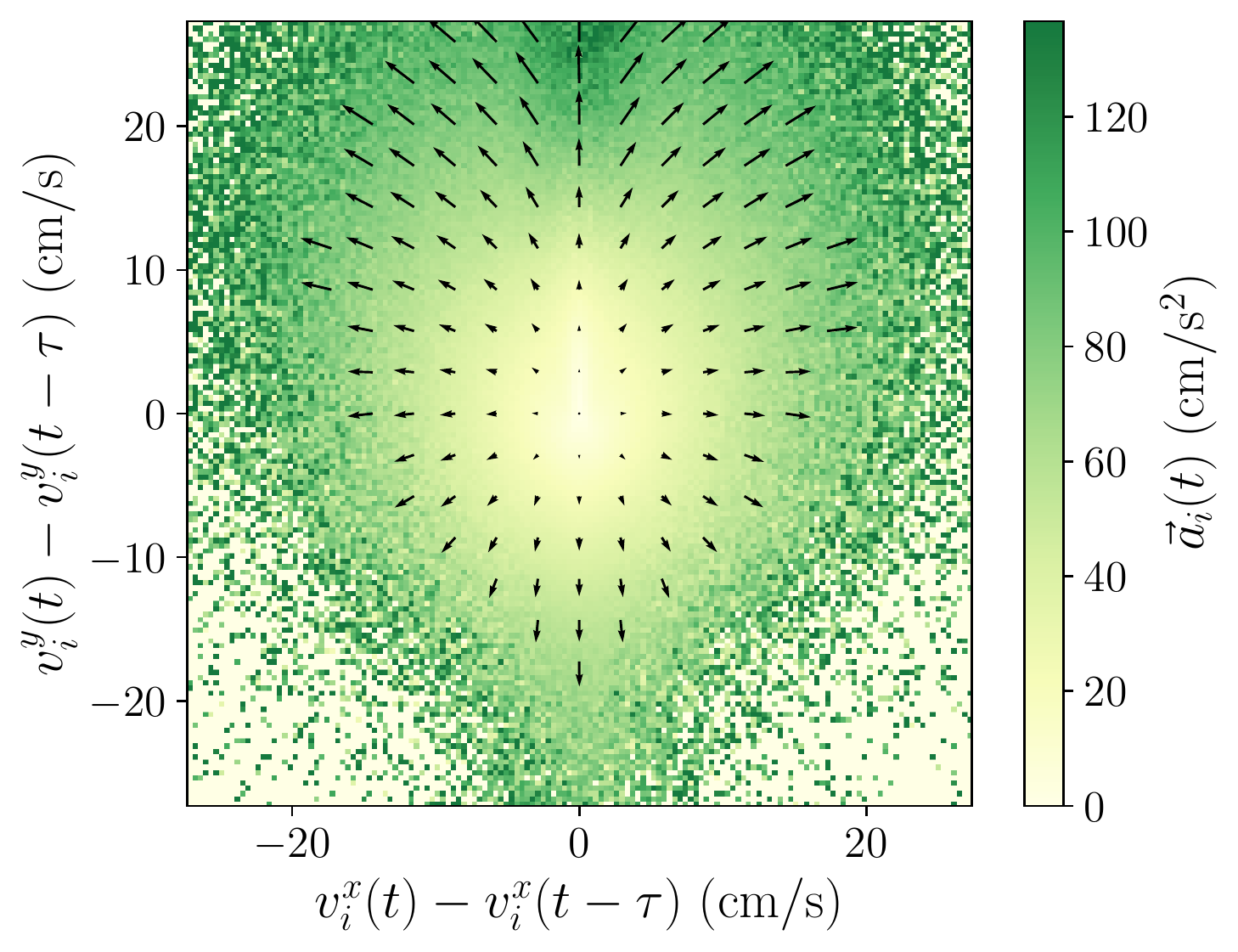}%
}
\caption{\textbf{Surrogate data of the experimental school with known leader-follower interactions}. (a) We compare the individual $i$ with itself at a delayed time $\tau$. (b)-(c): Alignment force maps comparing the individual $i$ with itself at a (b) positive and (c) negative delayed time $\tau=0.2$~s. When the delay is positive, the individual at present acts as a follower and we see pure alignment. When the delay is negative, the individual at present acts as a leader and we see pure anti-alignment.}\label{fig:delayedAlignmentForceMap}
\end{figure*}

We start by constructing surrogate data where we know a priori the flow of information between leaders and followers, and use the alignment force map to explore it. To achieve this, we use the following trick in the experimental data: for the focal individual we take the individual $i$ at the present time $t$, while for the neighbour we consider the same individual $i$ at a delayed time $t+\tau$. The situation is depicted in Fig.~\ref{fig:delayedAlignmentForceMap:diagram}. In the case of a positive delay $\tau$, the direction of the individual at present will follow that of the individual in the future after the delayed time, such that the individual at present can be interpreted as the follower within this pair. Now, when we observe the alignment force map for the individual at present (Fig.~\ref{fig:delayedAlignmentForceMap:pos} and Supplementary Figs.~\ref{supp:fig:counts_delayedAlignmentForceMap} and~\ref{supp:fig:radialForceMap:experimentalPositiveDelay}), we find alignment in all regions. This suggests a follower aligns with a leader. In the case of a negative delay, we have the opposite situation and the direction of the individual at present is adopted by the individual in the past after the delayed time and thus the individual at present can be interpreted as the leader. In the alignment force map (Fig.~\ref{fig:delayedAlignmentForceMap:neg} and Supplementary Figs.~\ref{supp:fig:counts_delayedAlignmentForceMap} and \ref{supp:fig:radialForceMap:experimentalNegativeDelay}), this results in anti-alignment in all regions, i.e. a leader appears to anti-align with a follower. However, this does not necessarily imply explicit anti-alignment, but originates rather from the leader ignoring the follower and moving in a different direction.

\begin{figure*}[t!p]
\centering
\subfloat[\label{fig:velocityCorrelations:direction}]{%
  \includegraphics[width=0.35\textwidth]{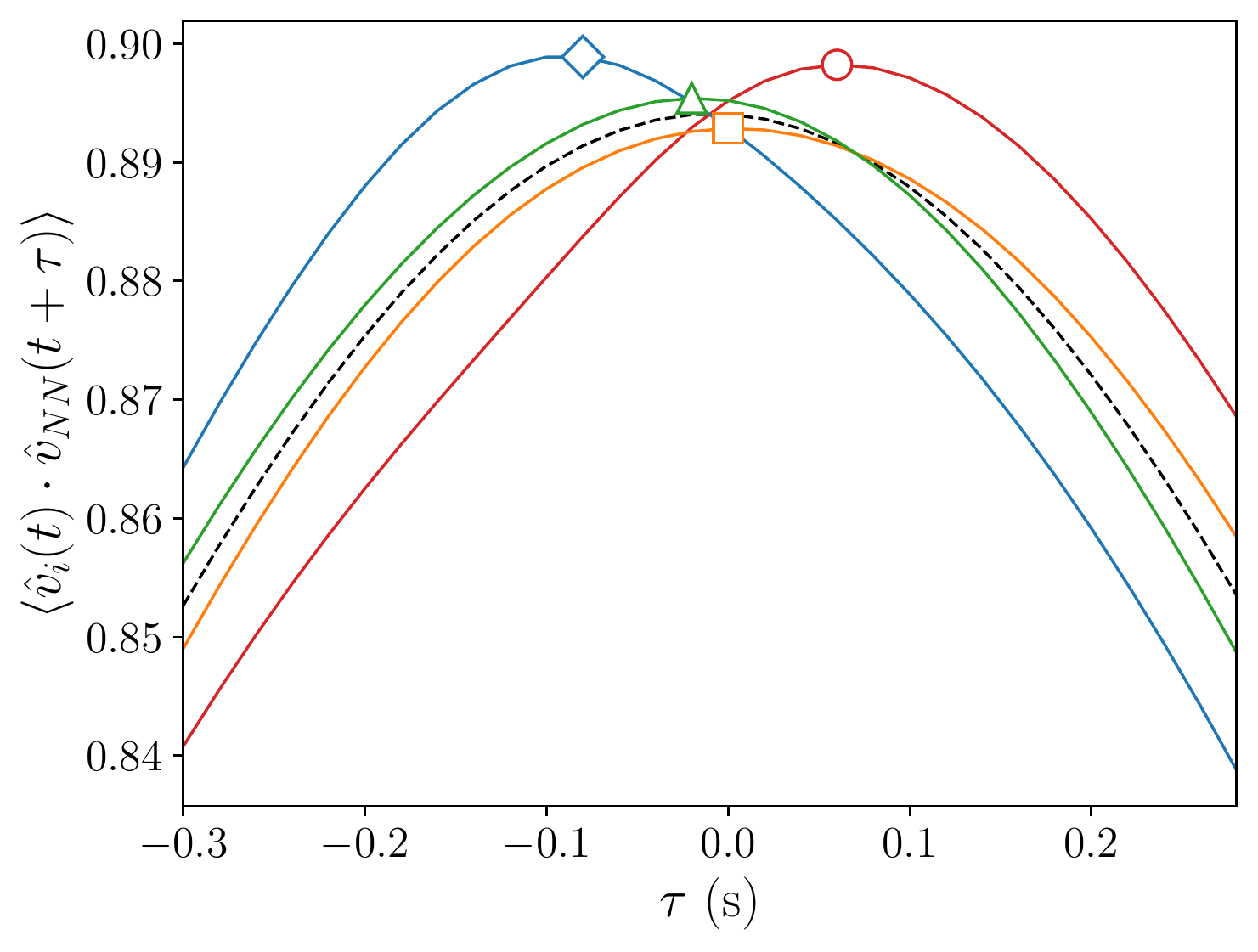}%
}
\hspace{0.01\textwidth}
\subfloat[\label{fig:velocityCorrelations:modulus}]{%
  \includegraphics[width=0.45\textwidth]{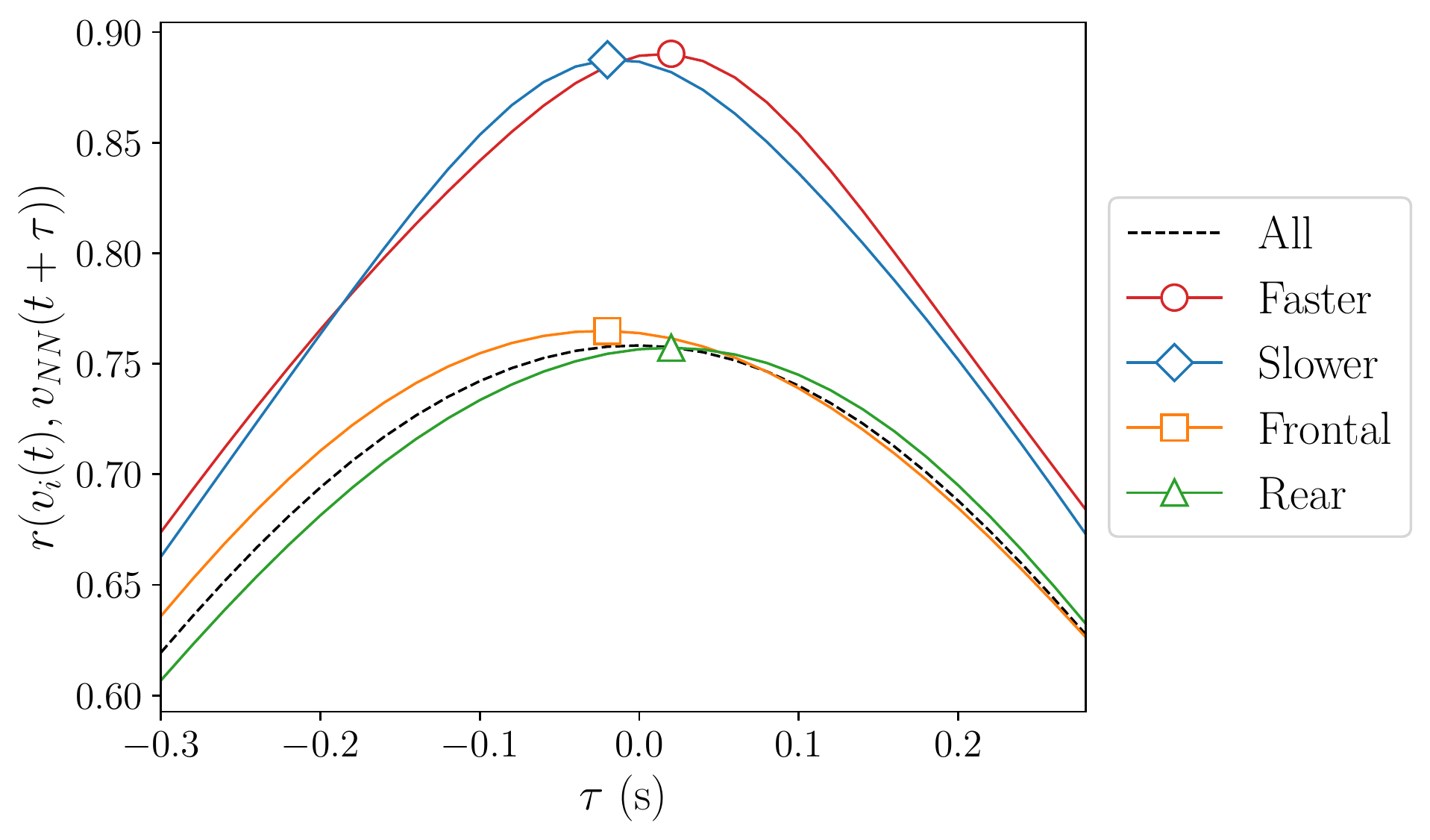}%
}
\caption{\textbf{Relative speeds in neighbours describe better leadership in schooling fish than frontal distances along the direction of motion of the individual.} Correlations in (a) the orientation and (b) speed (Pearson correlation coefficient $r$) between an individual $i$ and its nearest neighbour $NN$ at a delayed time $\tau$ for the experimental data. We filter for different states of the individual compared to the neighbour (see legend). The markers denote the maximum of each line. We find correlations are higher for faster and slower individuals than frontal and rear individuals. Frontal and rear individuals are more similar to the base line correlations given by all individuals.} \label{fig:velocityCorrelations}
\end{figure*}

Extrapolating this analysis to the experimental alignment force map (Fig.~\ref{fig:force_maps_exp:aligExp}) hints to the idea that individuals act as leaders when moving faster than their neighbours (anti-alignment region), while they act as followers when moving slower than their neighbours (alignment region). We verify this idea by calculating correlations between the velocities of the individual and the nearest neighbour at a delayed time in the experimental data. We consider two types of correlations, in the orientation given by $\left< \hat{v}_i (t) \cdot \hat{v}_{NN} (t+\tau) \right>$~\cite{nagyHierarchicalGroupDynamics2010} (Fig.~\ref{fig:velocityCorrelations:direction}) and in the speed given by the Pearson correlation coefficient $r(v_i(t), v_{NN}(t+\tau))$~\cite{katzInferringStructureDynamics2011} (Fig.~\ref{fig:velocityCorrelations:modulus}). Both measures are computed employing all time frames $t$. In the presence of leadership interactions, a follower requires some time to react and copy the leader. This results in these correlations having a higher value at that delayed time, provided the frame rate is higher than the reaction time of the follower. If the maximum occurs for a positive delay $\tau$, this implies the individual is the leader. While for a negative delay, the individual is the follower. Moreover, the value of the maximum reflects the amount of information transmitted by the leadership flow.

We are interested in studying correlations differentiating by the state of the individual. In Fig.~\ref{fig:velocityCorrelations}, the red and blue lines display the correlations performed for individuals faster and slower than their neighbour, respectively. Indeed, we see in both correlations that faster individuals have a maximum for positive delays, which means they are leaders. The opposite occurs with slower individuals, which act as followers. We also compare this measure of leadership with the commonly used measure of frontal distances along the direction of motion of the individual~\cite{nagyHierarchicalGroupDynamics2010, chenSwitchingHierarchicalLeadership2016}, filtering for frontal and rear individuals (orange and green lines). In fact, we find correlations have higher values for faster and slower individuals than for frontal and rear individuals, while frontal and rear individuals are more similar to the base line correlations considering all individuals (black dashed line). This indicates leadership in neighbours is better described by relative speeds rather than with frontal distances along the direction of motion. Note also that the faster and slower curves are not necessarily symmetrical along the $y$-axis, as several individuals can have the same nearest neighbour. The same occurs for frontal and rear, but in addition, a frontal and rear individual in the direction of motion of the individual can also be frontal and rear respectively in the direction of motion of the neighbour.

As an example of this phenomenon, in Supplementary Video S5 we display
  a sample of a recording where we observe how the focal individual (green)
  tends to align with faster neighbours (blue) and ignore slower neighbours
  (red).

\section*{Discussion}

In this paper we have introduced a force map to study the alignment force. Using this force map together with the  attraction-repulsion force map, we have been able to capture and distinguish the attraction-repulsion and alignment forces in a simple model of schooling fish. Applying it to experimental data, we found a robust signature of alignment of the focal individual with faster neighbours and anti-alignment with slower neighbours. Instead of explicit anti-aligning behavior, we suggest that the signatures of anti-alignment are the result of a selective attention mechanism, where fish pay less attention and, thus, have weaker interactions with slower neighbours. We were able to find support for this hypothesis by analysing the effective interactions in an agent-based model with selective social forces. While inferences from interactions in force maps are unlikely to be unique~\cite{escobedoDatadrivenMethodReconstructing2020}, we tested the hypothesis building a connection between the alignment patterns and leadership interactions. We show that followers display only alignment force and leaders only anti-alignment force. We then verified in the experimental schooling fish that faster and slower individuals are leaders and followers respectively with their nearest neighbours by calculating
correlations between the velocities of the individual and the nearest neighbour at a delayed
time. This measure of leadership is found to be more statistically significant than the traditionally used in the literature of frontal distances along the direction of motion~\cite{bumannFrontIndividualsLead1993, nagyHierarchicalGroupDynamics2010, chenSwitchingHierarchicalLeadership2016}.

Our findings of selective social interactions based on relative
  individual speeds provide further evidence of attention switch mechanisms
  previously described in collective
  behaviour~\cite{romanczukCollectiveMotionDue2009,
    lemassonCollectiveMotionAnimal2009,
    lemassonMotionguidedAttentionPromotes2013,
    sridharGeometryDecisionmakingIndividuals2021,
    oscarSimpleCognitiveModel2023}, as well as in the similar field of animal
  contests over resources~\cite{halutsSpatiotemporalDynamicsAnimal2021,
    halutsModellingAnimalContests2023}. Attention switch mechanisms are believed
  to reduce the cognitive load of individuals and facilitate efficient and
  accurate decision-making processes~\cite{lemassonCollectiveMotionAnimal2009,
    lemassonMotionguidedAttentionPromotes2013,
    sridharGeometryDecisionmakingIndividuals2021}. Our experimental species
  black neon tetra displays a strong schooling behavior. Thus we expect our
  results to be representative of other fish species forming highly polarized
  schools. In any case, the methods introduced in this work are general and
  therefore, applicable to other scenarios.


If we interpret social interactions between individuals as a method to transfer information across the group, these results suggest higher relevance of information provided by faster moving individuals~\cite{romero-ferreroStudyTransferInformation2023}. This makes biological sense as informed individuals acting on privately acquired information, such as perception of a predator or food, are expected to move at higher speeds \cite{rosenthalRevealingHiddenNetworks2015}. More generally, motion cues are believed to play a major role in visual sensory processes of organisms, conveying information that can directly impact the individual, such as in navigation, foraging, courtship or predator avoidance~\cite{lemassonMotionguidedAttentionPromotes2013, shettleworthCognitionEvolutionBehavior2010}; but also as faster movement cues, embedded in a background of slower moving ones, are perceptually more salient~\cite{ivryAsymmetryVisualSearch1992, wolfeWhatAttributesGuide2004, lemassonCollectiveMotionAnimal2009}. Apart from vision, the lateral line, with which fish detect movement, vibration, and pressure gradients in the surrounding water, is also reactive to neighbouring individual speeds~\cite{partridgeSensoryBasisFish1980, coombsLateralLineSystem2014}. It is important to emphasize that, in the context of these latter hypotheses, the reported selective interactions would not be attributed to an adaptive selective attention mechanism, but would instead emerge from cognitive constraints related to receiving information from slower-moving individuals.

While previous research suggests that individual speeds play an important role in shaping collective movement, and that leader-follower interactions are mediated by speeds, the proposed mechanism was based on individuals with different average speeds, assuming homogeneous and constant social interactions~\cite{pettitSpeedDeterminesLeadership2015,
gimenoLeadershipCollectiveMotion2018, jollesConsistentIndividualDifferences2017, jollesGrouplevelPatternsEmerge2020}. Having different average speeds induce sorting of individuals and ultimately give rise to the well-established leadership of frontal distances along the direction of motion~\cite{pettitInteractionRulesUnderlying2013}. With the current work we take a step forward connecting speed and leadership. The presence of selective social interactions based on instantaneous relative speeds implies a dynamic modulation of social interactions, which not only amplifies the importance of individual speed, but also provides a mechanism for dynamic adaptation of leader-follower relationships on short time-scales independently of the average preferred speed of individuals.

In the calculation of force maps, we considered only the relative position and velocity of a single, nearest neighbour. While this is supported by 1-2 influential neighbours identified in previous work on fish schools \cite{herbert-readInferringRulesInteraction2011, jiangIdentifyingInfluentialNeighbors2017, leiComputationalRoboticModeling2020}, one could, in principle, also use the relative position and velocity of all Voronoi neighbours (Supplementary Fig.~\ref{supp:fig:force_maps_Voronoi}). While we observe partial screening of attractive interactions in the attraction-repulsion force map, the force maps with Voronoi neighbours resemble qualitatively the force maps with the nearest neighbour. This suggests that the nearest neighbour dominates the interactions. We also find support for this idea when we explored simulations with individuals interacting only with the nearest neighbour and found similar results (Supplementary Fig.~\ref{supp:fig:force_maps_simulation_Nn}); and for simulations with individuals interacting with Voronoi neighbours without weights that resulted in more inadequate force maps with some screening effects. 

Force map techniques to infer underlying interactions are straightforward and easy to apply. However, interpreting the results requires caution, and implementing certain validation procedures is advised. In this work, we have presented a force map to extract the alignment force between individuals in experimental data of schooling fish. A first step in validating the results involves ensuring the force map is unaffected by other interactions. We approached this by identifying potential confounding factors such as the influence of tank walls, the attraction-repulsion force, and the group's density. We then demonstrated that the results were consistent under changing conditions of distances to the walls, distances and orientations to neighbours, and employing schools with differing numbers of fish. A secondary validation measure consists of applying the force map to simulations of models with controlled interactions and comparing the findings. We found that a simple model of schooling fish does not reproduce the experimental signatures and some modifications had to be introduced. Finally, to conclusively validate the results, it is recommended to employ other independent methods to infer interactions. In our study, we accomplished this by analyzing leader-follower relationships. Other possible methods include fitting simple expressions from symmetrical constraints and simplifying hypotheses \cite{caloviDisentanglingModelingInteractions2018}, and training neural networks \cite{herasDeepAttentionNetworks2019}.

Our results showcase how selective social interactions may lead to unexpected signatures at the level of averaged behavioral responses to social cues. It is of relevance for quantitative analysis of social interactions across taxa, as it demonstrates that extreme caution is needed when extracting behavioral algorithms from aggregated data such as force maps, in particular in the context of complex self-organized collective behaviors. Here, a combination of quantitative data as well as theoretical modelling taking into account properties and constraints of perception will provide deeper insights into social interactions and functional aspects of collective behavior with respect to transfer and distributed processing of information. More specifically, selective social interactions depending on relative speeds likely play an important role for different grouping phenomena~\cite{romanczukCollectiveMotionDue2009}, and provide an explicit mechanism on how speed may be a core variable at the individual level affecting the dynamics and the structure of animal collectives \cite{jollesConsistentIndividualDifferences2017, kentSpeedmediatedPropertiesSchooling2019, jollesGrouplevelPatternsEmerge2020,jollesRoleIndividualHeterogeneity2020,klamserCollectivePredatorEvasion2021}.

\section*{Data availability}

The experimental datasets can be accessed on Zenodo~\cite{puySelectiveSocialInteractions2024a}: \url{https://zenodo.org/records/10890112}. The code to run the models and reproduce the figures is available on GitHub: \url{https://github.com/Puco4/SelectiveSocialInteractions}.

\section*{Materials and methods}

\subsection*{Standard model}\label{sec:methods:standardModel}
We consider a model of schooling fish in two-dimensions with $N$ individuals
where each individual $i$ is subject to a total force $\vec{F}_i$ given by
social forces, represented by attraction-repulsion $\vec{F}^\text{att-rep}_i$
and alignment $\vec{F}^\text{alig}_i$ forces, and individual forces, consisting
of a friction-propulsion force $\vec{F}^\text{fric-prop}_i$ and noise
$\vec{F}^\text{noise}_i$ that capture other unknown forces
\cite{romanczukActiveBrownianParticles2012}:
\begin{equation*}
    \vec{F}_i = \vec{F}^\text{att-rep}_i + \vec{F}^\text{alig}_i + \vec{F}^\text{fric-prop}_i + \vec{F}^\text{noise}_i.
\end{equation*}
For simplicity, we use a first order approximation for the different force
terms.

The attraction-repulsion force acting on an individual $i$ due to the neighbour $j$ reads:
\begin{equation}
    \vec{F}^\text{att-rep}_{ij} = - k \left ( d_{ij} - d_{0} \right) \frac{\vec{x}_i - \vec{x}_j}{d_{ij}},\label{eq:attRepForce}
\end{equation}
where $d_{0}$ is the equilibrium distance between neighbours, $d_{ij} = \left | \vec{x}_j - \vec{x}_i \right|$ is the distance between individuals $i$ and $j$, $\vec{x}_i$ is the position of individual $i$ and $k=k_\text{rep}$ for $d_{ij}\le d_{0}$ and $k=k_\text{att}$ for $d_{ij}>d_{0}$. This force acts as a restoring force towards the equilibrium distance between the neighbours.

The alignment force experienced by an individual $i$ due to its neighbour $j$ is given by~\cite{klamserImpactVariableSpeed2021}:
\begin{equation}
    \vec{F}^\text{alig}_{ij} = -\mu(\vec{v}_i - \vec{v}_j),\label{eq:aligForce}
\end{equation}
where $\mu$ is the alignment constant and $\Vec{v}_{k}$ are individual velocities. This force acts as a restoring force for the focal individual towards reaching the same speed and heading as its neighbour.

We use interactions with Voronoi neighbours, which are a good approximation to visual interaction networks  and can be calculated efficiently~\cite{strandburg-peshkinVisualSensoryNetworks2013, gautraisDecipheringInteractionsMoving2012, ginelliIntermittentCollectiveDynamics2015, klamserImpactVariableSpeed2021}, and we average the social forces, $\vec{F}^\text{social}_{ij} = \vec{F}^\text{att-rep}_{ij} + \vec{F}^\text{alig}_{ij}$, for the different neighbours with weights given by the inverse of their metric distance $d_{ij}$:
\begin{equation*}
    \vec{F}^\text{social}_i = \sum_j w_{ij} \vec{F}^\text{social}_{ij},
\end{equation*}
where $w_{ij} = \frac{1}{d_{ij}}$. With this choice of weights, closer neighbours exert more influence, a result observed in previous studies~\cite{herbert-readInferringRulesInteraction2011, rosenthalRevealingHiddenNetworks2015, sosnaIndividualCollectiveEncoding2019, herasDeepAttentionNetworks2019}.

For the individual forces, we use a friction-propulsion term~\cite{klamserImpactVariableSpeed2021}:
\begin{equation*}
    \vec{F}^\text{fric-prop}_i = -\frac{v_i-v_0}{\tau} \hat{v}_i,
\end{equation*}
where $v_0$ is the fish preferred speed and $\tau$ is the speed relaxation time. This introduces a self-propulsion force for the fish to move with a speed around $v_0$, with $\tau$ being the typical timescale of velocity adaptation. The noise force corresponds to active fluctuations~\cite{romanczukBrownianMotionActive2011, klamserImpactVariableSpeed2021}, with independent noise terms acting on the direction of motion $\hat{v}_i$ and its perpendicular direction $\hat{\varphi}_i$:
\begin{equation*}
    \vec{F}^\text{noise}_i = \sigma_v \xi_v(t) \hat{v}_i + \sigma_\varphi \xi_\varphi(t) \hat{\varphi}_i,
\end{equation*}
where $\sigma_v$ and $\sigma_\varphi$ are the noise constants, and $\xi_v(t)$ and $\xi_\varphi(t)$ are independent Gaussian white noise processes with $\langle \xi_x(t)\xi_y(t')\rangle=\delta(t-t')\delta_{xy}$.

We perform simulations integrating numerically the equations of motion in order to obtain the trajectory $\vec{x}_i$:
\begin{equation*}
    \Vec{a}_i = \Vec{F}_i,\label{eq:newton2nLaw}
\end{equation*}
where $\Vec{a}_i = \frac{d^2\vec{x}_i}{dt^2}$ is the acceleration, and we work
with units of mass $m=1$. The values of the parameters in the model,
  shown in Table~\ref{tab:parameters}, have been selected in order to provide
  similar results to our experimental data, as described in the Supplementary
  Information.

Here we work with $N=39$ individuals, matching the maximum number of fish in our
experimental data. We solve the stochastic differential equation with the
Euler-Maruyama algorithm~\cite{toralStochasticNumericalMethods2014}.  We choose
random initial positions and velocities for the individuals such that the
extension of the group and speeds are similar to the experimental school, but
each individual is oriented uniformly at random. The simulation space
  is unbounded and positions can reach any value. We set a time step $\Delta t = 0.02$~s and save all frames, analogous to the experimental data. We discard as transient the
first 1000 frames and run simulations for 150000 frames. We show in
Supplementary Video S2 samples of trajectories generated by the
standard model.

\begin{table}[b]
\centering
\caption{Parameters used in the standard model. We work in units of mass $m=1$.}\label{tab:parameters}
\begin{tabular}{ l @{\qquad} l @{\qquad} l }
\hline
     Parameter & Description & Value \\
     \hline
     $k_\text{rep}$ & Repulsive constant &  $12.5~\textrm{s}^{-2}$ \\
     $k_\text{att}$ & Attractive constant &   $5~\textrm{s}^{-2}$  \\
     $d_{0}$ & Equilibrium distance &  $5.8~\textrm{cm}$ \\
     $\mu$ & Alignment constant &  $1.5~\textrm{s}^{-1}$ \\
     $v_0$ & Preferred speed &  $11~\textrm{cm}/\textrm{s}$ \\
     $\tau$ &  Speed relaxation time & $1.6~\textrm{s}$ \\
     $\sigma_v$ & Noise constant in $\hat{v}_i$ & $6.4~\textrm{cm}/\textrm{s}^{3/2}$ \\
     $\sigma_\varphi$ & Noise constant in $\hat{\varphi}_i$ &$2.6~\textrm{cm}/\textrm{s}^{3/2}$\\
\hline
\end{tabular}
\end{table}

\subsection*{Experimental data}\label{sec:methods:experimentalData}
Experiments, performed at the Scientific and Technological Centers UB (CCiTUB),
University of Barcelona (Spain), were made with black neon tetra
\textit{Hyphessobrycon herbertaxelrodi}, a small freshwater fish of average body
length 2.5 cm that has a strong tendency to form cohesive, highly polarized and planar
  schools~\cite{gimenoDifferencesShoalingBehavior2016}. Applicable
institutional guidelines for the care and use of animals were followed. All
procedures performed were in accordance with the standards approved by the
Ethics Committee of the University of Barcelona.

Experiments consisted in $N$ individuals freely swimming in a square tank of 100
cm of side with a water column of $5$~cm of depth, resulting in
  approximately two-dimensional movement. Videos of the fish movement were
recorded at 50 frames per second with a resolution of 5312×2988 pixels,
such that the side of the tank measures $2745$ pixels. Digitized
individual trajectories were obtained from the video recordings using the
software idtracker.ai~\cite{romero-ferreroIdtrackerAiTracking2019}.  Invalid
values returned by the program caused by occlusions were corrected in a
supervised way, semi-automatically interpolating with spline functions (now
incorporated in the Validator tool from version 5 of idtracker.ai). For better
accuracy, we projected the trajectories in the plane of the fish movement,
warping the tank walls of the image into a proper square (see Supplementary
Information). Moreover, we smooth the trajectories with a Gaussian filter with
$\sigma = 2$, truncating the filter at $5\sigma$. We calculate the velocities
and accelerations from the Gaussian filter using derivatives of the Gaussian
kernel.

We performed experiments with $N=39$ fish (2 recordings of 60 minutes) and $N=8$
(6 recordings of 30 minutes). In Supplementary Video S1 we show a rendering of
of the evolution of the $N=39$ school with the digitized trajectories overlapped
with the real video.


\begin{acknowledgments}
A.P. acknowledges a fellowship from the Secretaria d’Universitats i
  Recerca of the Departament d’Empresa i Coneixement, Generalitat de Catalunya,
  Catalonia, Spain and a scholarship Erasmus+. P.B. and P.R. acknowledge funding
  by the Deutsche Forschungsgemeinschaft (DFG, German Research Foundation) under
  Germany’s Excellence Strategy – EXC 2002/1 ``Science of Intelligence'' –
  project number 390523135. M.C.M. acknowledges financial support from
    the Spanish MCIN/AEI/10.13039/501100011033, under Projects
    No. PID2019-106290GB-C22 and No. PID2022-137505NB-C22. R.P.-S. acknowledges
    financial support from the Spanish MCIN/AEI/10.13039/501100011033, under
    Projects No. PID2019-106290GB-C21 and No. PID2022-137505NB-C21.   We  thank  J.  Mugica  for  insightful comments, and F. S. Beltr\'{a}n and V. Quera, from the Institute of Neurosciences,
  University of Barcelona (Spain), for invaluable help in the experimental
  process.
\end{acknowledgments}

\section*{Author contributions}
A.P., P.B., M.C.M., R.P.-S. and P.R. designed the study. E.G. and J.T. acquired the empirical data. J.T. and A.P. processed the empirical data. A.P. analysed the empirical data and performed the numerical simulations. A.P., M.C.M., R.P.-S. and P.R. analysed the results. A.P., R.P.-S. and P.R. wrote the paper. All authors commented on the manuscript.

\bibliography{all}


\clearpage

\renewcommand{\thepage}{\arabic{page}} 
\renewcommand{\theequation}{S\arabic{equation}} 
\renewcommand{\thesection}{\arabic{section}}  
\renewcommand{\thetable}{S\arabic{table}}  
\renewcommand{\thefigure}{S\arabic{figure}}
\renewcommand{\thevideo}{S\arabic{video}}


\setcounter{page}{1}
\setcounter{section}{0}
\setcounter{figure}{0}
\setcounter{equation}{0}
\onecolumngrid
\begin{center}
\textbf{\large SUPPLEMENTARY INFORMATION\\~\\}
\end{center}

\twocolumngrid

\section*{Standard model parameters selection}
Our aim with the standard model is to describe with few generic (linear)
interaction terms the qualitative behaviour of schooling fish, rather than
matching quantitative aspects. In this line of thought, instead of fitting the
parameters to some goal function we use values at the expected order of
magnitude comparing with the experimental data of schooling fish. We also
confirmed that our qualitative results remain unchanged upon variation of these
parameters.

The constants $k_\text{rep}$ and $k_{\text{att}}$ are chosen by comparing the pairwise attraction-repulsion force (Eq.~\ref{eq:attRepForce}) to the attraction-repulsion force map (Fig.~\ref{fig:force_maps_exp:exp_atracRepModel}). In particular, we find $k_\text{rep} > k_\text{att}$, as fish may prioritize repulsion to attraction~\cite{couzinCollectiveMemorySpatial2002}.

The equilibrium distance  $d_{0}$ is fixed such that numerical simulations match the average distance between nearest neighbors observed in experimental data  (Fig.~\ref{fig:observables:dNN}).

The alignment constant $\mu$ is taken analogously comparing the pairwise alignment force map (Eq.~\ref{eq:aligForce}) to the alignment region in the alignment force map (Fig.~\ref{fig:force_maps_exp:aligExp}).

\begin{figure*}[t!p]
\subfloat[\label{supp:fig:alig_modVel:modVelPDF}]{%
  \includegraphics[width=0.33\textwidth]{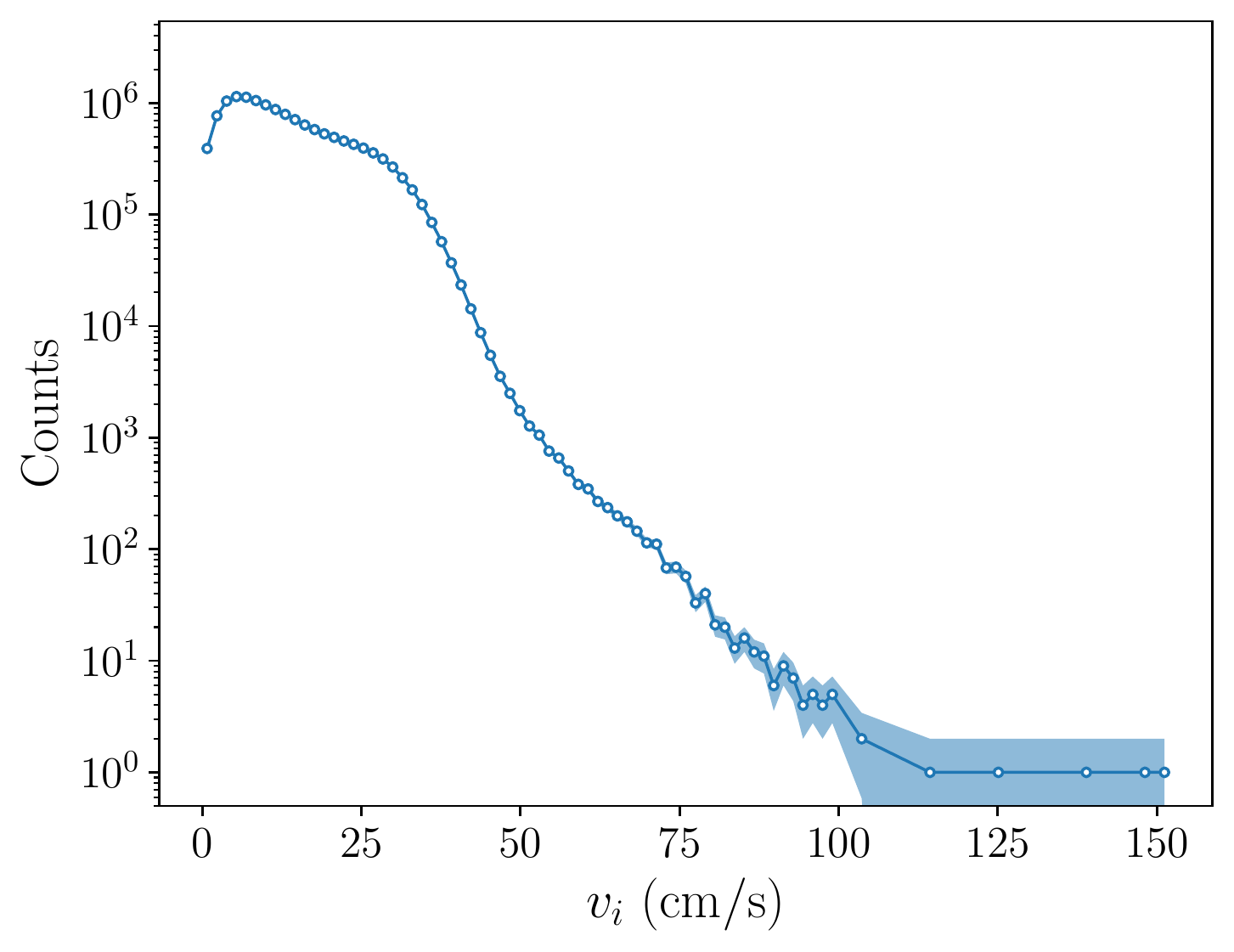}%
}

\subfloat[]{%
  \includegraphics[width=0.37\textwidth]{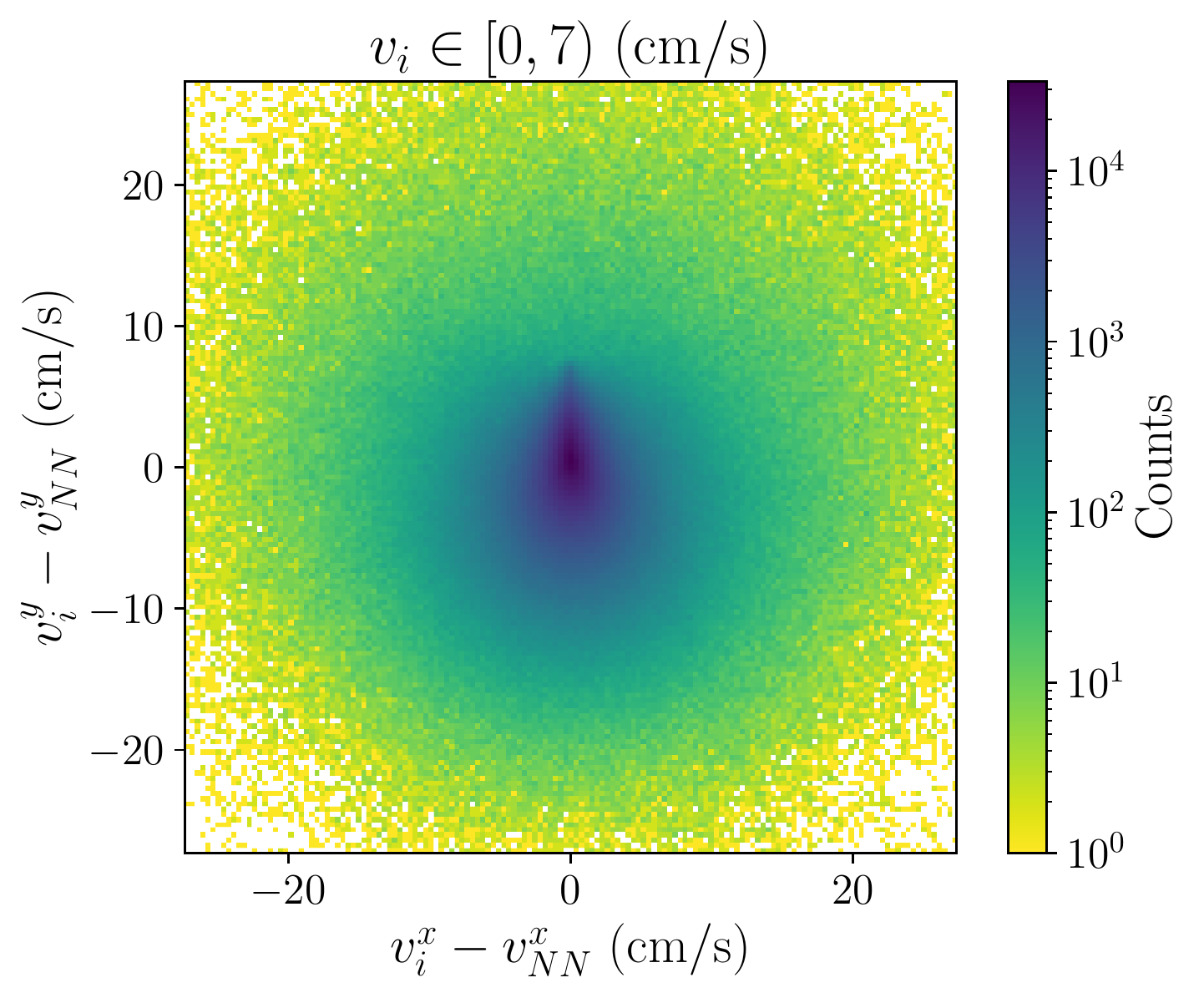}%
}
\hspace{0.01\textwidth}
\subfloat[]{%
  \includegraphics[width=0.37\textwidth]{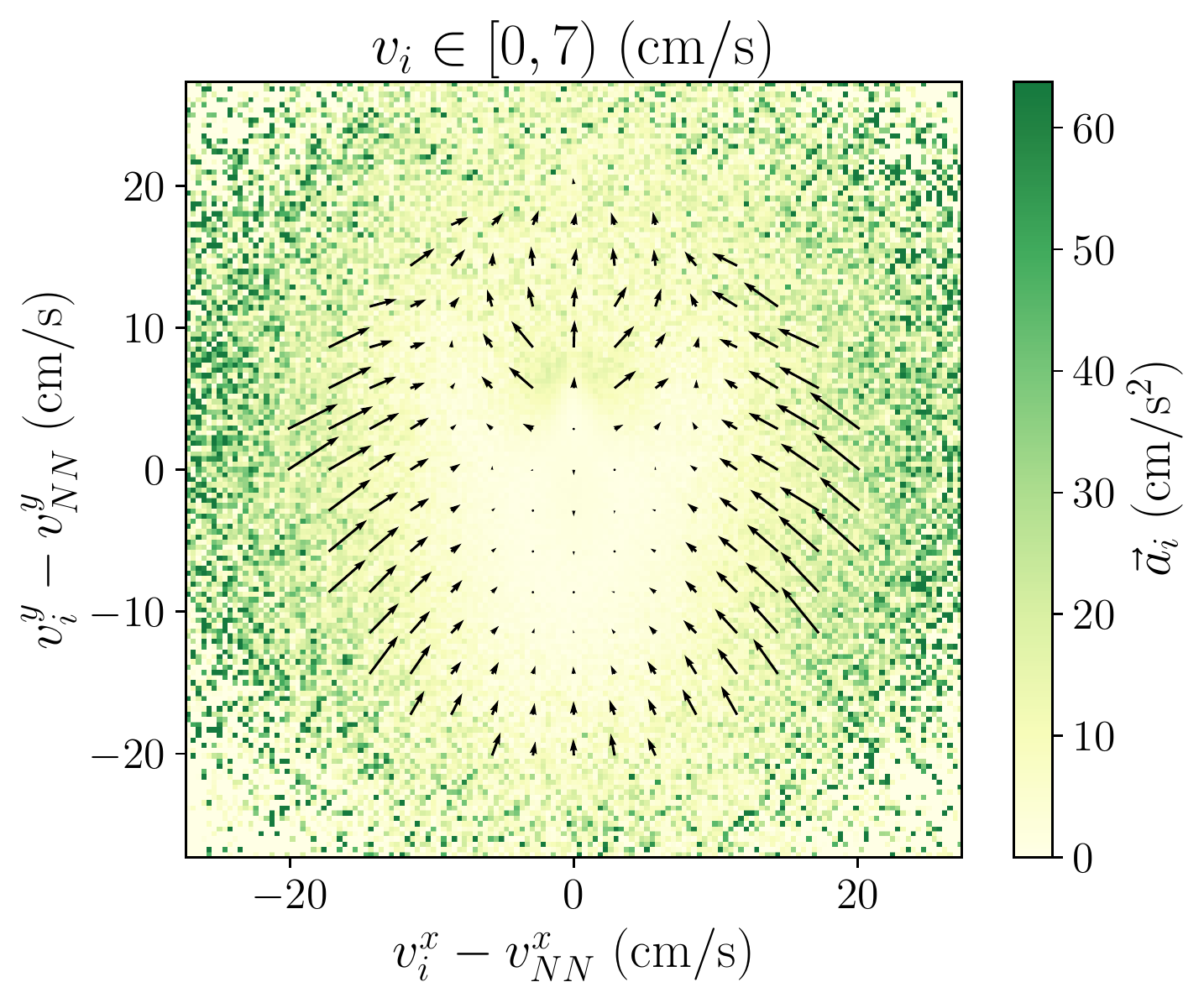}%
}

\subfloat[]{%
  \includegraphics[width=0.37\textwidth]{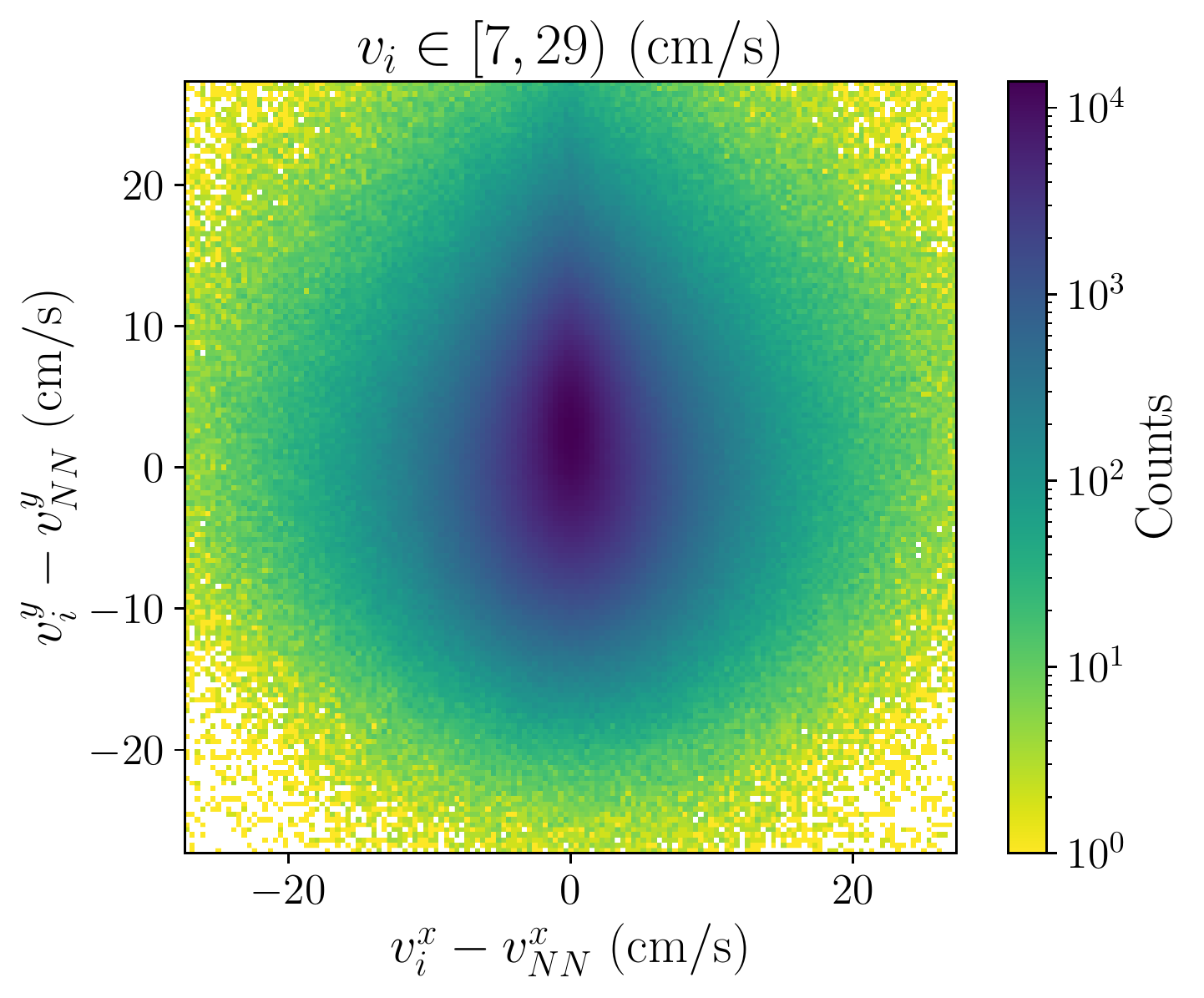}%
}
\hspace{0.01\textwidth}
\subfloat[]{%
  \includegraphics[width=0.37\textwidth]{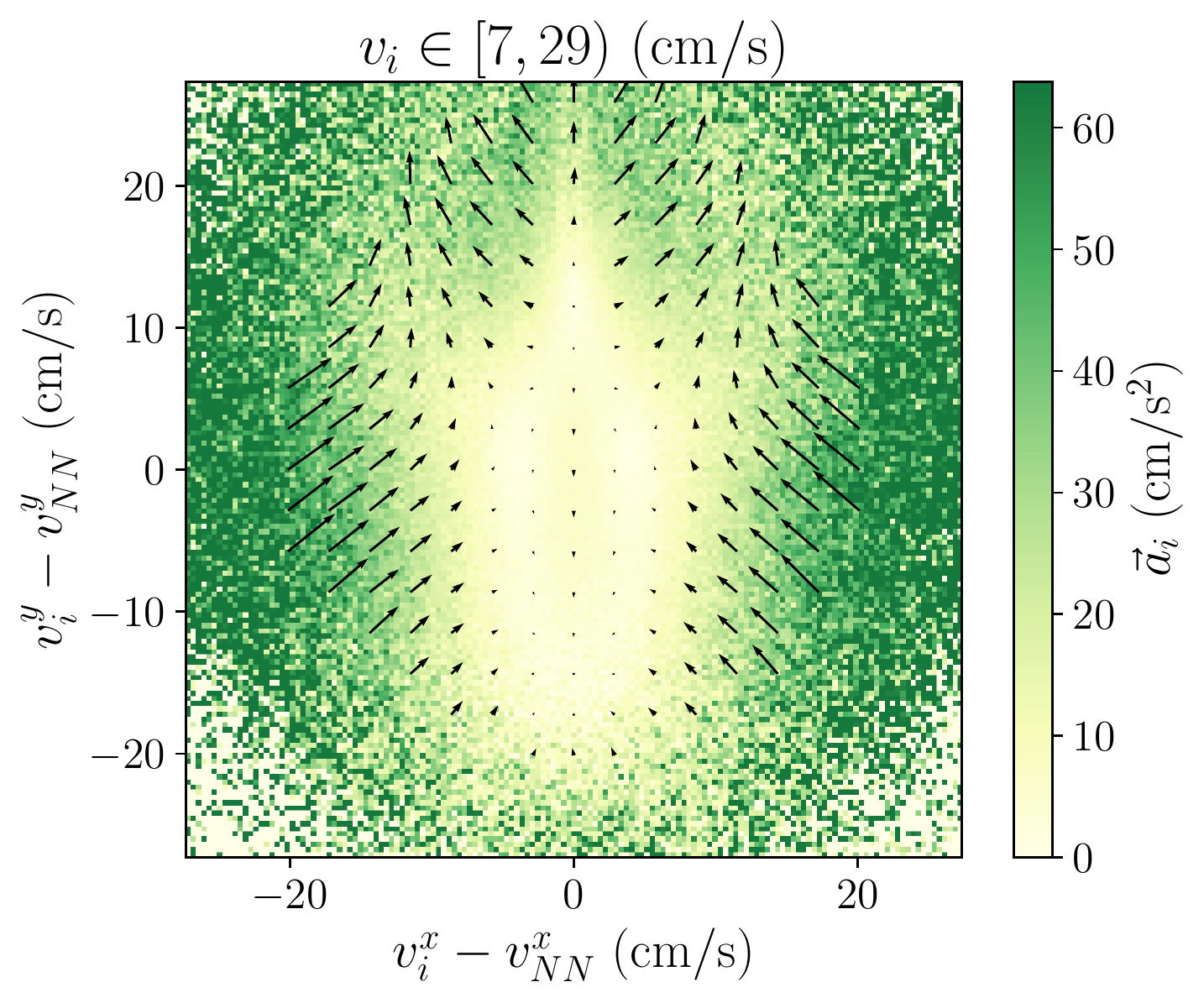}%
}

\subfloat[]{%
  \includegraphics[width=0.37\textwidth]{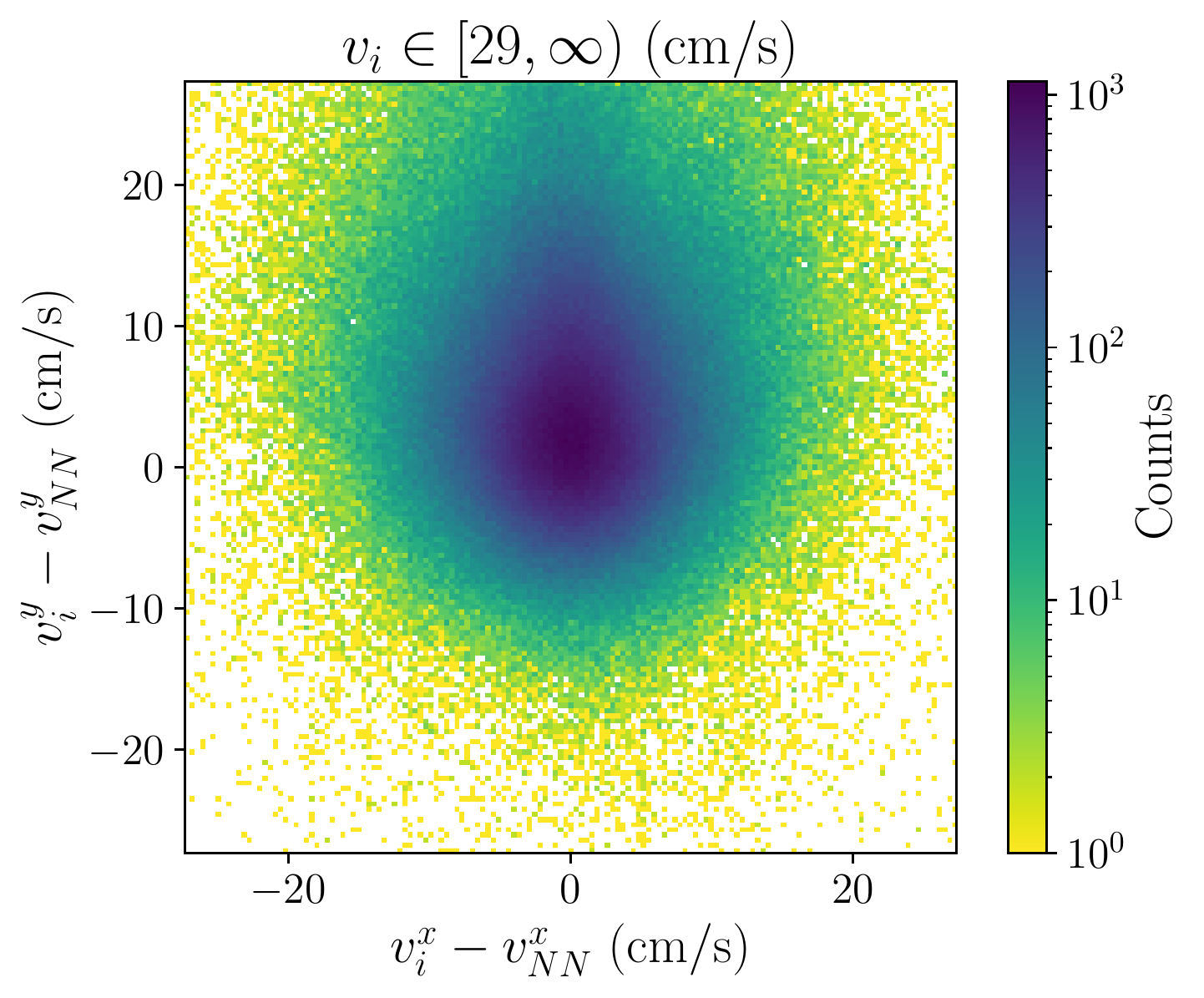}%
}
\hspace{0.01\textwidth}
\subfloat[]{%
  \includegraphics[width=0.37\textwidth]{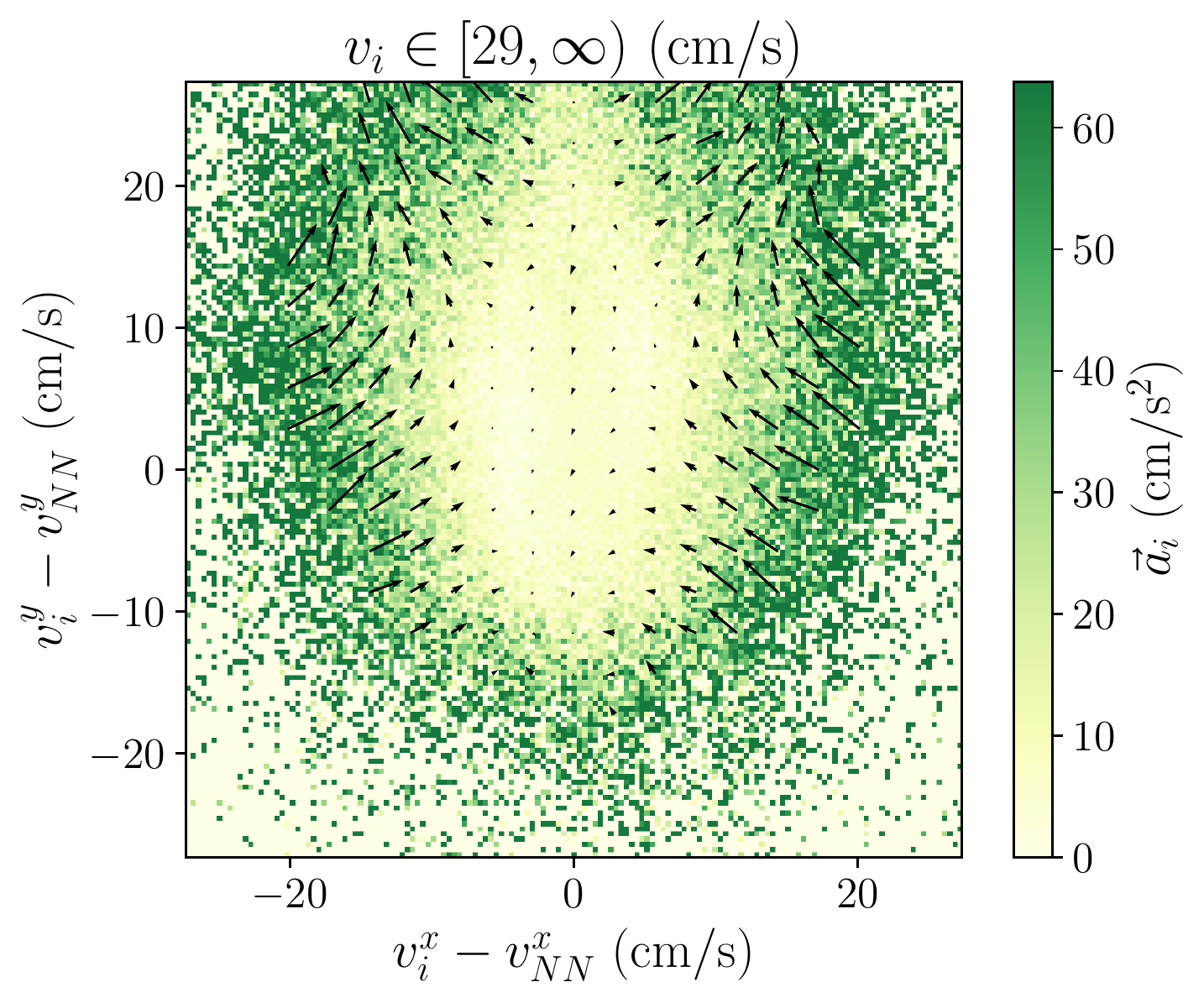}%
}
\caption{(a) Distribution of individual speeds $v_i$ and (b)-(g) alignment force maps for different $v_i$ for the experimental data. Notice how the strength of the force increases with the speed of the individual and the anti-alignment region shifts to the front. The error band in (a) is calculated from the standard deviation of a Bernoulli distribution with the probability given by the fraction of counts in each bin.} \label{supp:fig:alig_modVel}
\end{figure*}

The fish preferred speed $v_0$ is chosen from the average individual speeds (Supplementary Fig.~\ref{supp:fig:alig_modVel:modVelPDF}).

The speed relaxation time $\tau$ is obtained from the average time period of the the burst-and-coast oscillations observed in the individual speeds.

The noise constants $\sigma_v$ and $\sigma_\phi$ are chosen such that the explored regions in the force maps are comparable to the experimental data. Studying the PDFs of acceleration differences $\delta \vec{a}_i (t) \equiv \vec{a}_i (t+1) - \vec{a}_i (t)$, we found $\sigma_v > \sigma_\phi$, which we also preserve in the simulations.

\section*{Trajectories projection}
The perspective of the camera was not exactly perpendicular to the plane of motion of the fish. For better accuracy, we project the calculated trajectories to the plane parallel to the bottom of the tank. In order to do this, first we look for the pixels of the corners of the tank walls in the videos and calculate the height and base of the corresponding rectangle in pixels. Because the tank is actually a square, we define the side of the tank $L$ in pixels as the average value of the base and height of the rectangle image. We define the new tank corners: $(0, 0)$, $(0, L)$, $(L, 0)$ and $(L, L)$ and we calculate the perspective transform matrix $M$ from the original and new tank corners using the function \texttt{getPerspectiveTransform()} from the OpenCV library~\cite{bradskiOpenCVLibrary2000}. Finally, we obtain the transformed trajectories $(x', y')$ applying a perspective transformation to the original trajectories  $(x, y)$, as explained in the OpenCV documentation:
\begin{equation*}
 (x', y') = \left(\frac{M_{11}x +  M_{12}y + M_{13}}{M_{31}x +  M_{32}y + M_{33}}, \frac{M_{21}x +  M_{22}y + M_{23}}{M_{31}x +  M_{32}y + M_{33}} \right).
\end{equation*}


\begin{video*}
\caption{Rendering of the movement of fish in a experimental school of size $N=39$, overlapped with the digitized trajectories, displayed in blue.}
\end{video*}

\begin{video*}
\caption{Dynamics for the standard model of schooling fish. The reference frame is comoving with the center of mass of the group. The position of the individuals and their velocities are shown in blue. We also display the different force terms in the model (legend): the alignment force (magenta), the attraction-repulsion force (red) and the friction-propulsion force (cyan). Each individual is linked with an edge to their Voronoi neighbours: in blue when they have attraction and in orange when they have repulsion.}
\end{video*}

\begin{video*}
\caption{Dynamics for the explicit anti-alignment model. The reference frame is comoving with the center of mass of the group. The position of the individuals and their velocities are shown in blue. We also display the different force terms in the model (legend): the alignment force (magenta), the attraction-repulsion force (red) and the friction-propulsion force (cyan). Each individual is linked with an edge to their Voronoi neighbours: in blue when they have attraction and in orange when they have repulsion.}
\end{video*}

\begin{video*}
\caption{Dynamics for the selective interactions model. The reference frame is comoving with the center of mass of the group. The position of the individuals and their velocities are shown in blue when they interact with other individuals and in black when they do not interact. We also display the different force terms in the model (legend): the alignment force (magenta), the attraction-repulsion force (red) and the friction-propulsion force (cyan). Each individual is linked to their Voronoi neighbours with an edge reaching half their distance if they are interacting: in blue when they have attraction and in orange when they have repulsion.}
\end{video*}

\begin{video*}
\caption{Sample of a recording where the focal fish $i$ (green) tends to align with faster neighbours (blue) and ignore slower neighbours (red). Arrows indicate the velocity of individuals. Voronoi neighbours are coloured by their relative speed with the focal individual, $v_{jV}-v_i$.}
\end{video*}

\begin{figure*}[t!p]
\subfloat[]{%
  \includegraphics[width=0.4\textwidth]{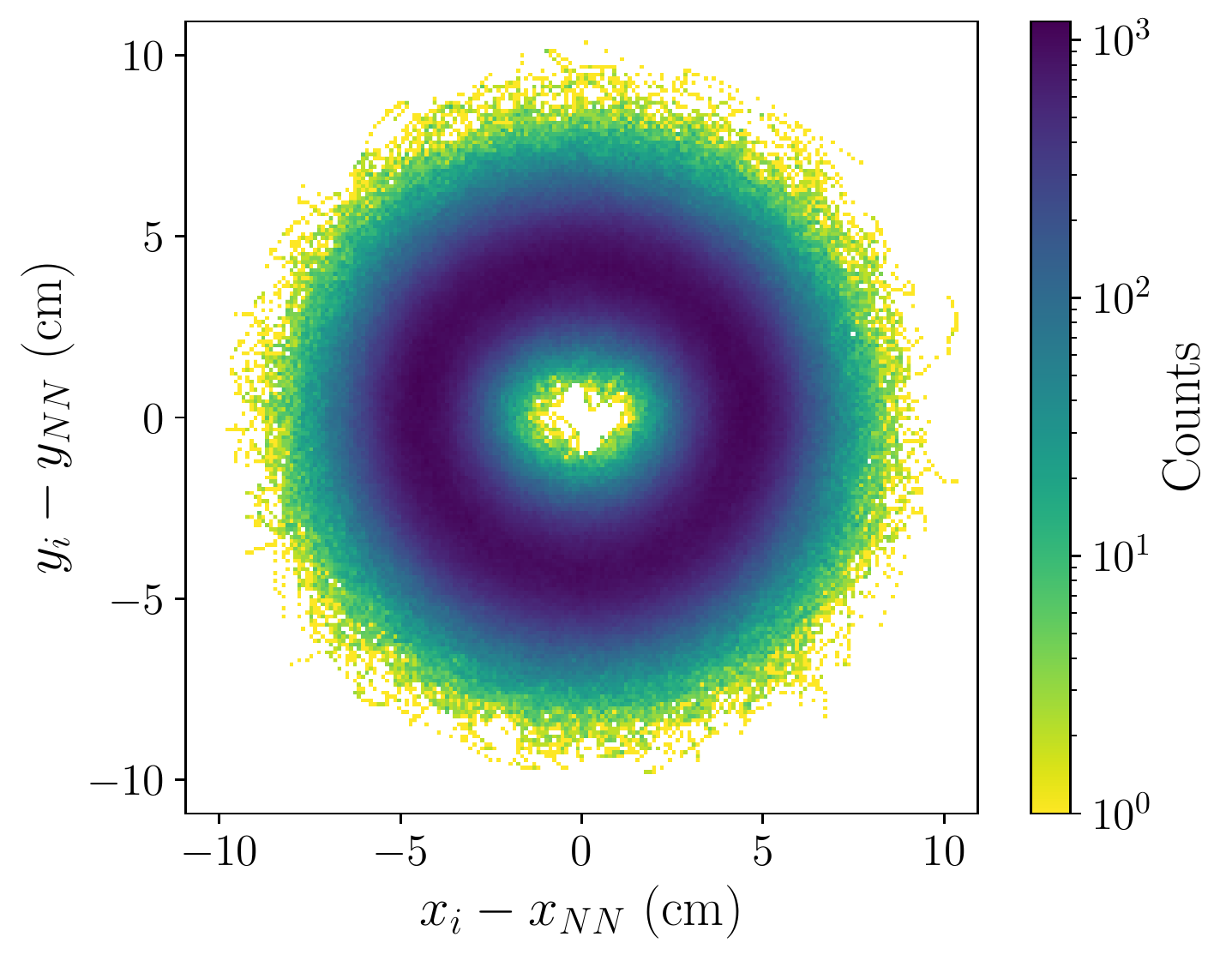}%
}

\subfloat[]{%
  \includegraphics[width=0.4\textwidth]{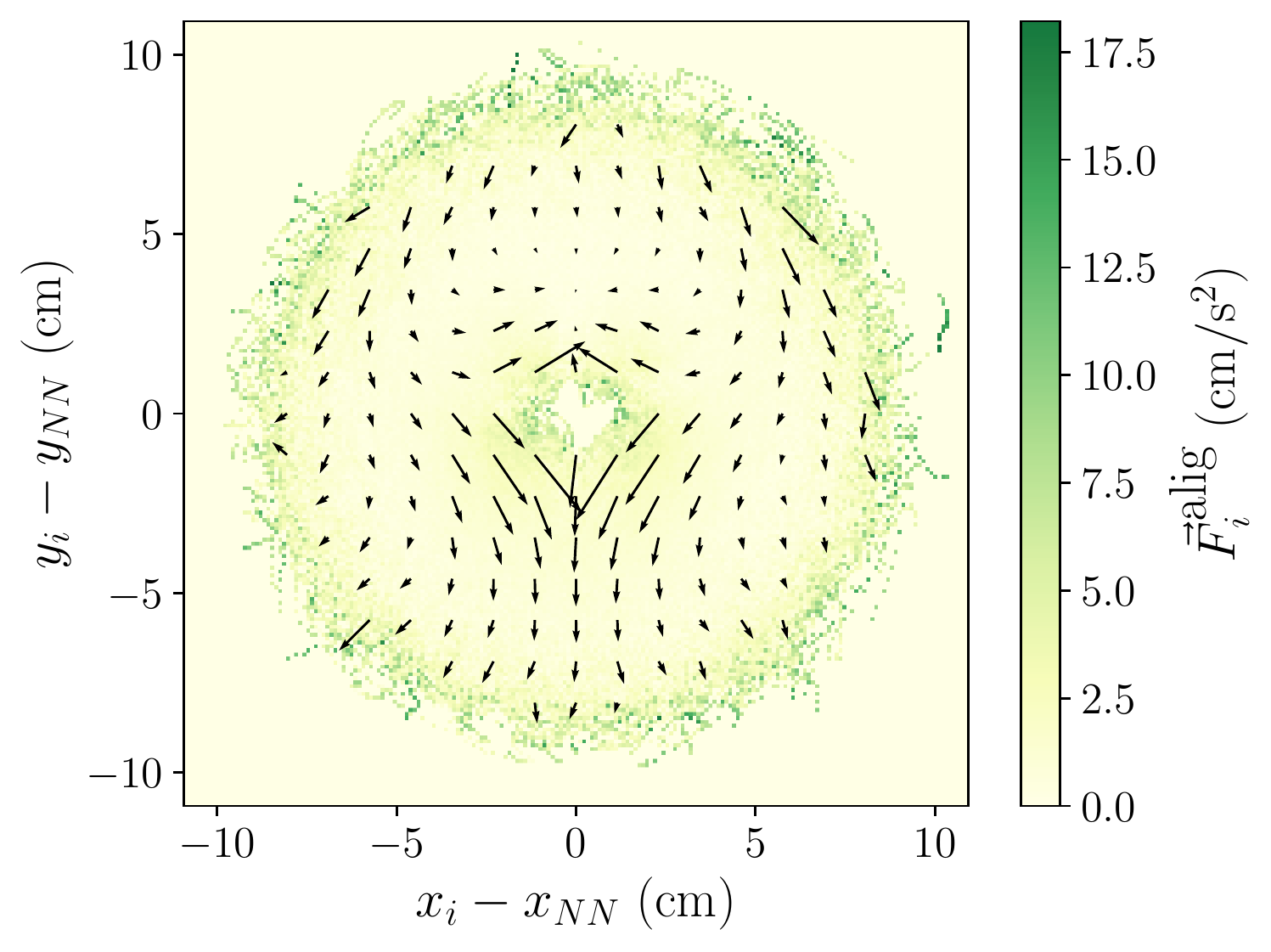}%
}
\hspace{0.01\textwidth}
\subfloat[]{%
  \includegraphics[width=0.4\textwidth]{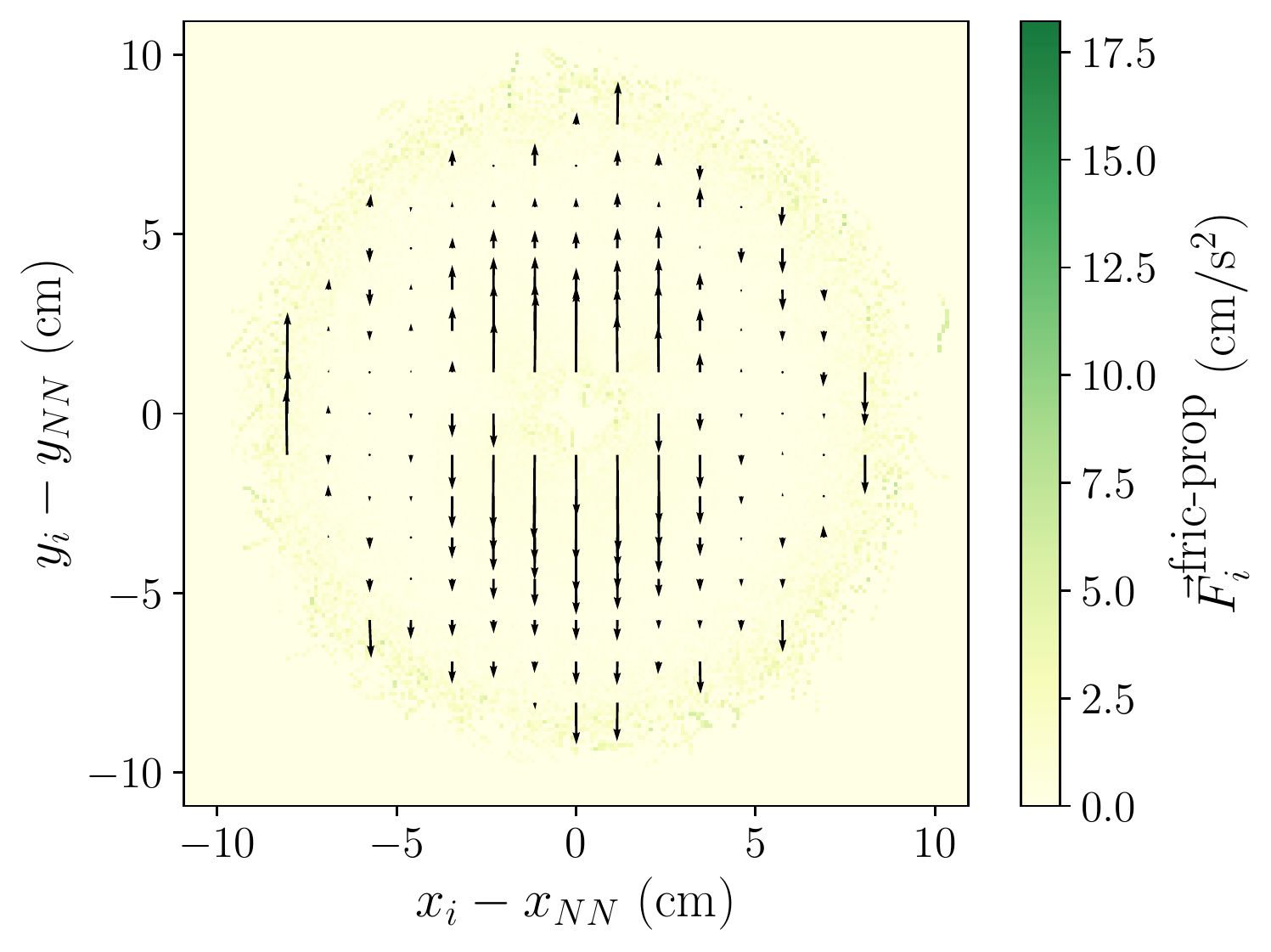}%
}
\caption{(a) Counts, (b) average alignment and (c) average friction-propulsion force acting on an individual $i$ depending on its relative position with nearest neighbour $NN$ for the standard model.  The $y$-coordinate is along the direction of motion of the individual.  Forces are expressed in units of mass $m=1$.}\label{supp:fig:basicModel_attRep}
\end{figure*}

\begin{figure*}[t!p]
\subfloat[]{%
  \includegraphics[width=0.4\textwidth]{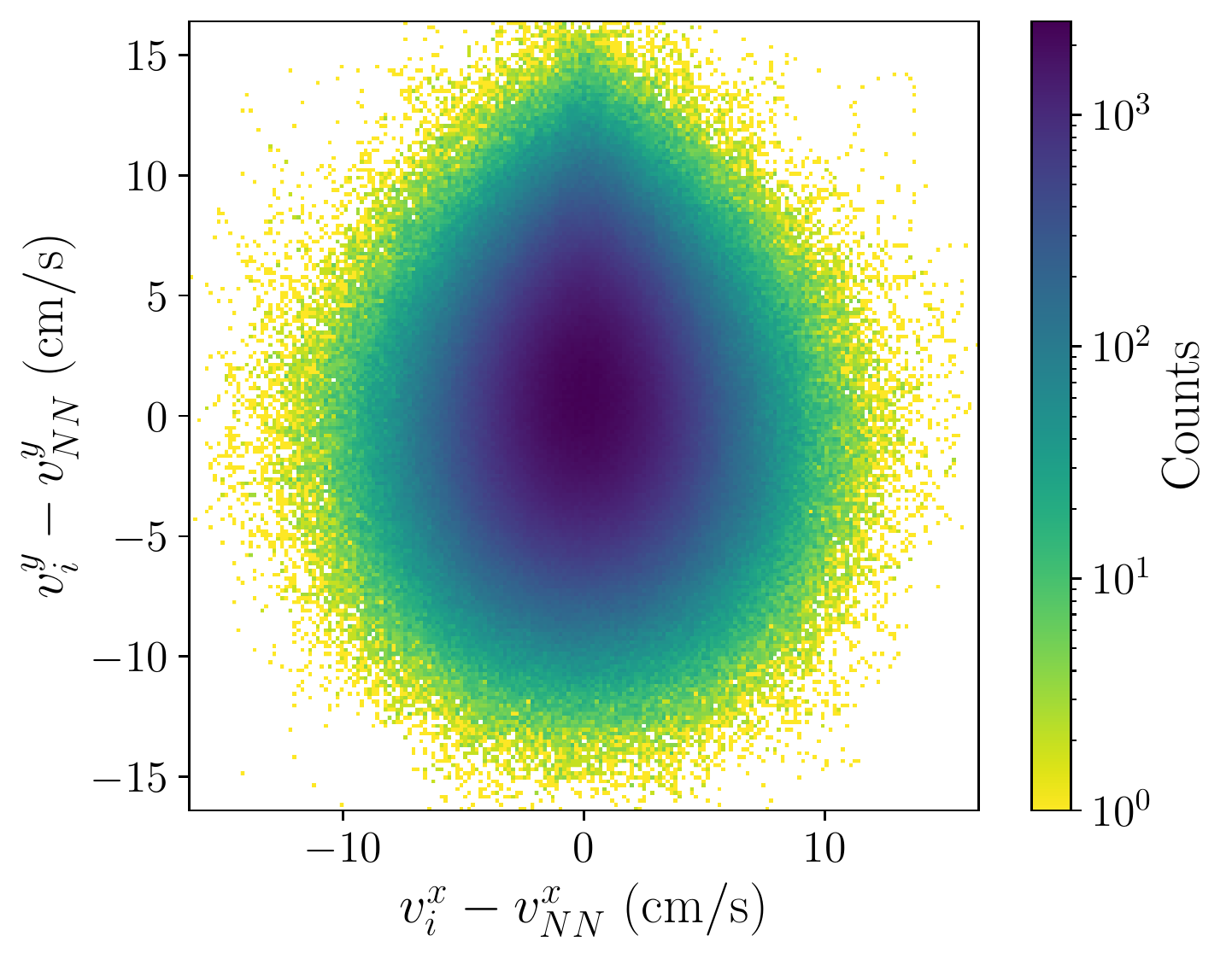}%
}

\subfloat[]{%
  \includegraphics[width=0.4\textwidth]{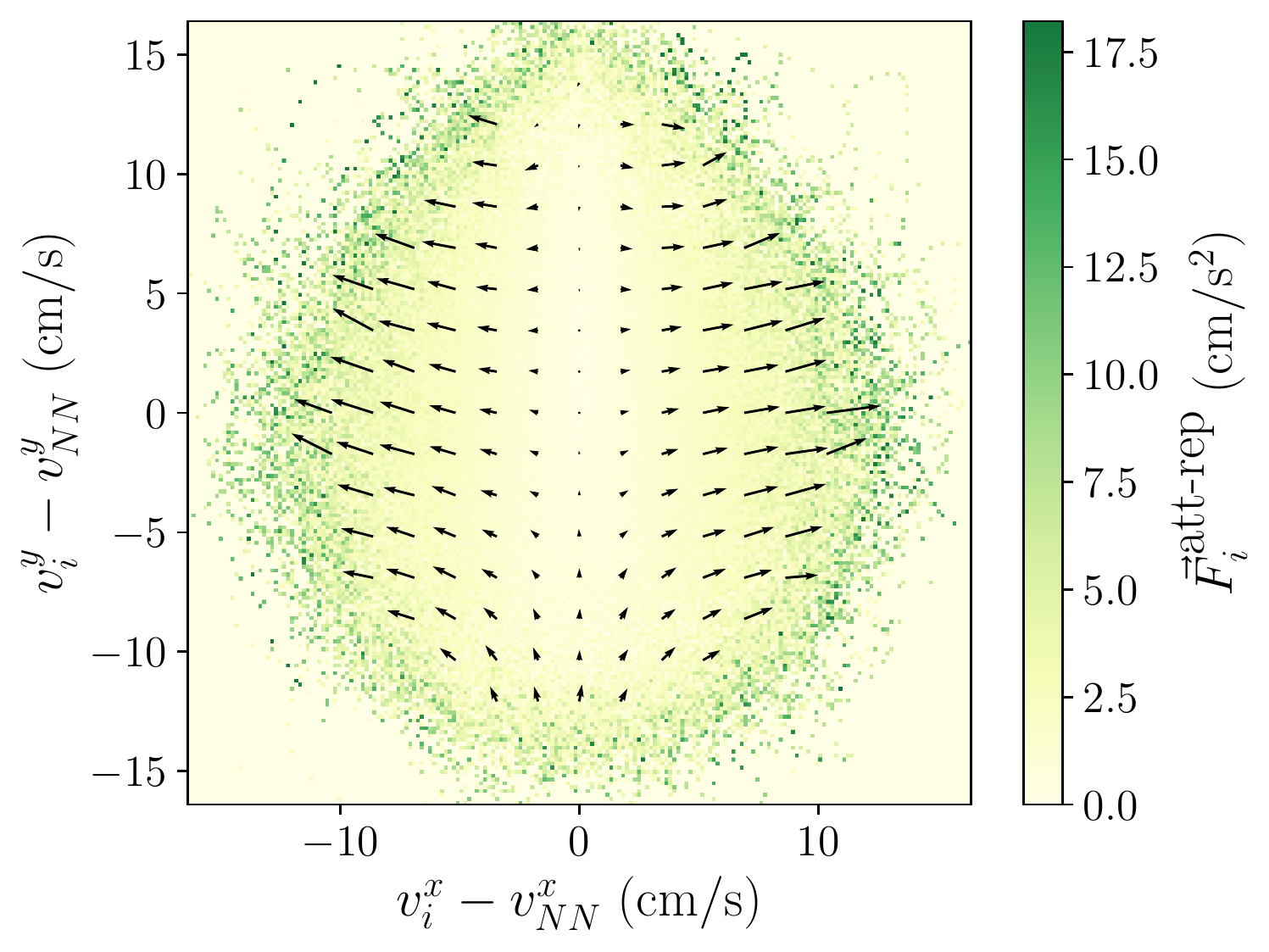}%
}
\hspace{0.01\textwidth}
\subfloat[]{%
  \includegraphics[width=0.4\textwidth]{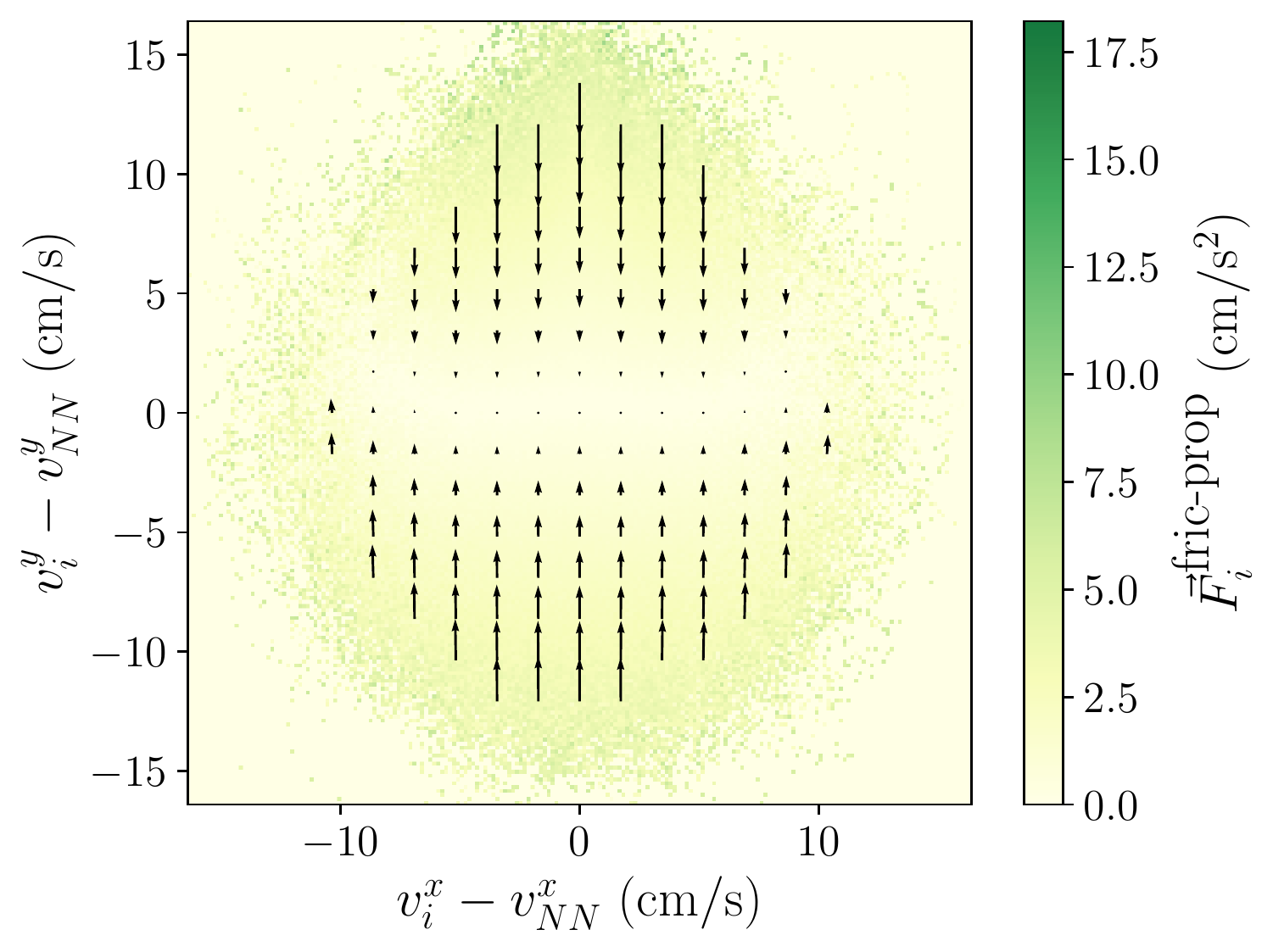}%
}
\caption{(a) Counts, (b) average attraction-repulsion and (c) average friction-propulsion force acting on an individual $i$ depending on its relative velocity with the nearest neighbour $NN$ for the standard model.  The $y$-coordinate is along the direction of motion of the individual. Forces are expressed in units of mass $m=1$. } \label{supp:fig:basicModel_align}
\end{figure*}

\begin{figure*}[t!p]
\subfloat[\label{supp:fig:radialForceMap:standardModel}]{%
  \includegraphics[width=0.4\textwidth]{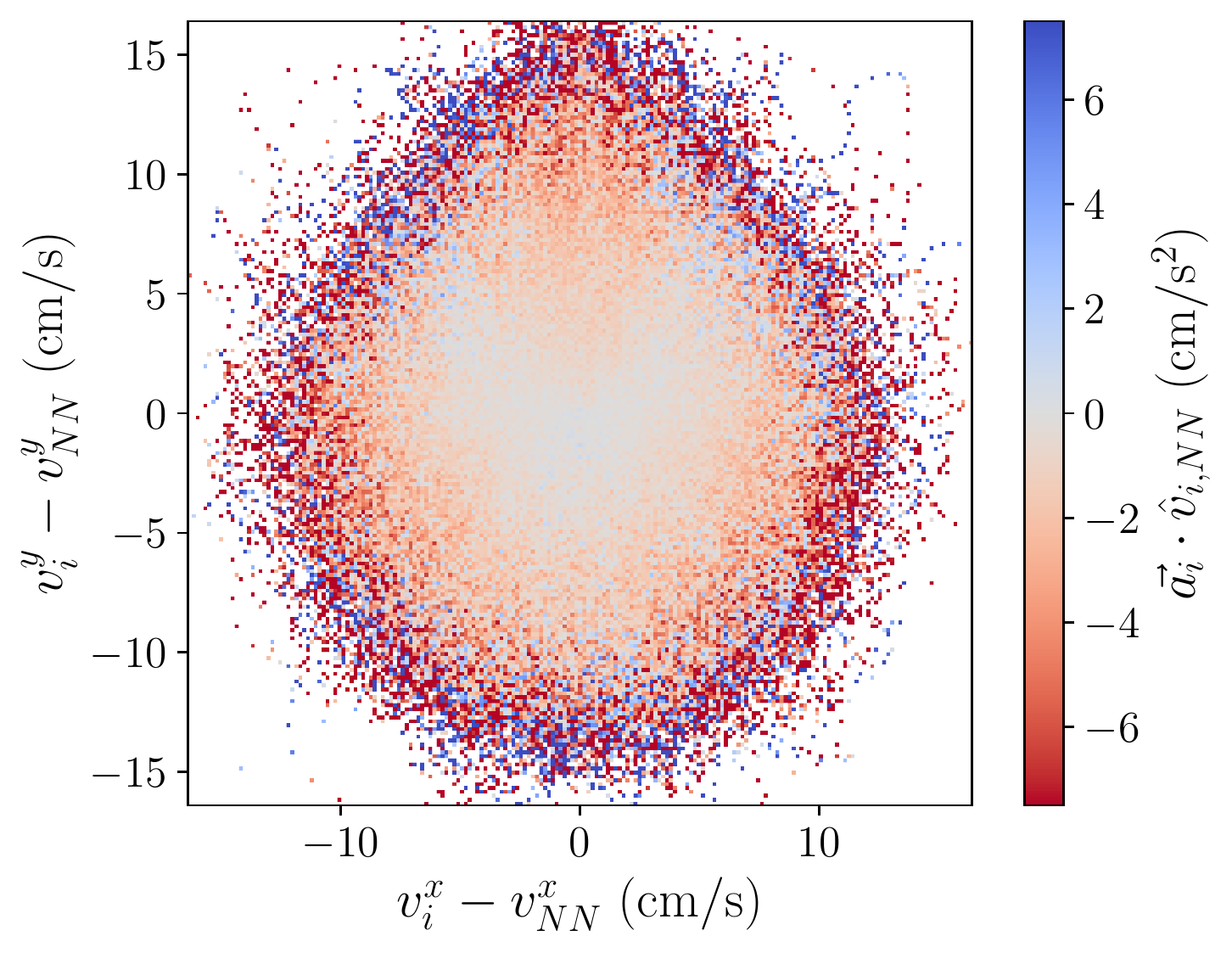}%
}
\hspace{0.01\textwidth}
\subfloat[\label{supp:fig:radialForceMap:experimentalData}]{%
  \includegraphics[width=0.4\textwidth]{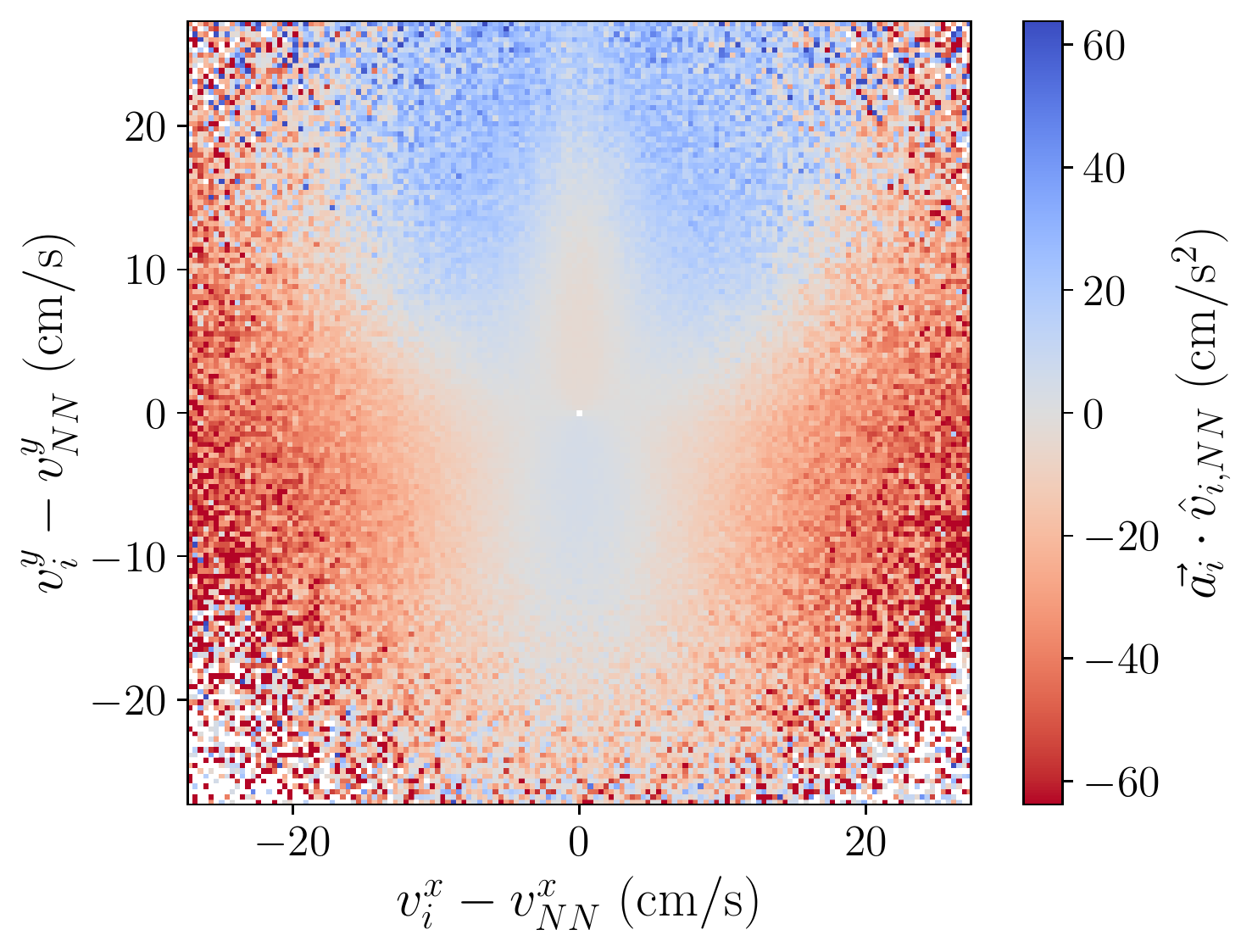}%
}

\subfloat[\label{supp:fig:radialForceMap:explicitAntiAlignment}]{%
  \includegraphics[width=0.4\textwidth]{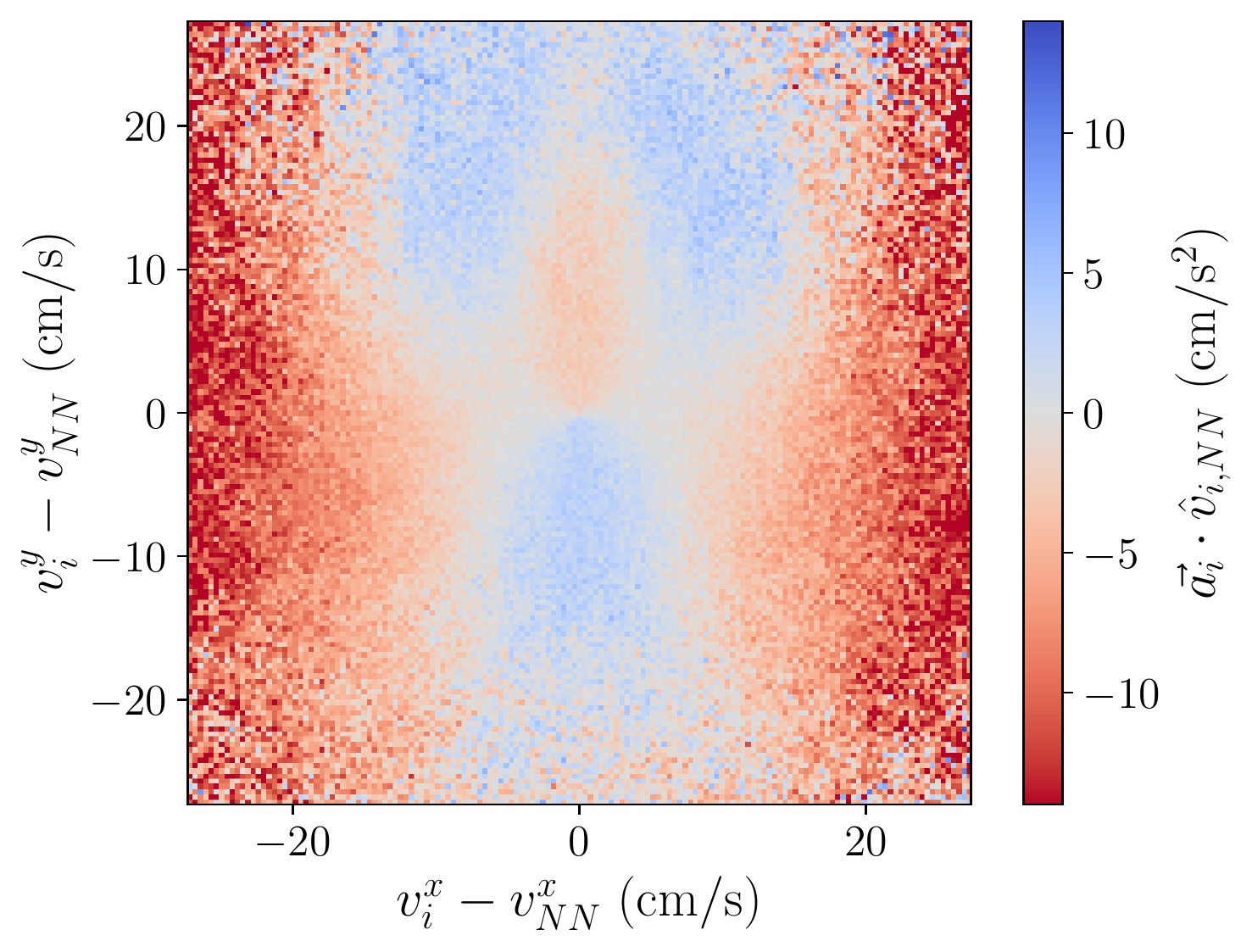}%
}
\hspace{0.01\textwidth}
\subfloat[\label{supp:fig:radialForceMap:selectiveInteractions}]{%
  \includegraphics[width=0.4\textwidth]{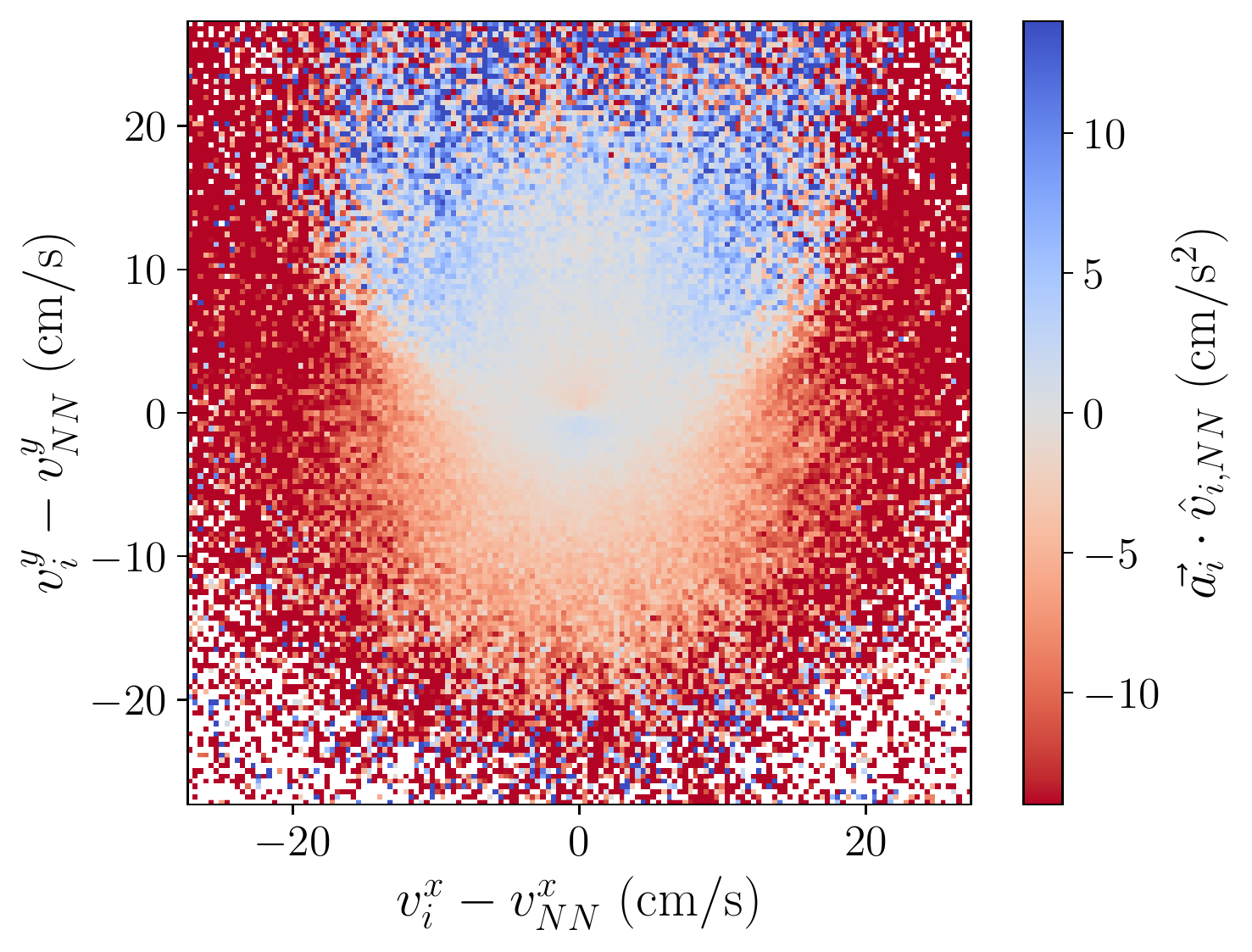}%
}

\subfloat[\label{supp:fig:radialForceMap:experimentalPositiveDelay}]{%
  \includegraphics[width=0.4\textwidth]{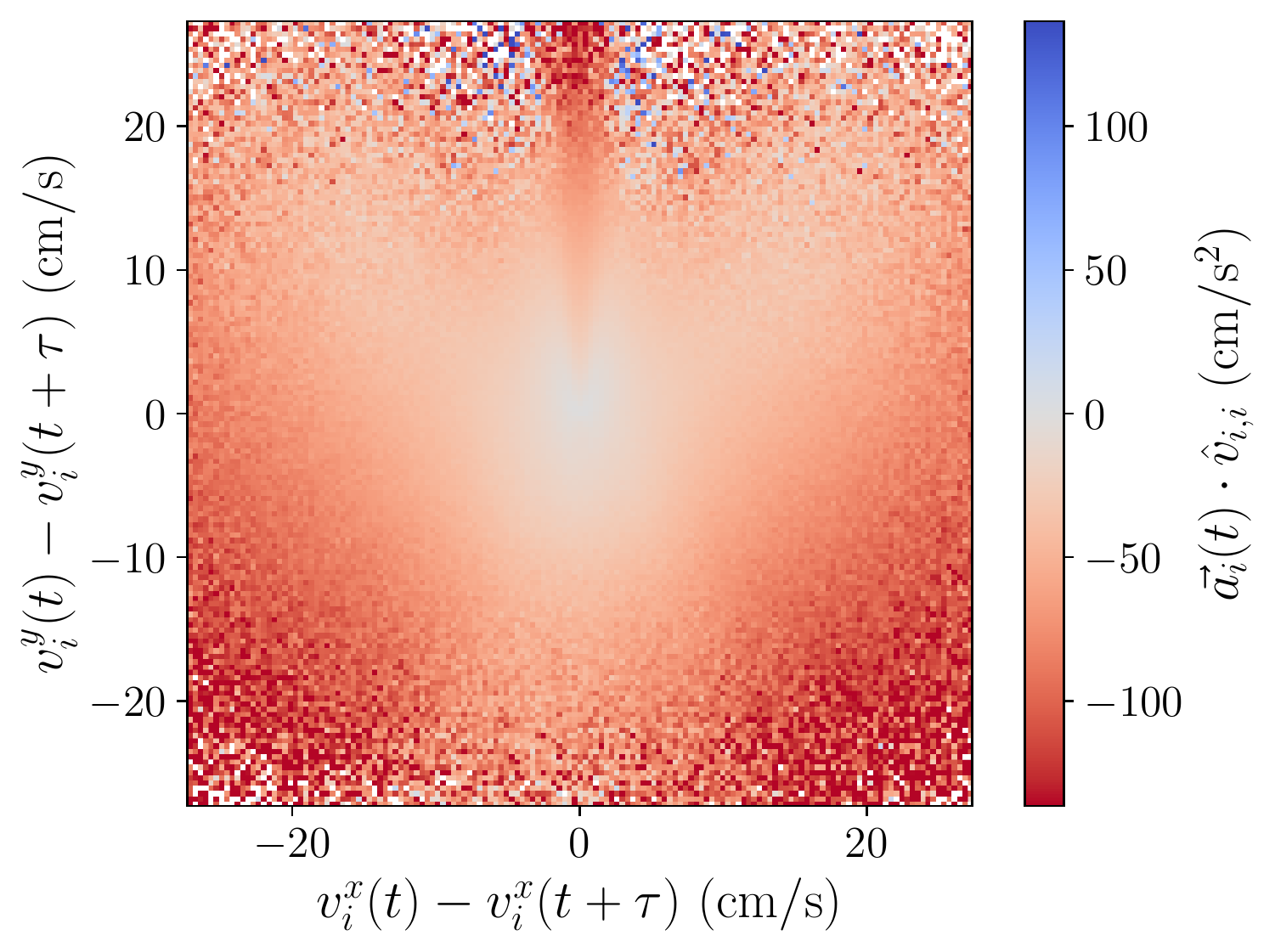}%
}
\hspace{0.01\textwidth}
\subfloat[\label{supp:fig:radialForceMap:experimentalNegativeDelay}]{%
  \includegraphics[width=0.4\textwidth]{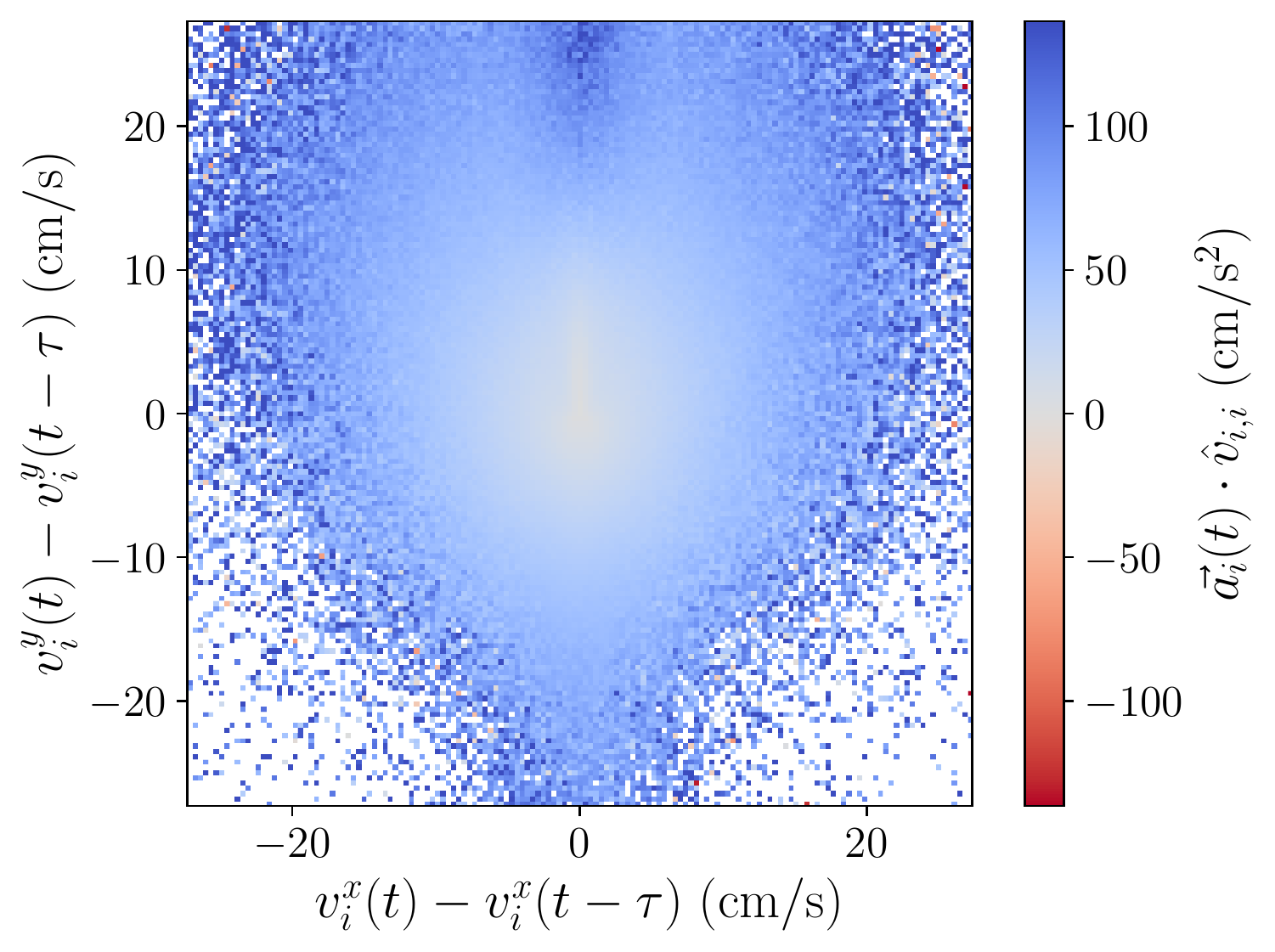}%
}
\caption{Projection of the average acceleration along the radial direction of the alignment force map for (a) the standard model (Fig~\ref{fig:force_maps:align_basicModel}), (b) the experimental data (Fig.~\ref{fig:force_maps_exp:aligExp}), (c) the explicit anti-alignment model (Fig.~\ref{fig:alignmentForceMap_variations:expAlignModel_alignmentForceMap}), (d) the selective interactions model (Fig.~\ref{fig:alignmentForceMap_variations:selecModel_alignmentForceMap}) and for the experimental data comparing the individual $i$ with itself at a (e) positive (Fig.~\ref{fig:delayedAlignmentForceMap:pos}) and (f) negative (Fig.~\ref{fig:delayedAlignmentForceMap:neg}) delay of $\tau = 0.2$~s. In (a)-(d) the radial direction is given by the relative velocity of the individual with the nearest neighbour $\hat{v}_{i,NN} \equiv \frac{\vec{v}_i - \vec{v}_{NN}}{\left| \vec{v}_i - \vec{v}_{NN} \right|}$, while in (e)-(f) instead of the nearest neighbour we use the velocity of the individual $i$ at the delayed time. In these plots, blue colors correspond to outward arrows (anti-alignment) in the alignment force map, and red colors to inward arrows (alignment).}\label{supp:fig:radialForceMap}
\end{figure*}

\begin{figure*}[t!p]
\subfloat[\label{supp:fig:force_maps_counts_exp:atracRep}]{%
  \includegraphics[width=0.4\textwidth]{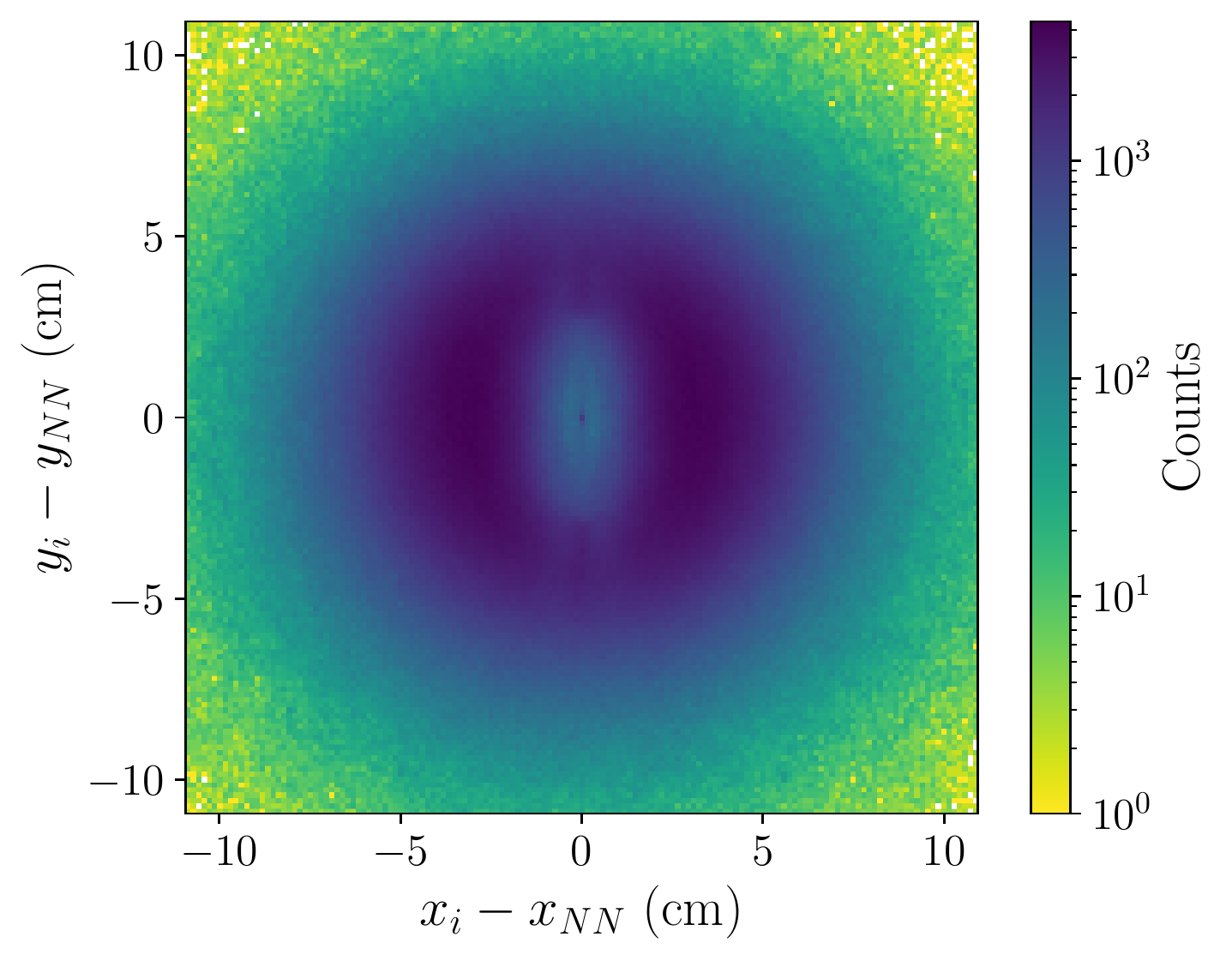}%
}
\hspace{0.01\textwidth}
\subfloat[\label{supp:fig:force_maps_counts_exp:alig}]{%
  \includegraphics[width=0.4\textwidth]{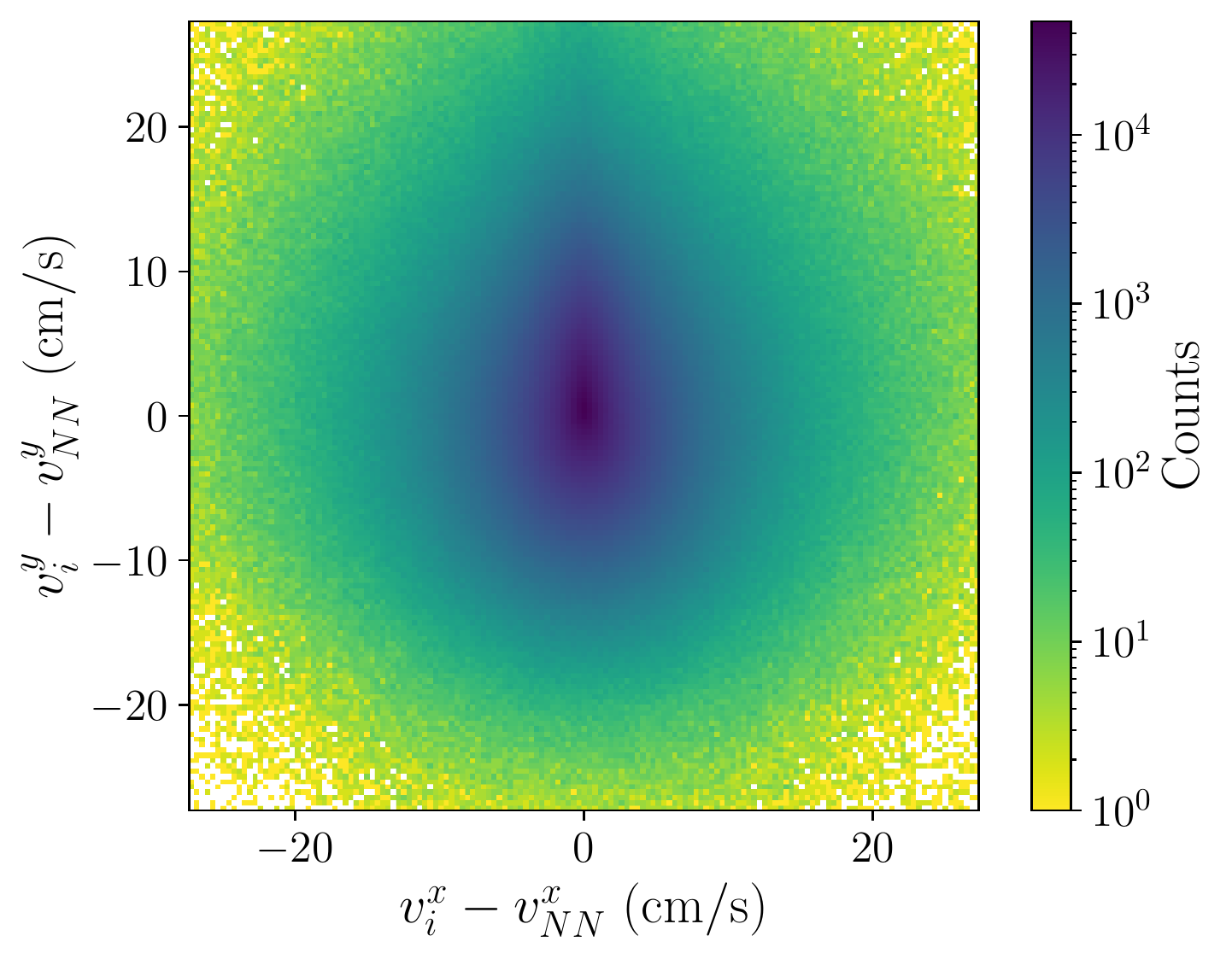}%
}
\caption{Counts for (a) attraction-repulsion and (b) alignment force maps for the experimental data.} \label{supp:fig:force_maps_counts_exp}
\end{figure*}

\begin{figure*}[t!p]

\subfloat[]{%
  \includegraphics[width=0.36\textwidth]{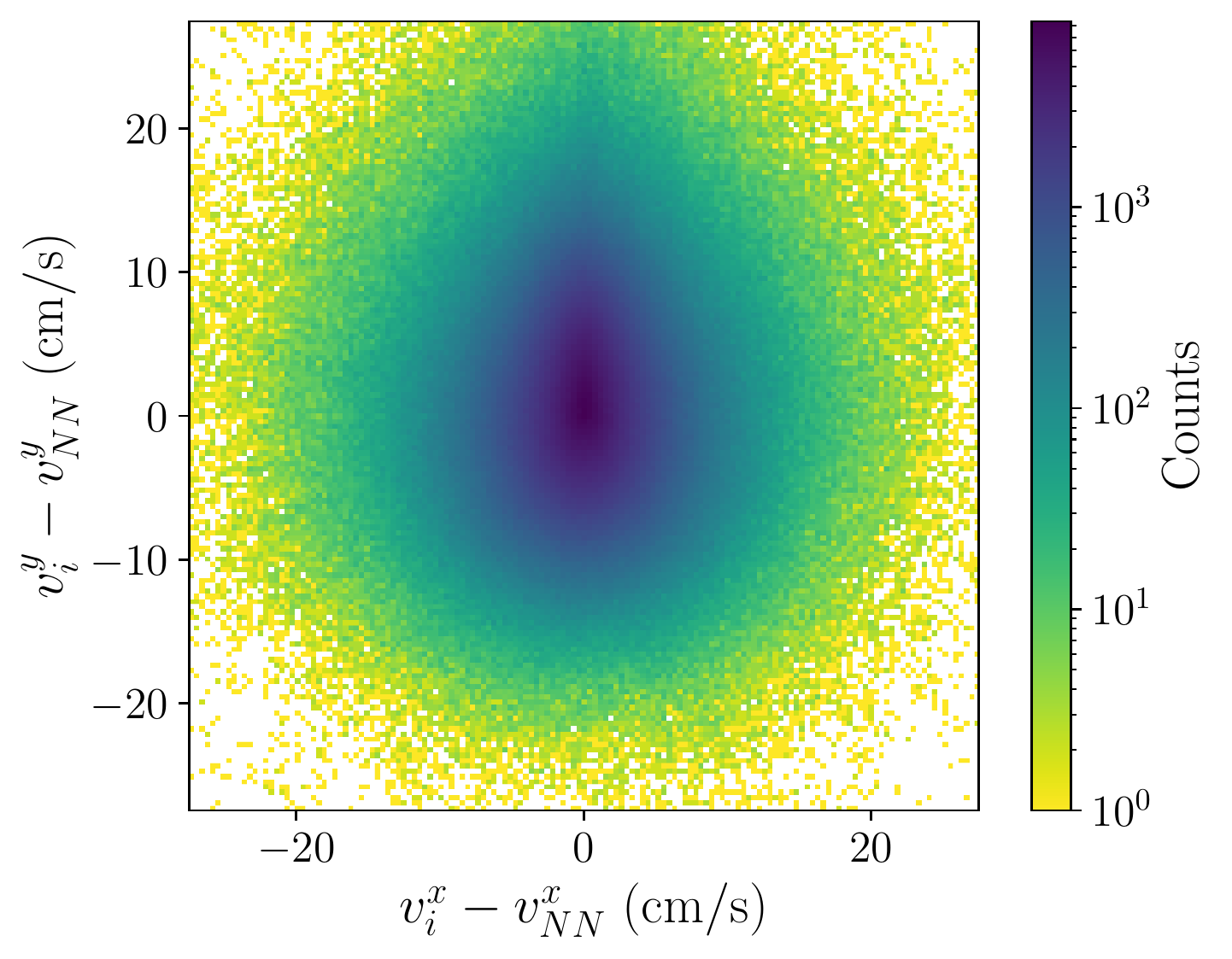}%
}
\hspace{0.01\textwidth}
\subfloat[]{%
  \includegraphics[width=0.36\textwidth]{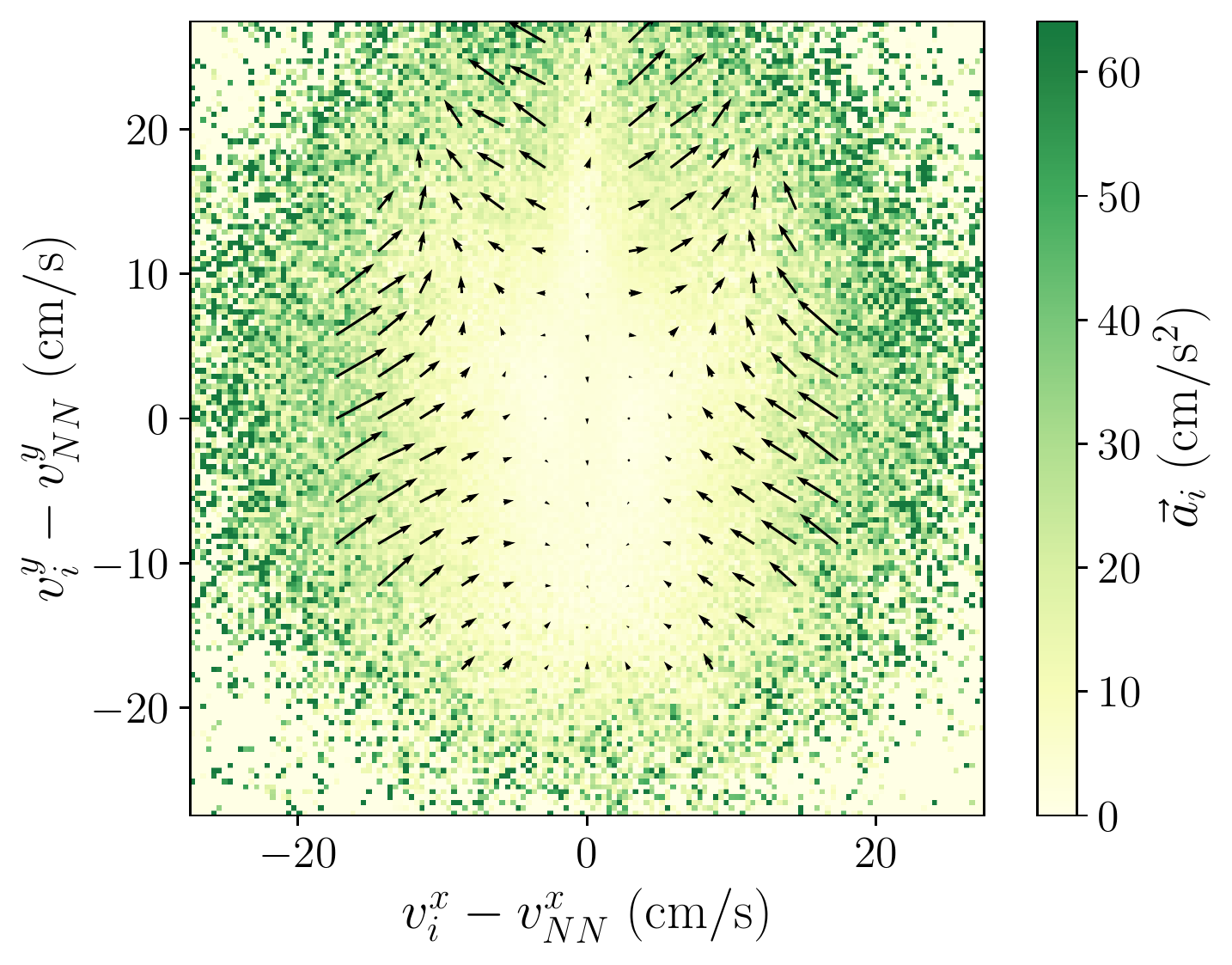}%
}

\caption{Alignment force maps for experimental data for a smaller number of fish $N=8$.} \label{supp:fig:alig_different_N}
\end{figure*}

\begin{figure*}[t!p]
\subfloat[]{%
  \includegraphics[width=0.35\textwidth]{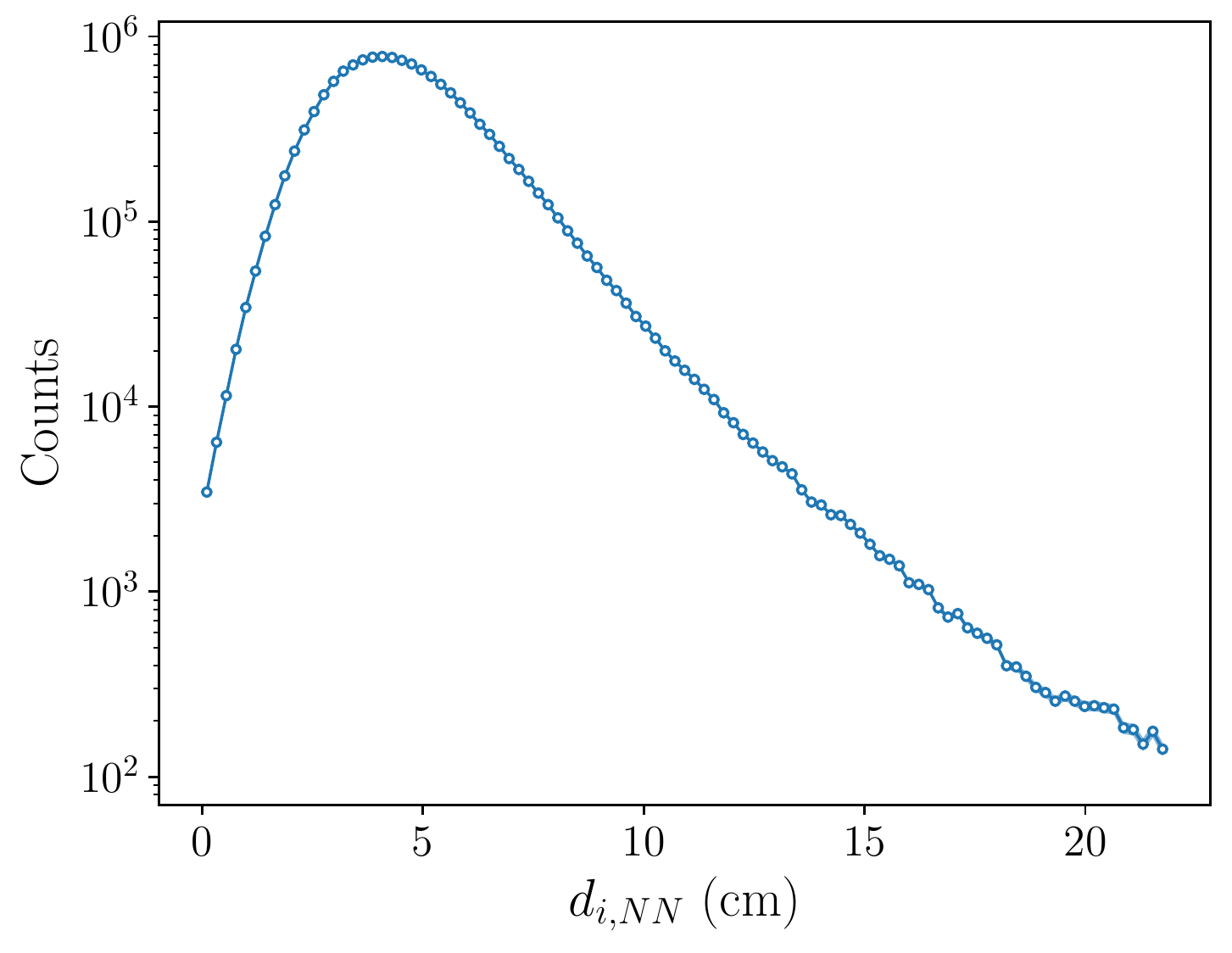}%
}

\subfloat[]{%
  \includegraphics[width=0.36\textwidth]{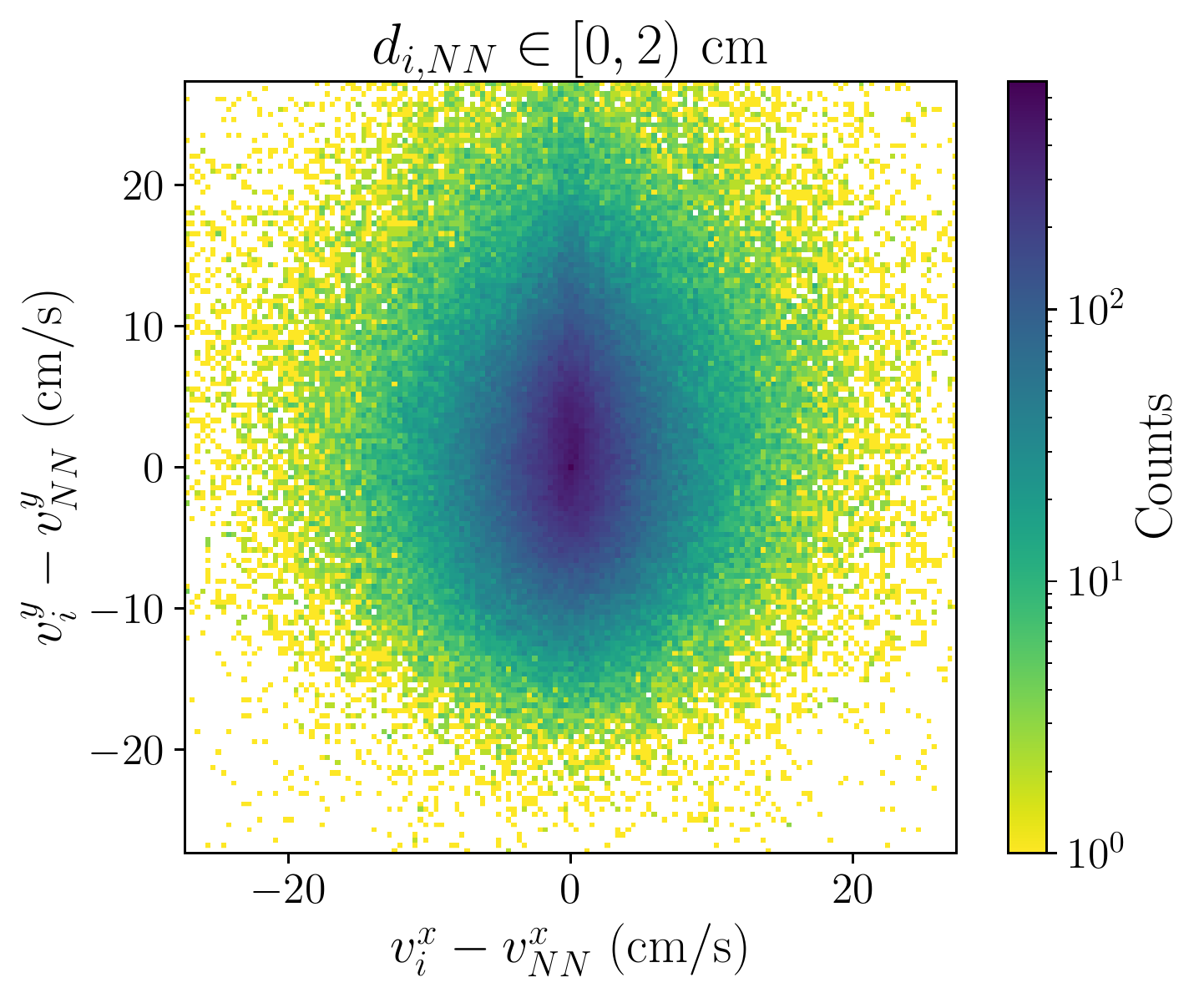}%
}
\hspace{0.01\textwidth}
\subfloat[]{%
  \includegraphics[width=0.36\textwidth]{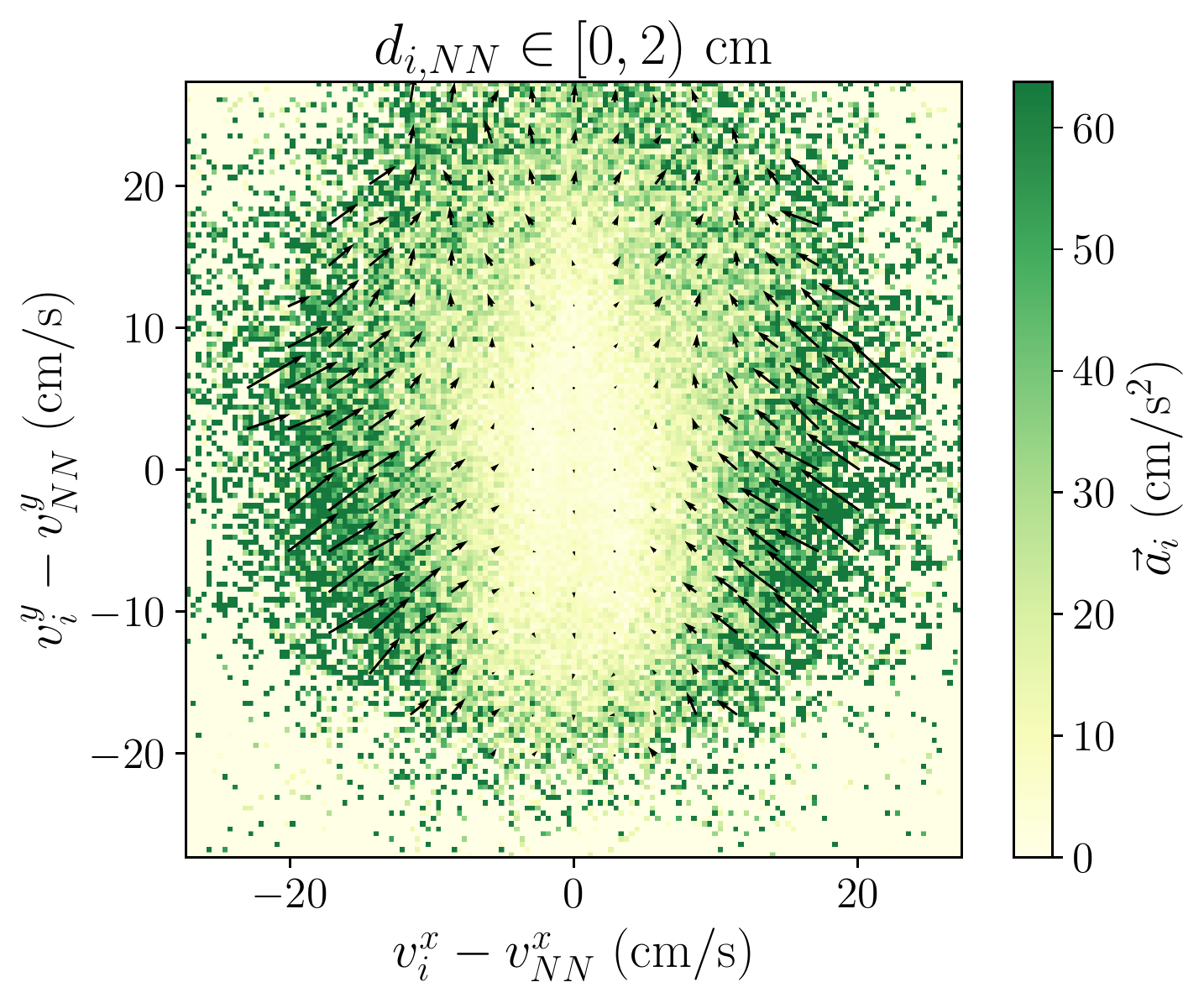}%
}

\subfloat[]{%
  \includegraphics[width=0.36\textwidth]{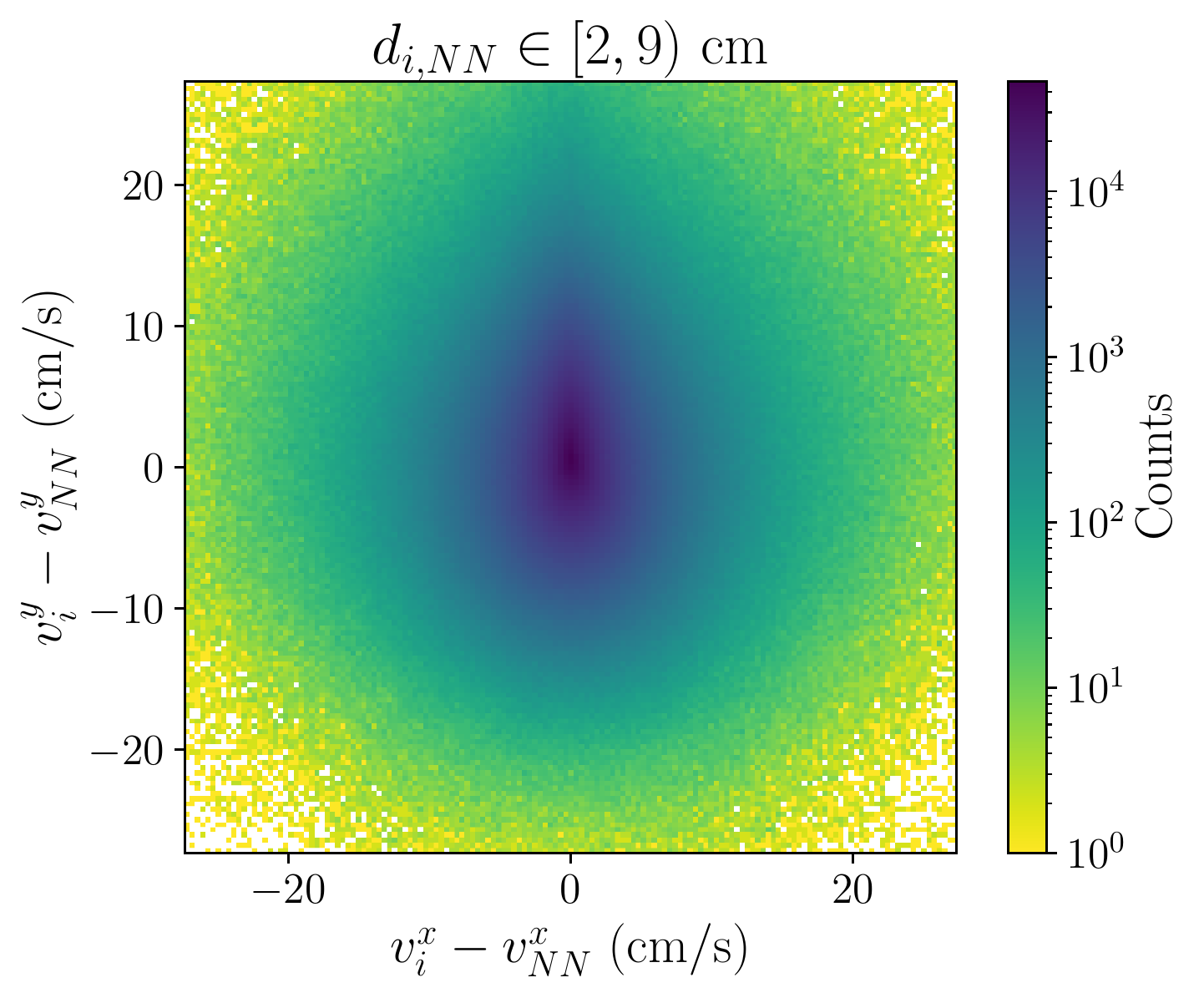}%
}
\hspace{0.01\textwidth}
\subfloat[]{%
  \includegraphics[width=0.36\textwidth]{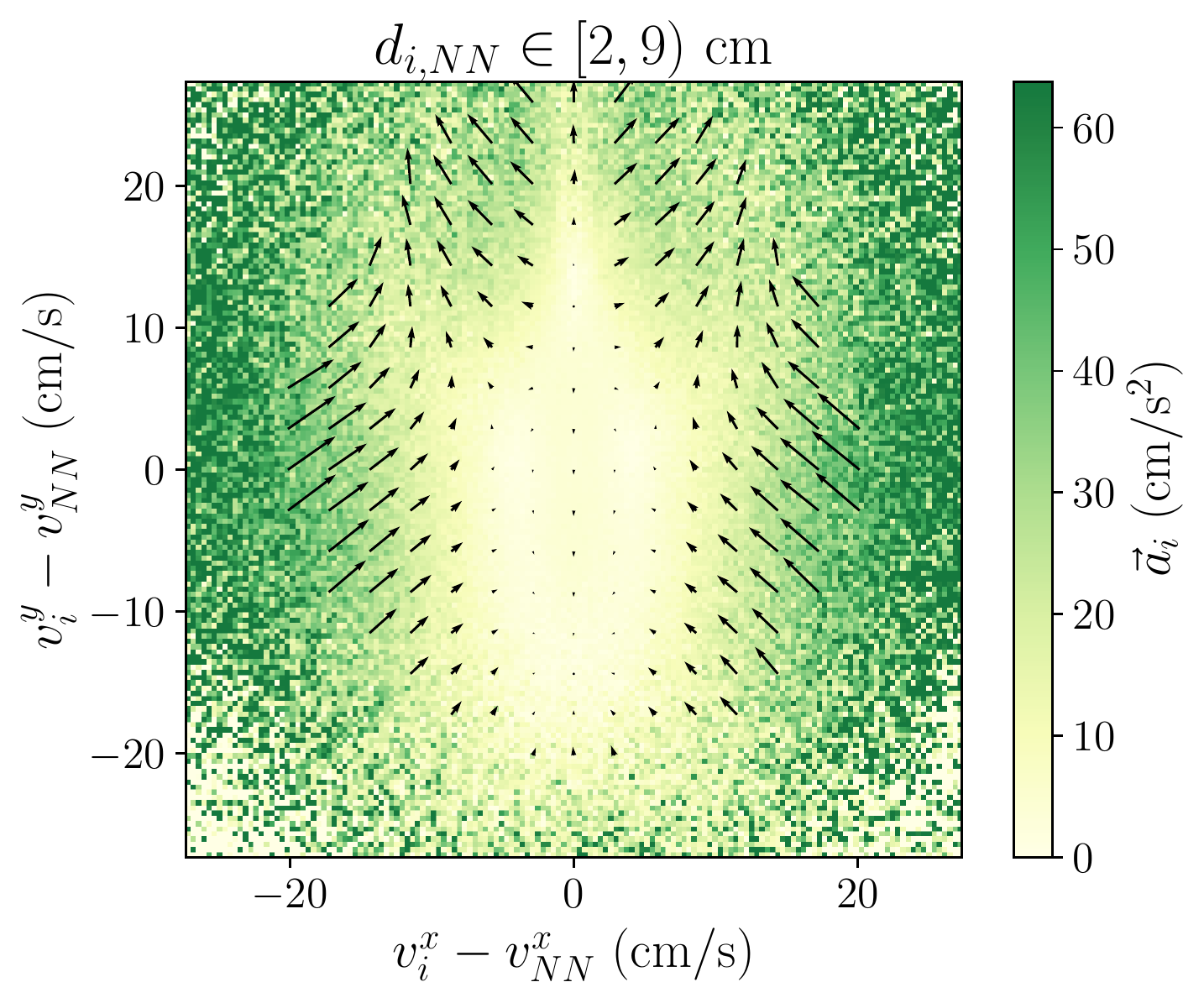}%
}

\subfloat[]{%
  \includegraphics[width=0.36\textwidth]{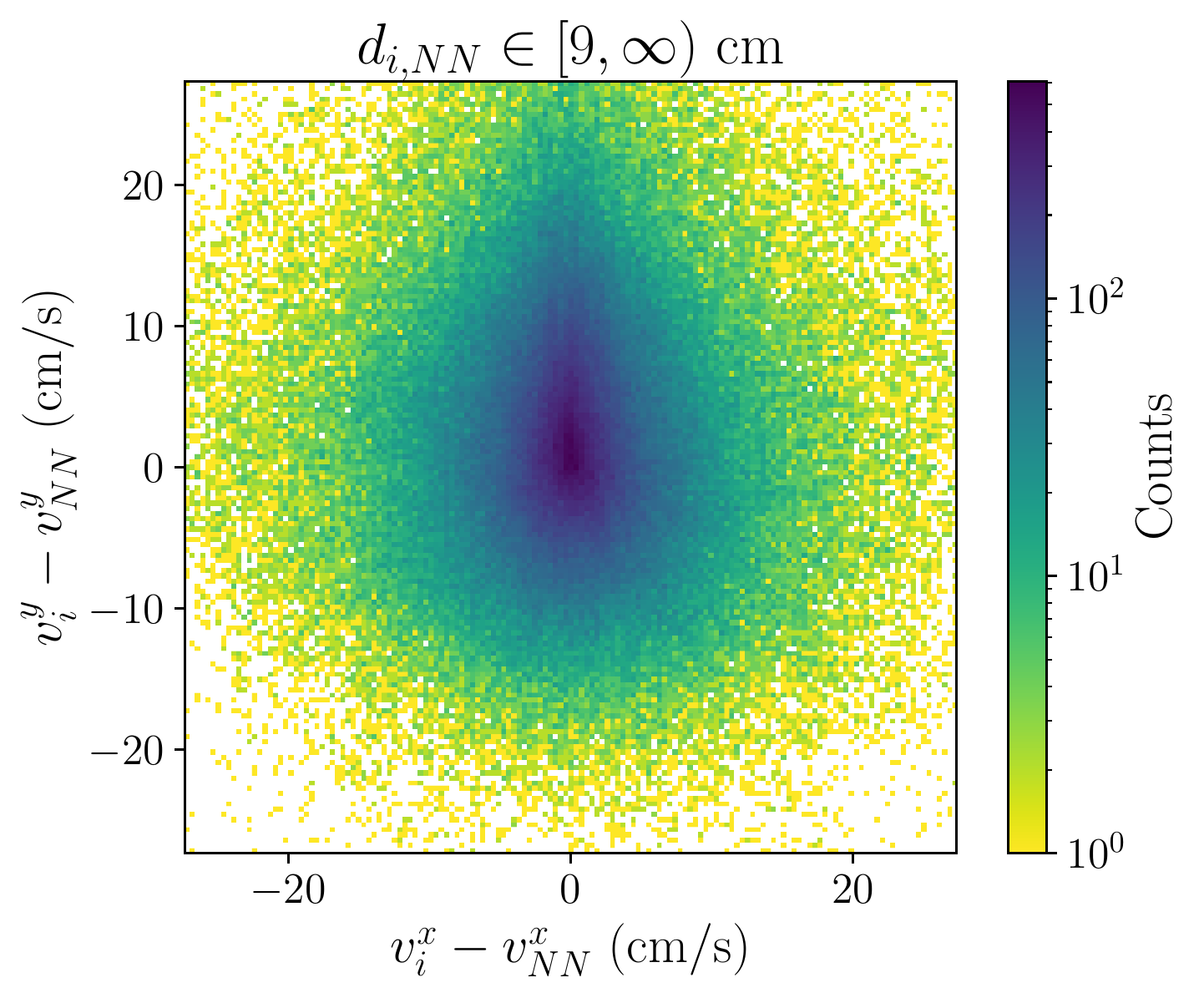}%
}
\hspace{0.01\textwidth}
\subfloat[]{%
  \includegraphics[width=0.36\textwidth]{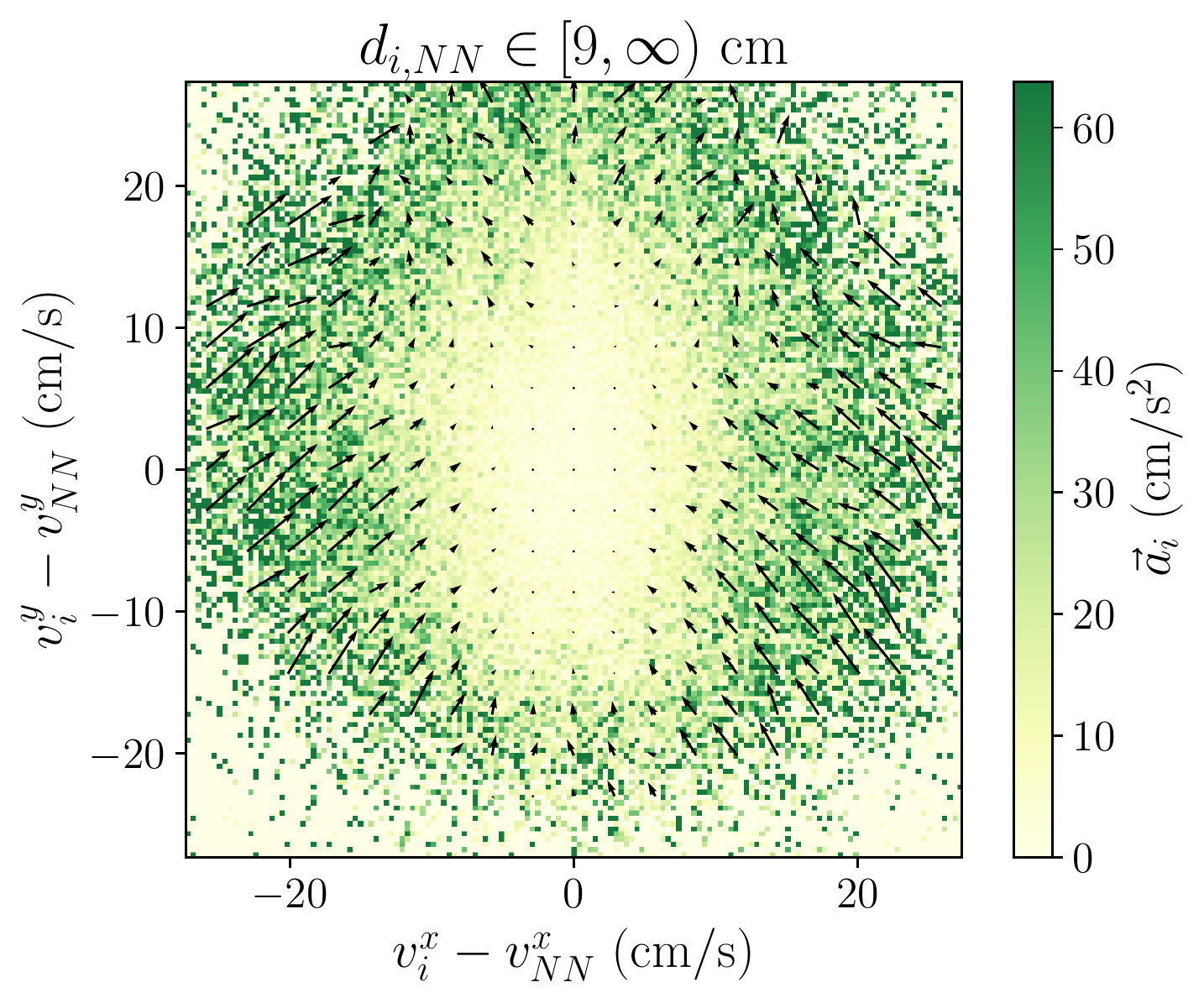}%
}
\caption{(a) Distribution of distances between nearest neighbours $d_{i,NN}$  and (b)-(g) alignment force maps for different $d_{i,NN}$ for the experimental data.} \label{supp:fig:alig_Nn_dij}
\end{figure*}

\begin{figure*}[t!p]
\subfloat[]{%
  \includegraphics[width=0.35\textwidth]{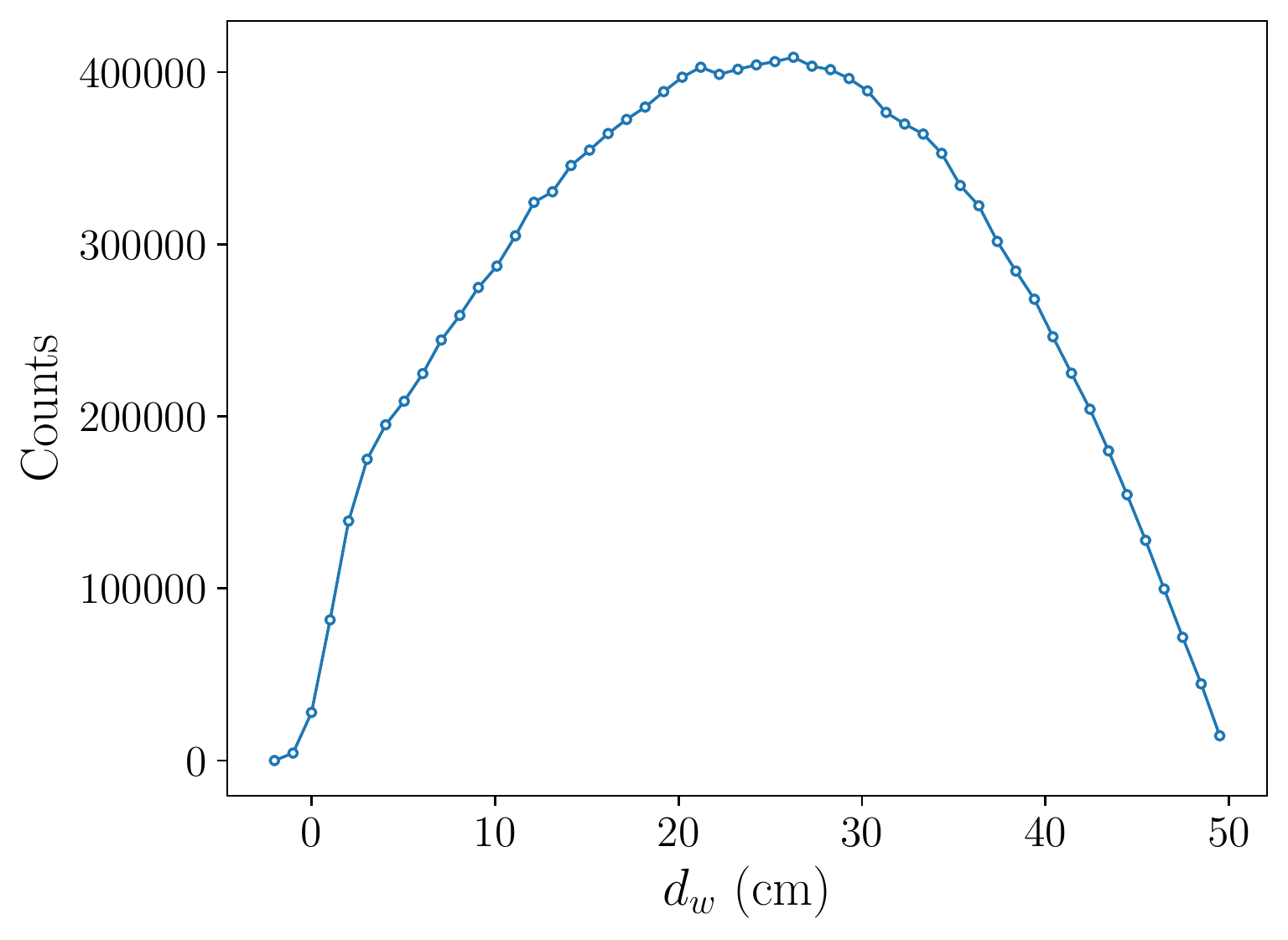}%
}

\subfloat[]{%
  \includegraphics[width=0.36\textwidth]{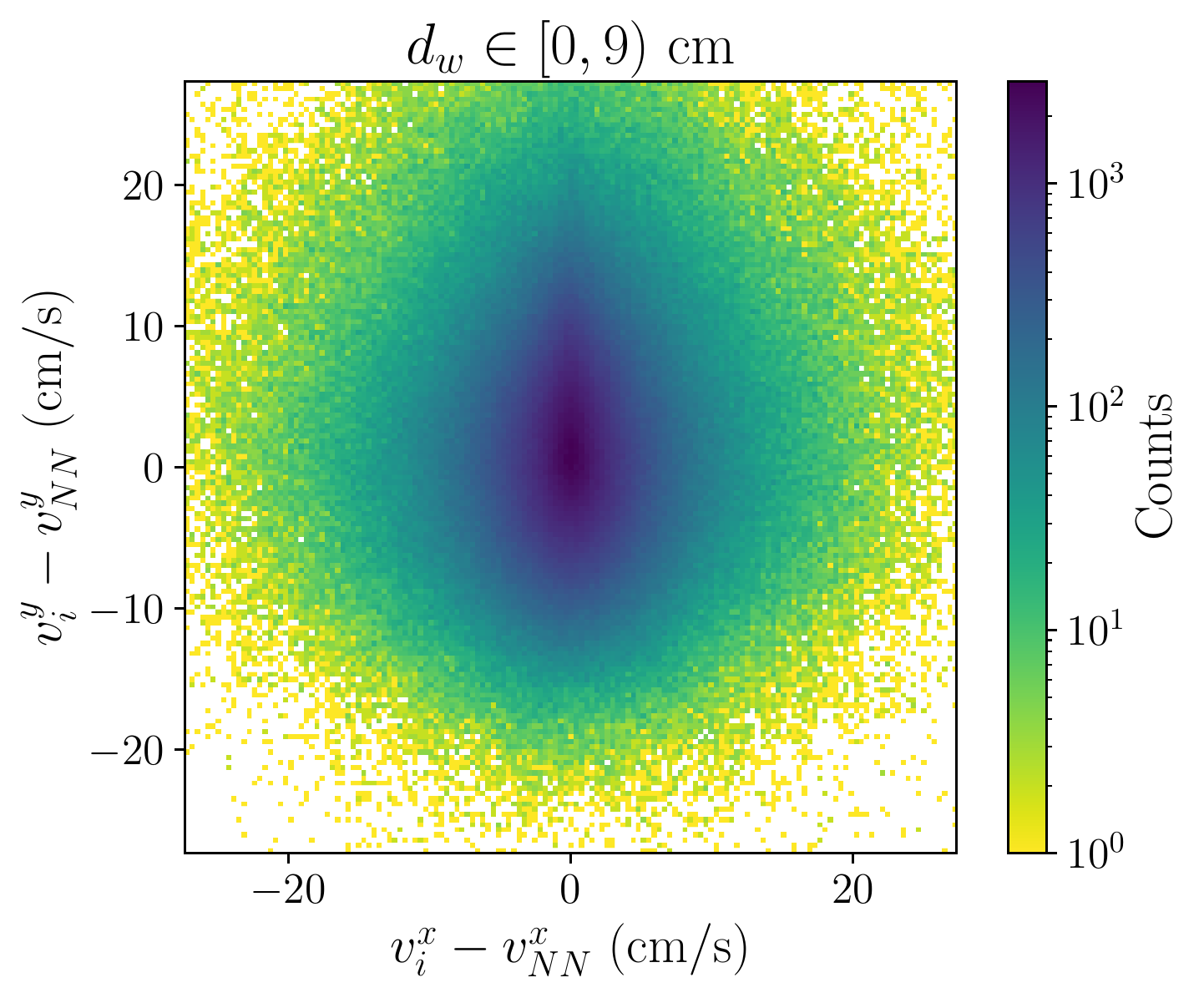}%
}
\hspace{0.01\textwidth}
\subfloat[]{%
  \includegraphics[width=0.36\textwidth]{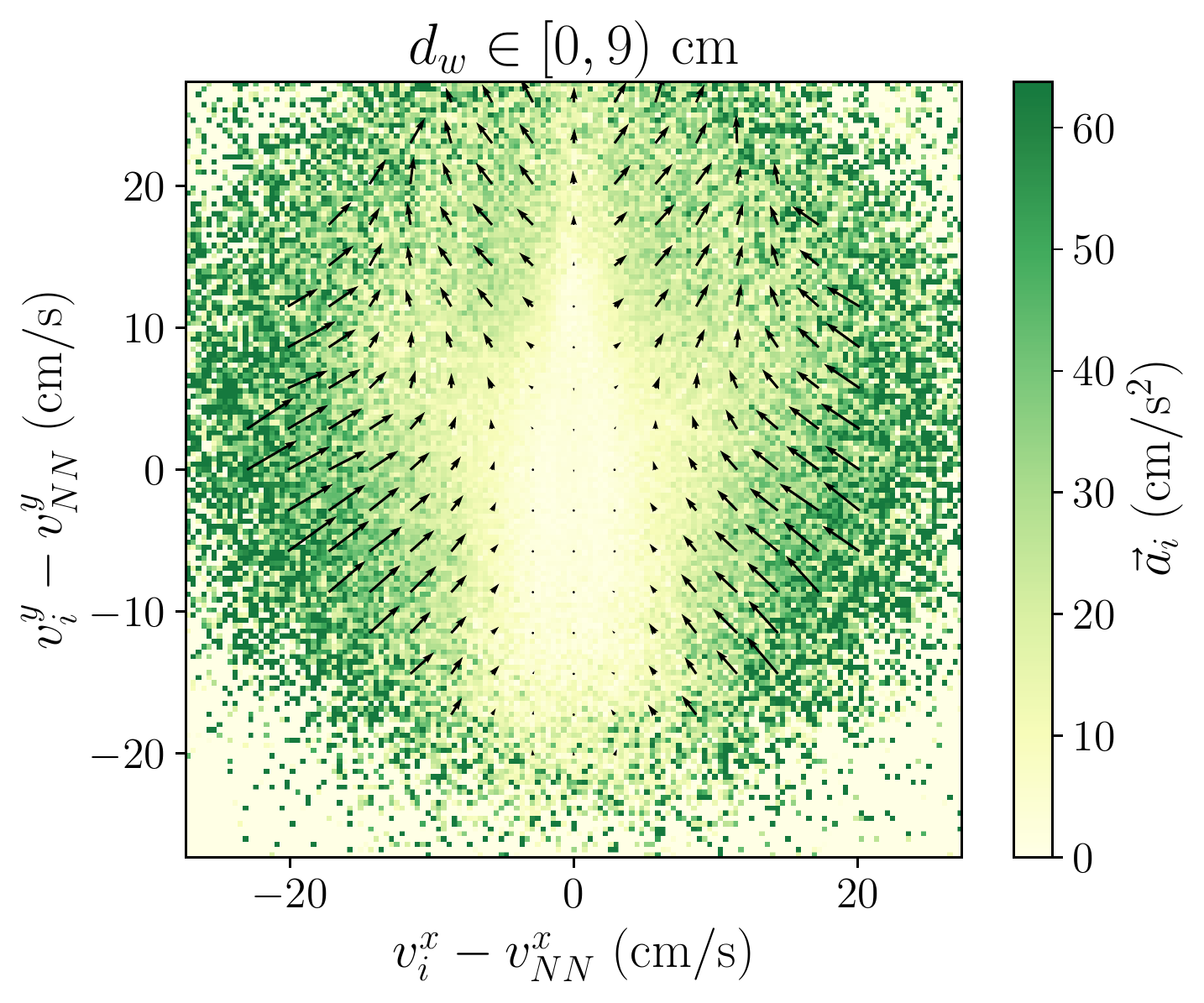}%
}

\subfloat[]{%
  \includegraphics[width=0.36\textwidth]{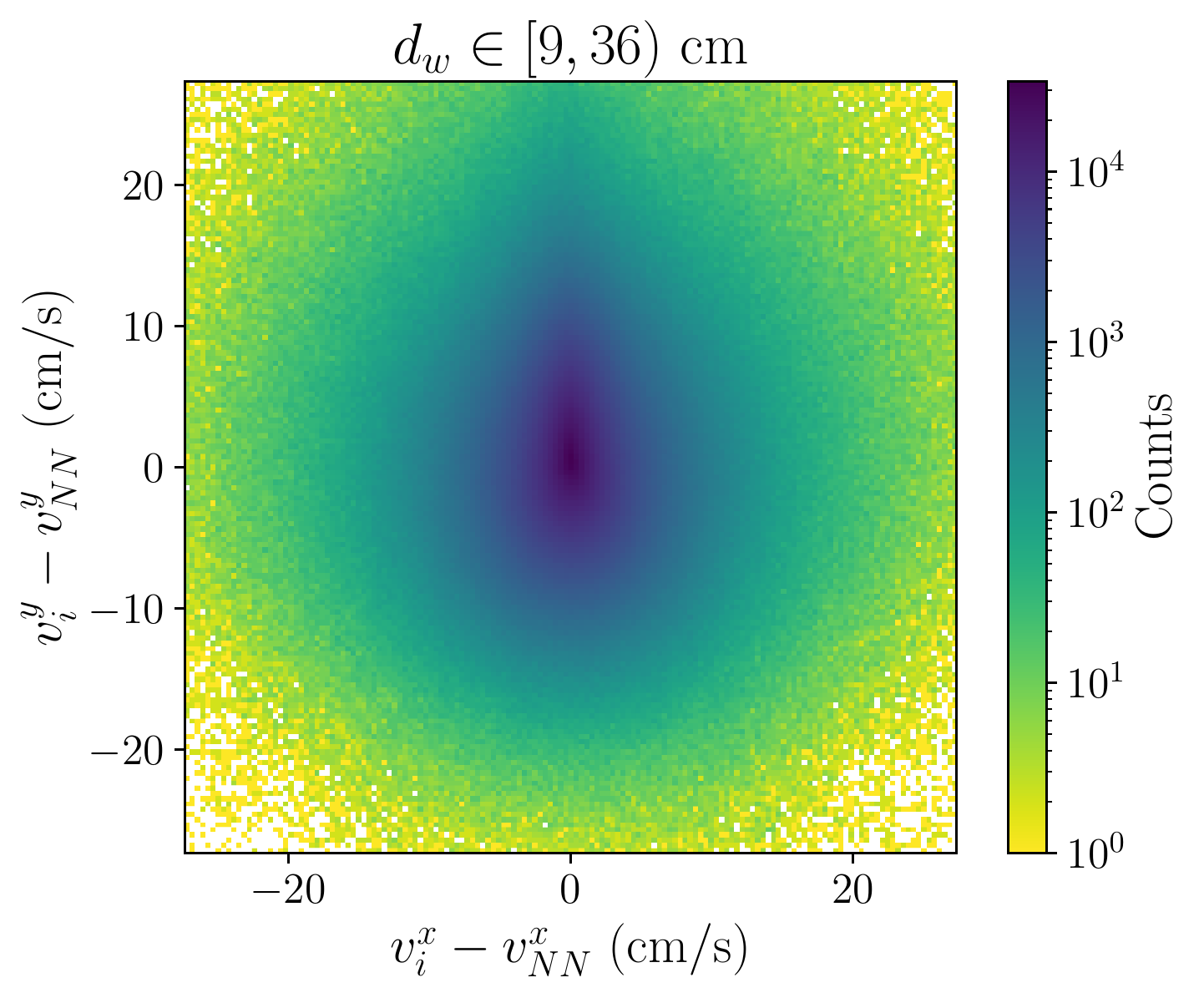}%
}
\hspace{0.01\textwidth}
\subfloat[]{%
  \includegraphics[width=0.36\textwidth]{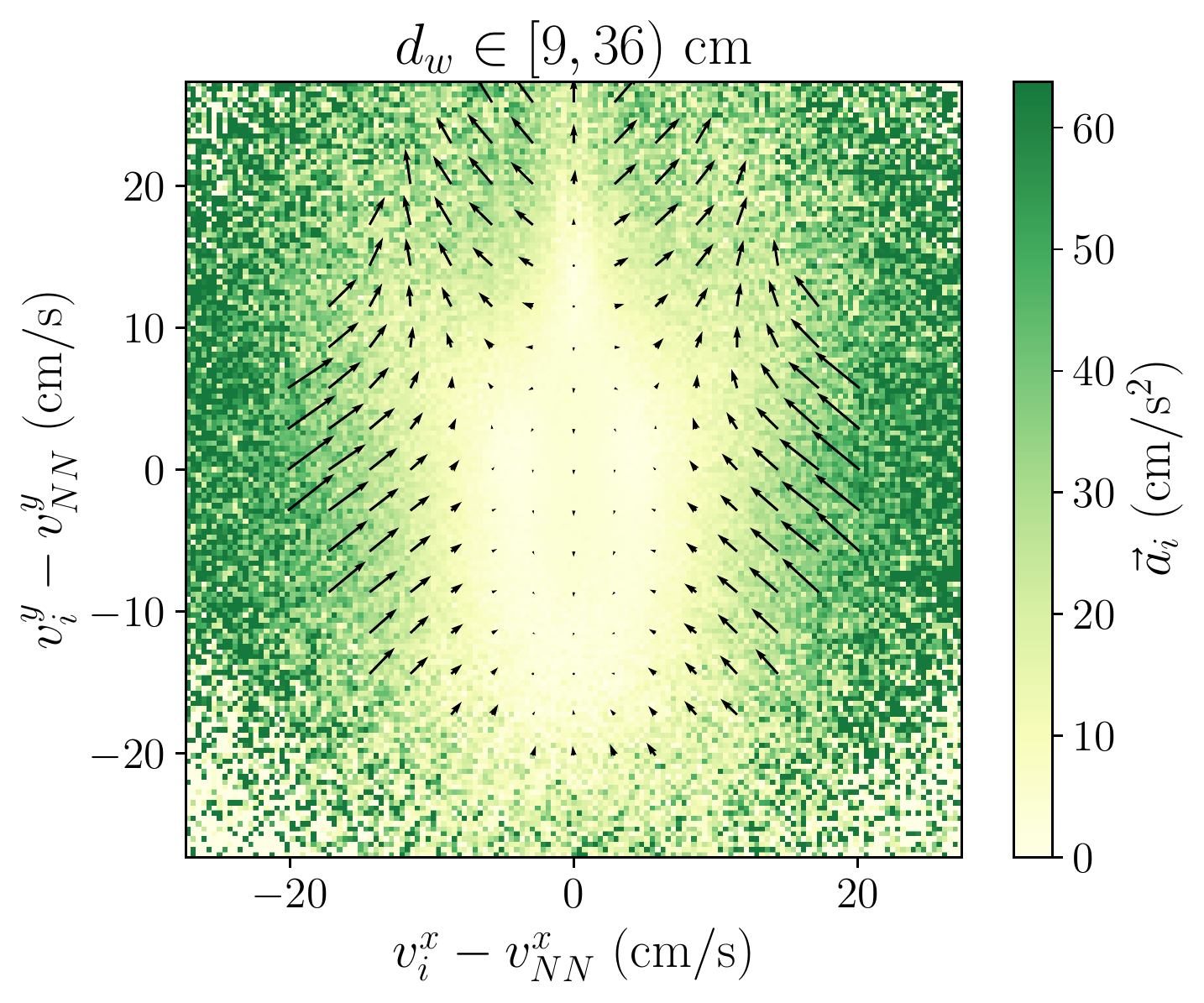}%
}

\subfloat[]{%
  \includegraphics[width=0.36\textwidth]{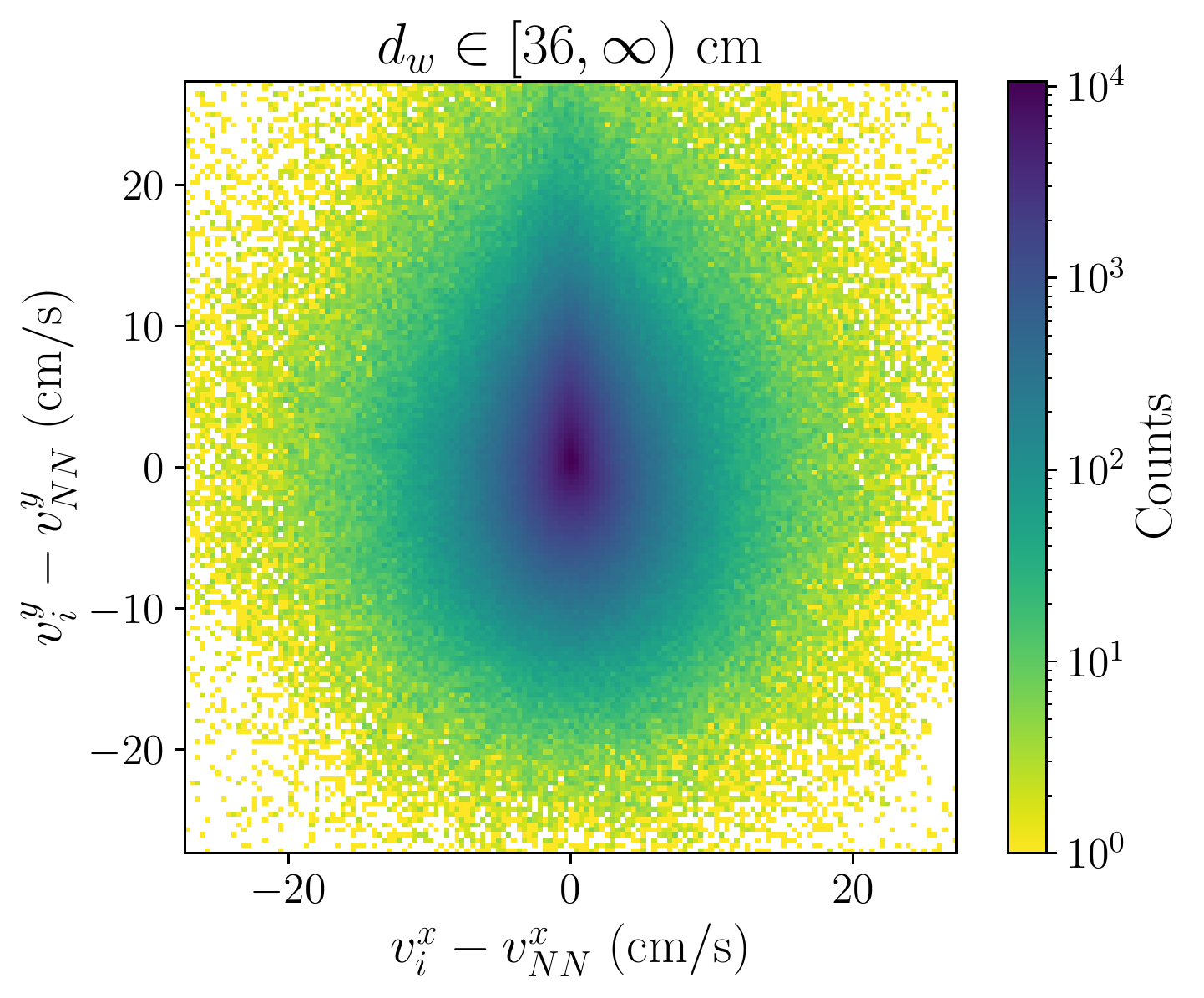}%
}
\hspace{0.01\textwidth}
\subfloat[]{%
  \includegraphics[width=0.36\textwidth]{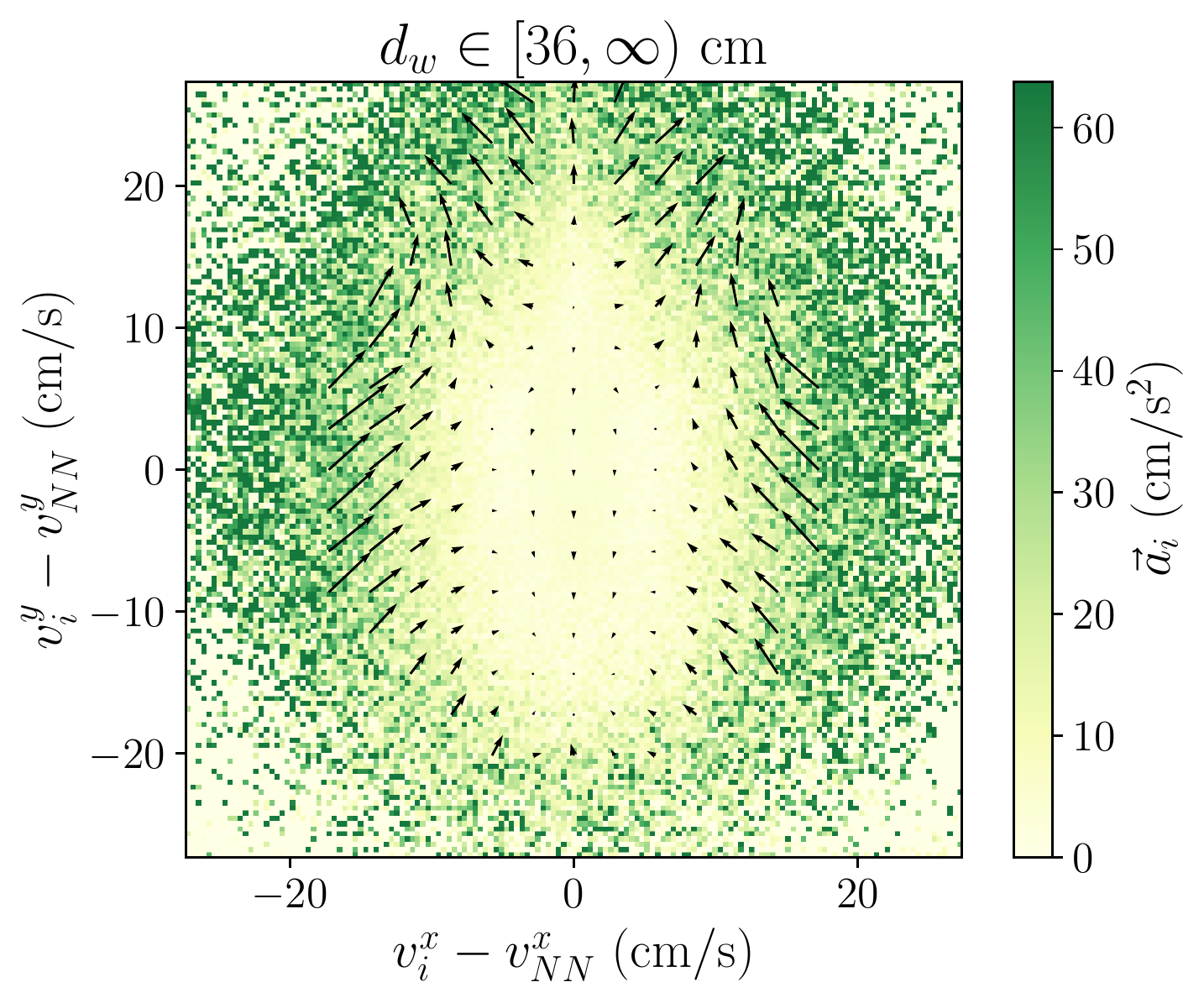}%
}
\caption{(a) Distribution of minimum distances to the walls $d_{w}\equiv \min ( x, L-x, y, L-y)$, for $L$ the side of the tank, and (b)-(g) alignment force maps for different $d_{w}$ of the individual $i$ for the experimental data.} \label{supp:fig:alig_dWall}
\end{figure*}

\begin{figure*}[t!p]
\subfloat[]{%
  \includegraphics[width=0.4\textwidth]{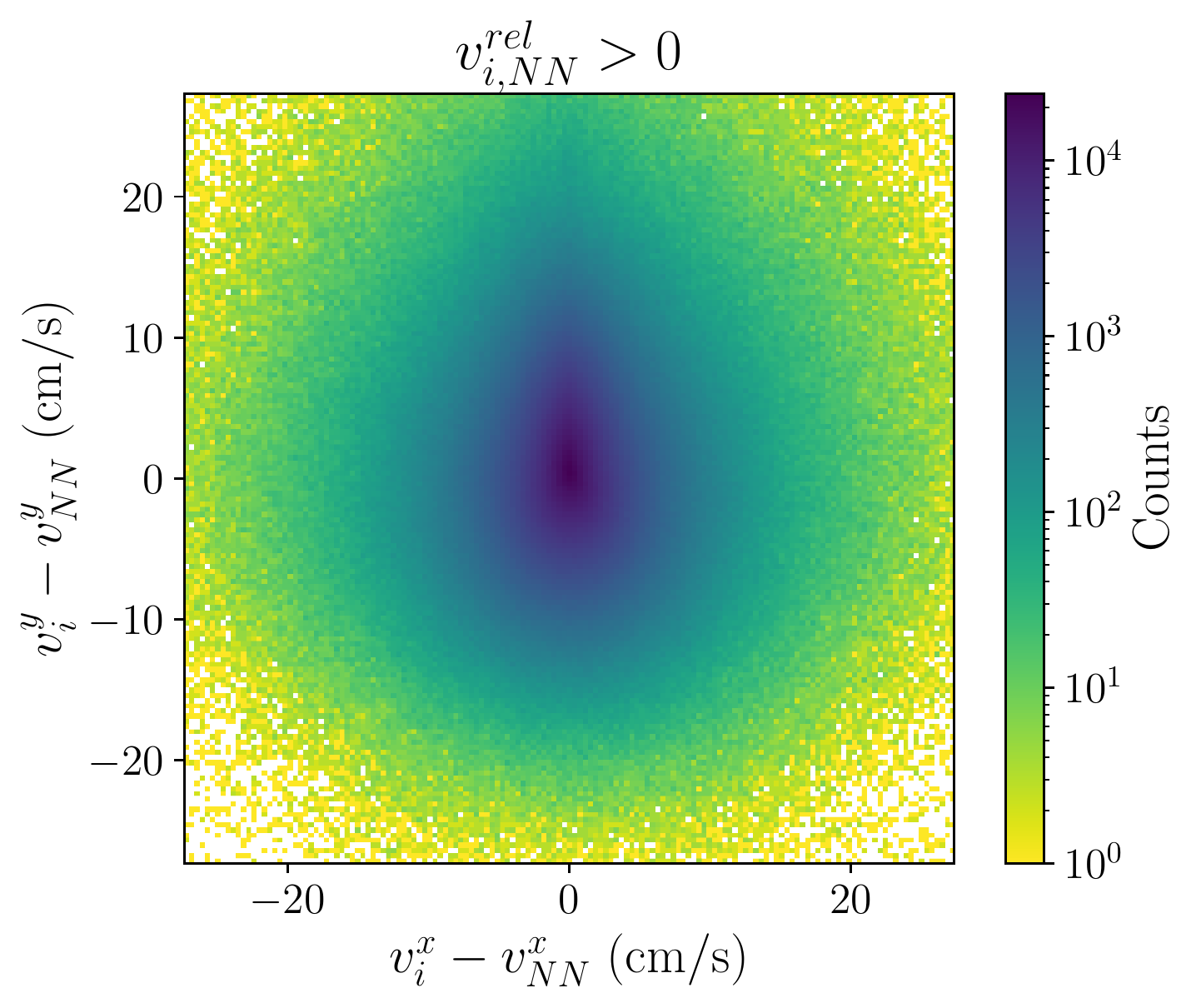}%
}
\hspace{0.01\textwidth}
\subfloat[]{%
  \includegraphics[width=0.4\textwidth]{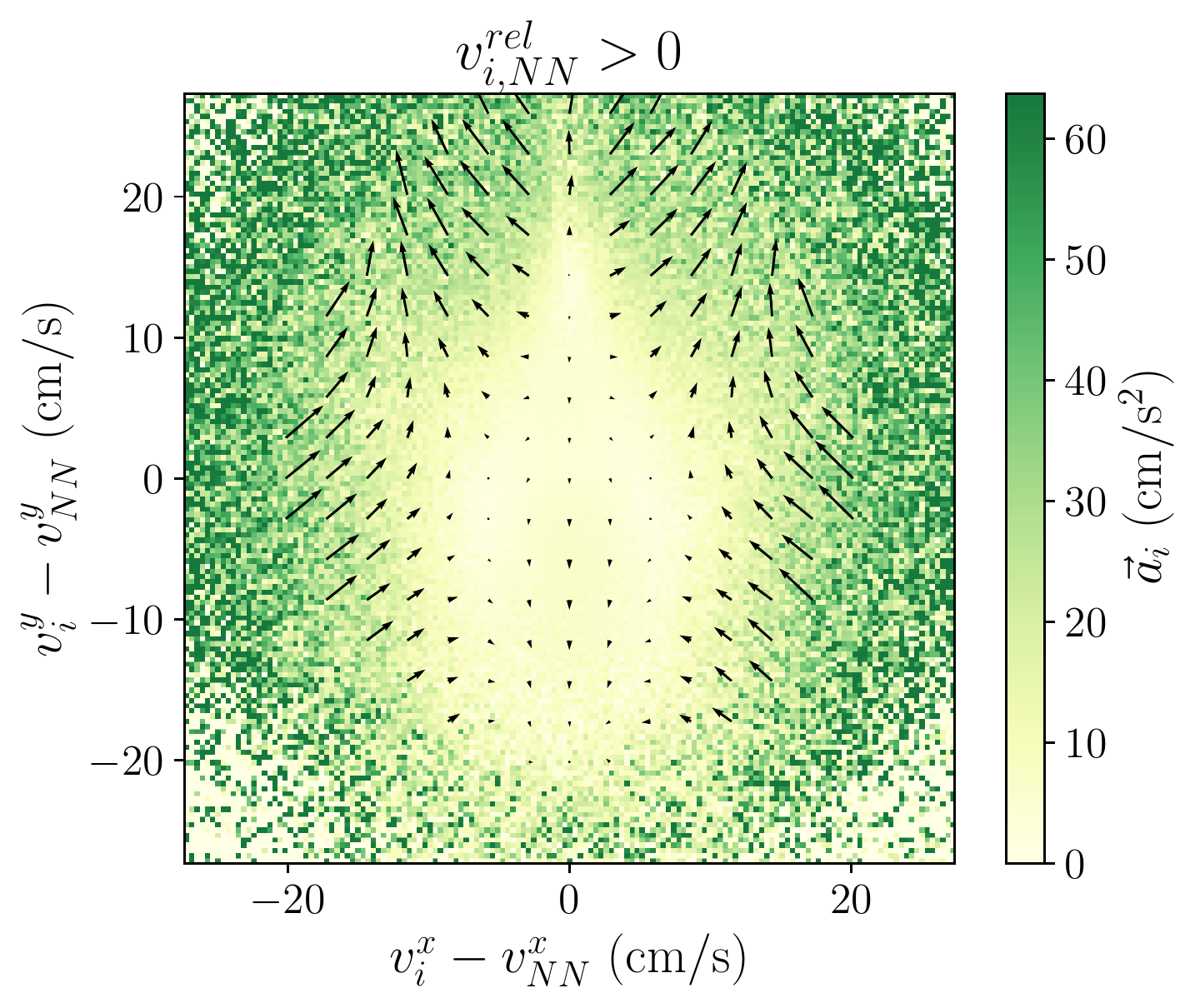}%
}

\subfloat[]{%
  \includegraphics[width=0.4\textwidth]{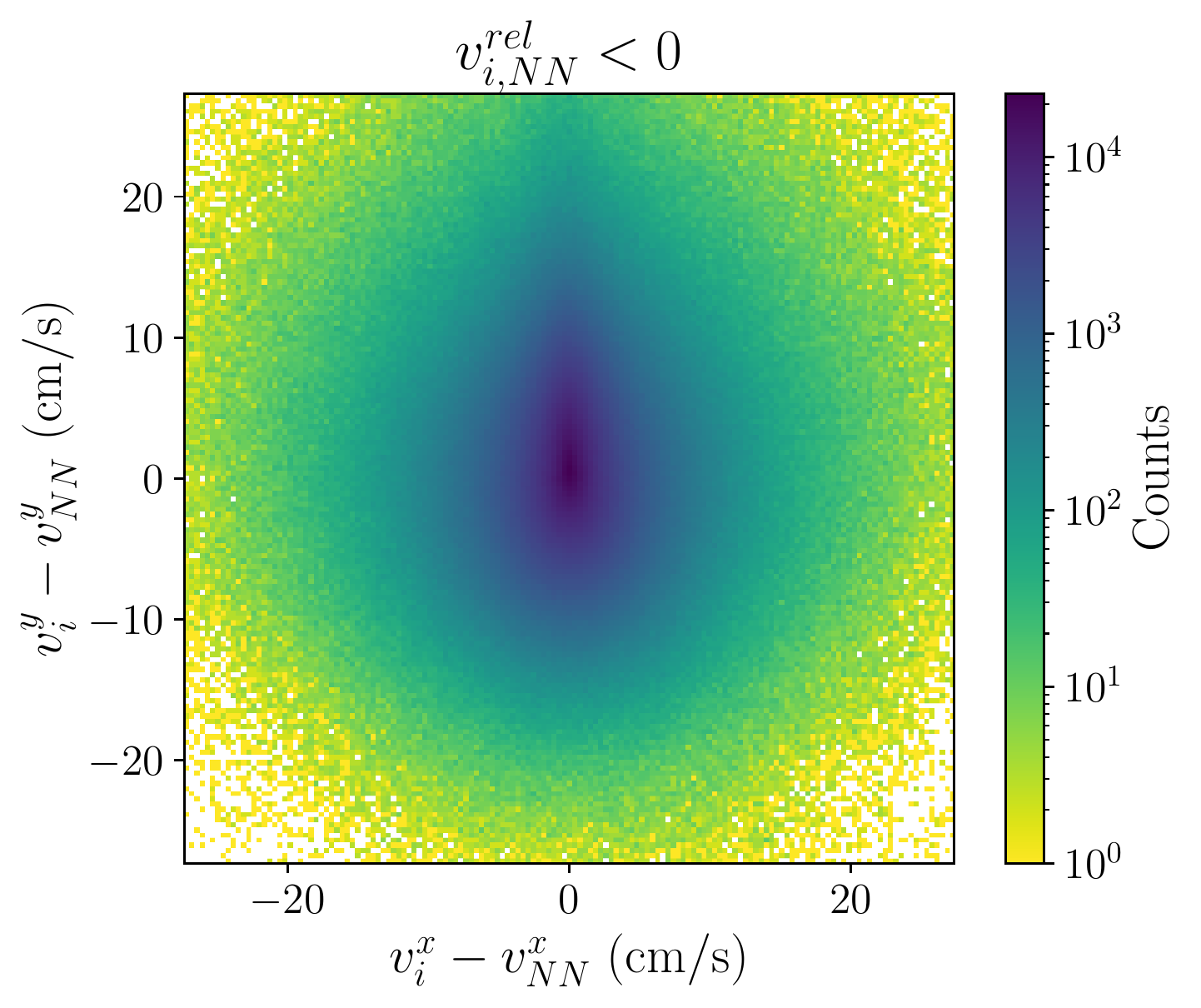}%
}
\hspace{0.01\textwidth}
\subfloat[]{%
  \includegraphics[width=0.4\textwidth]{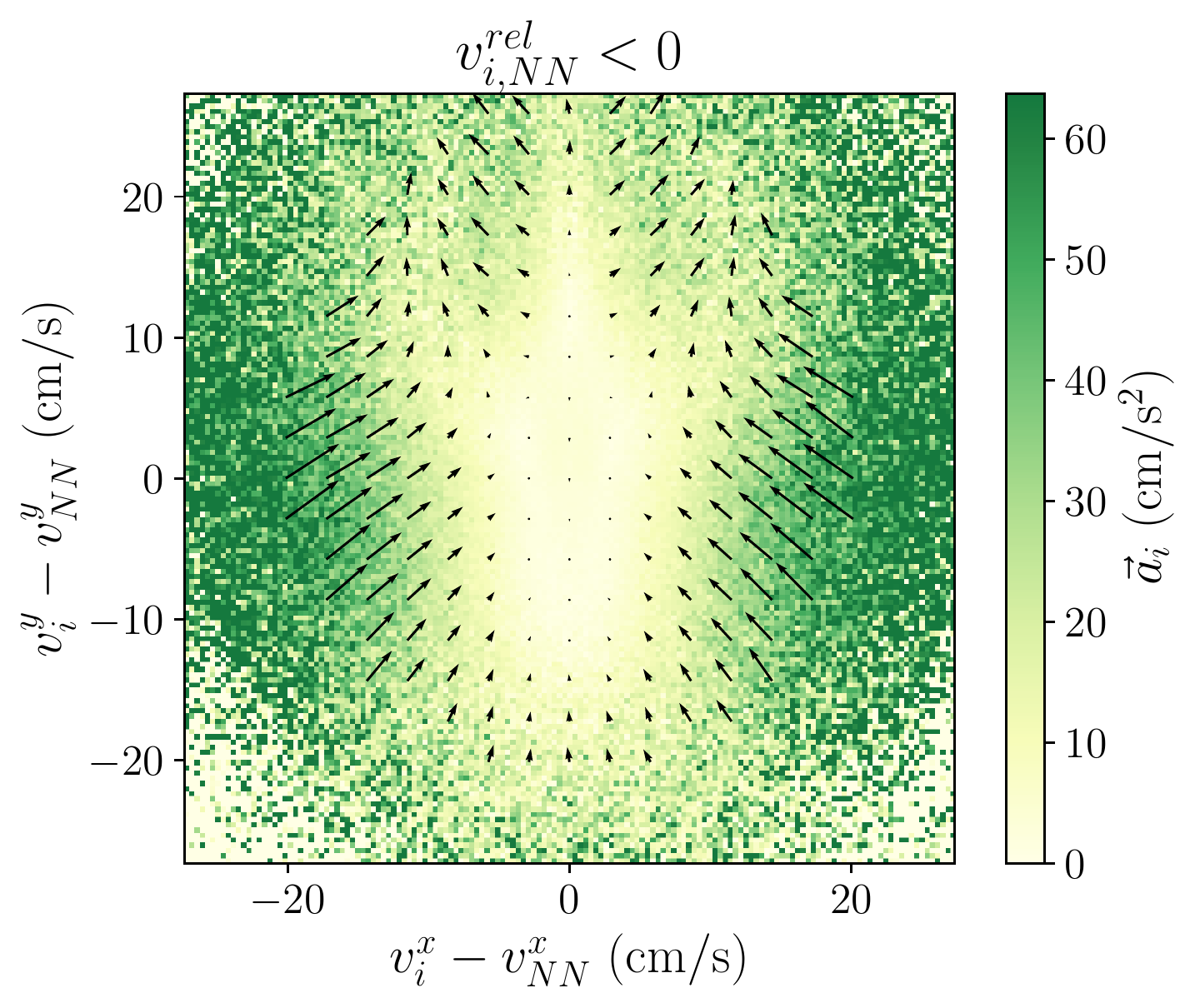}%
}
\caption{Alignment force maps in the case the individual is approaching or moving away from the neighbour for the experimental data. We employ the relative velocity of the individual $i$ with the nearest neighbour $NN$ projected along the direction of their relative position, $v_{i,NN}^{rel} \equiv (\vec{v}_{i} -\vec{v}_{NN} ) \cdot \frac{\vec{x}_{i} -\vec{x}_{NN}}{\left| \vec{x}_{i} -\vec{x}_{NN} \right| }$. In (a), (b) $v_{i,NN}^{rel} > 0$ and the individual is moving away from the neighbour. In (c), (d) $v_{i,NN}^{rel} < 0$ and the individual is approaching the neighbour. We notice in this last case the alignment region is slightly stronger.}\label{supp:fig:alig_approachingLeaving}
\end{figure*}

\begin{figure*}[t!p]
\subfloat[]{%
  \includegraphics[width=0.4\textwidth]{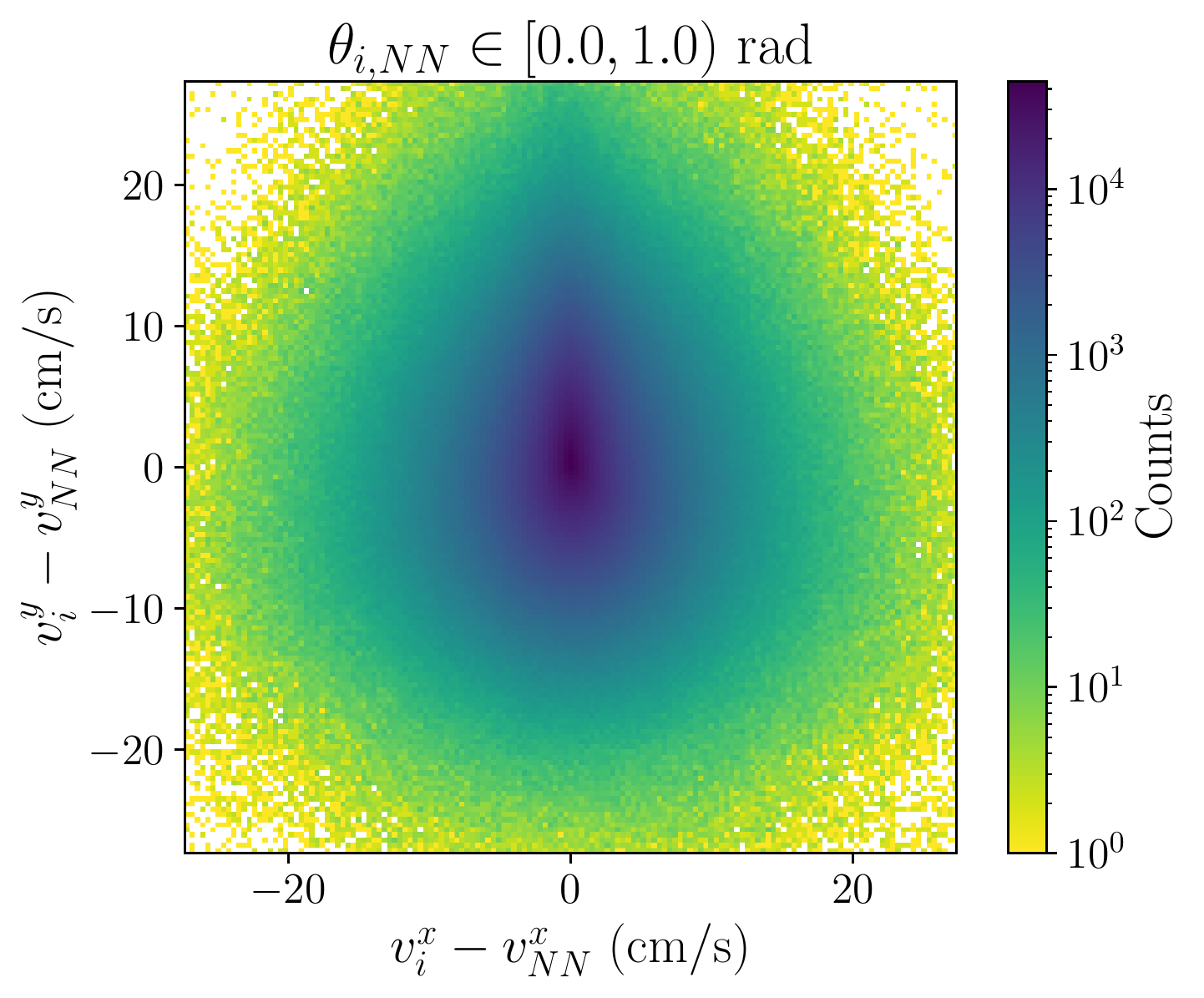}%
}
\hspace{0.01\textwidth}
\subfloat[]{%
  \includegraphics[width=0.4\textwidth]{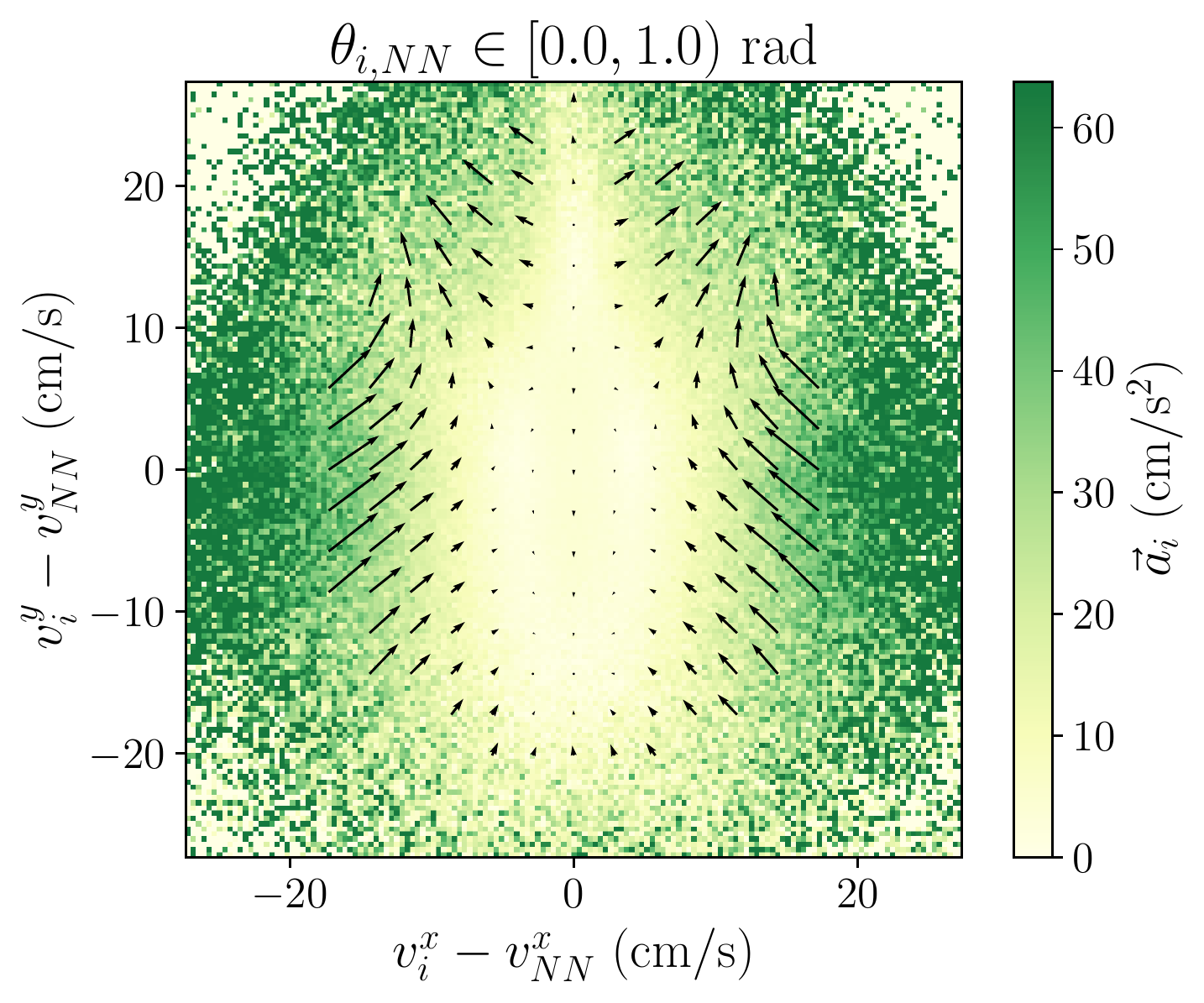}%
}

\subfloat[]{%
  \includegraphics[width=0.4\textwidth]{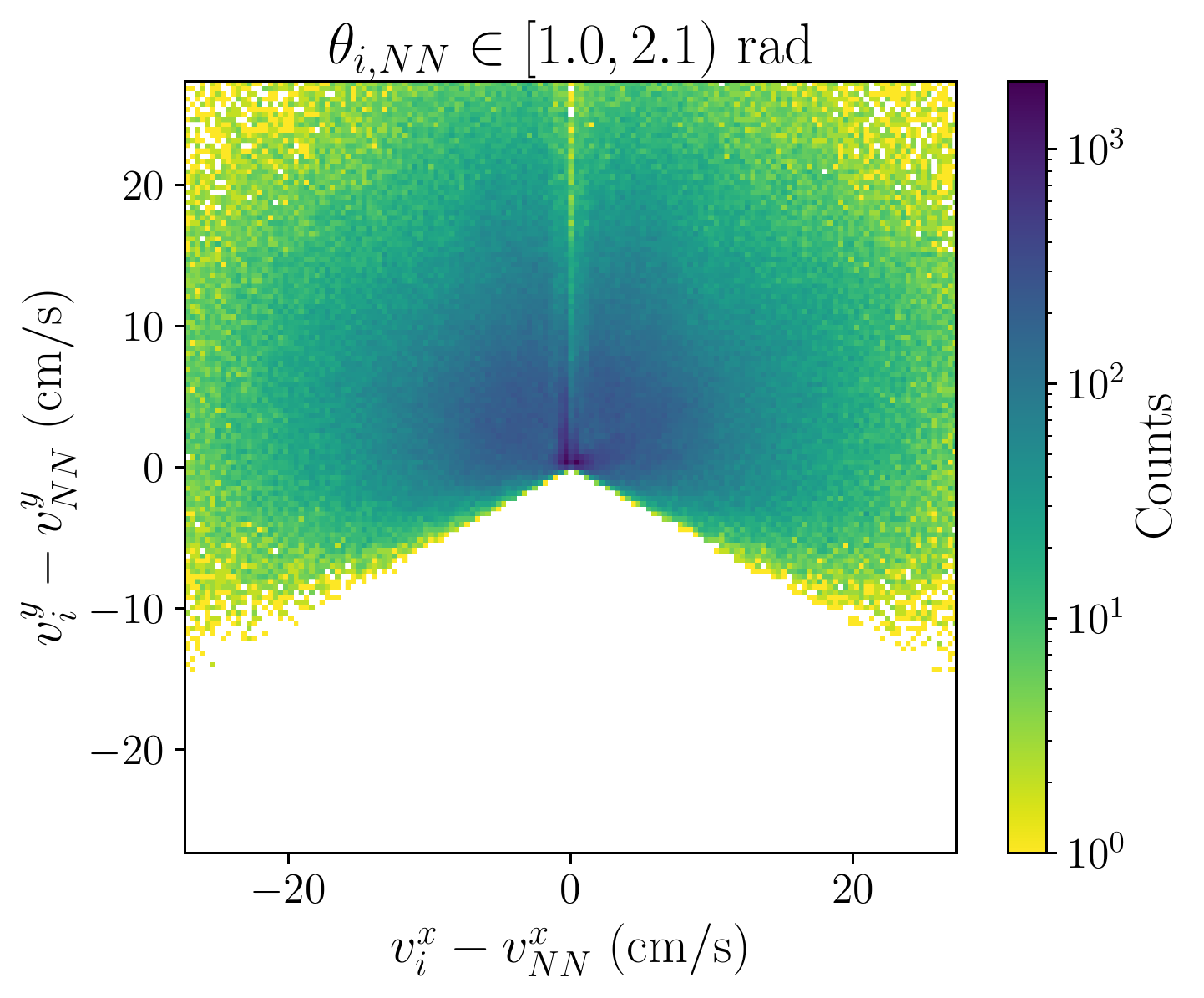}%
}
\hspace{0.01\textwidth}
\subfloat[]{%
  \includegraphics[width=0.4\textwidth]{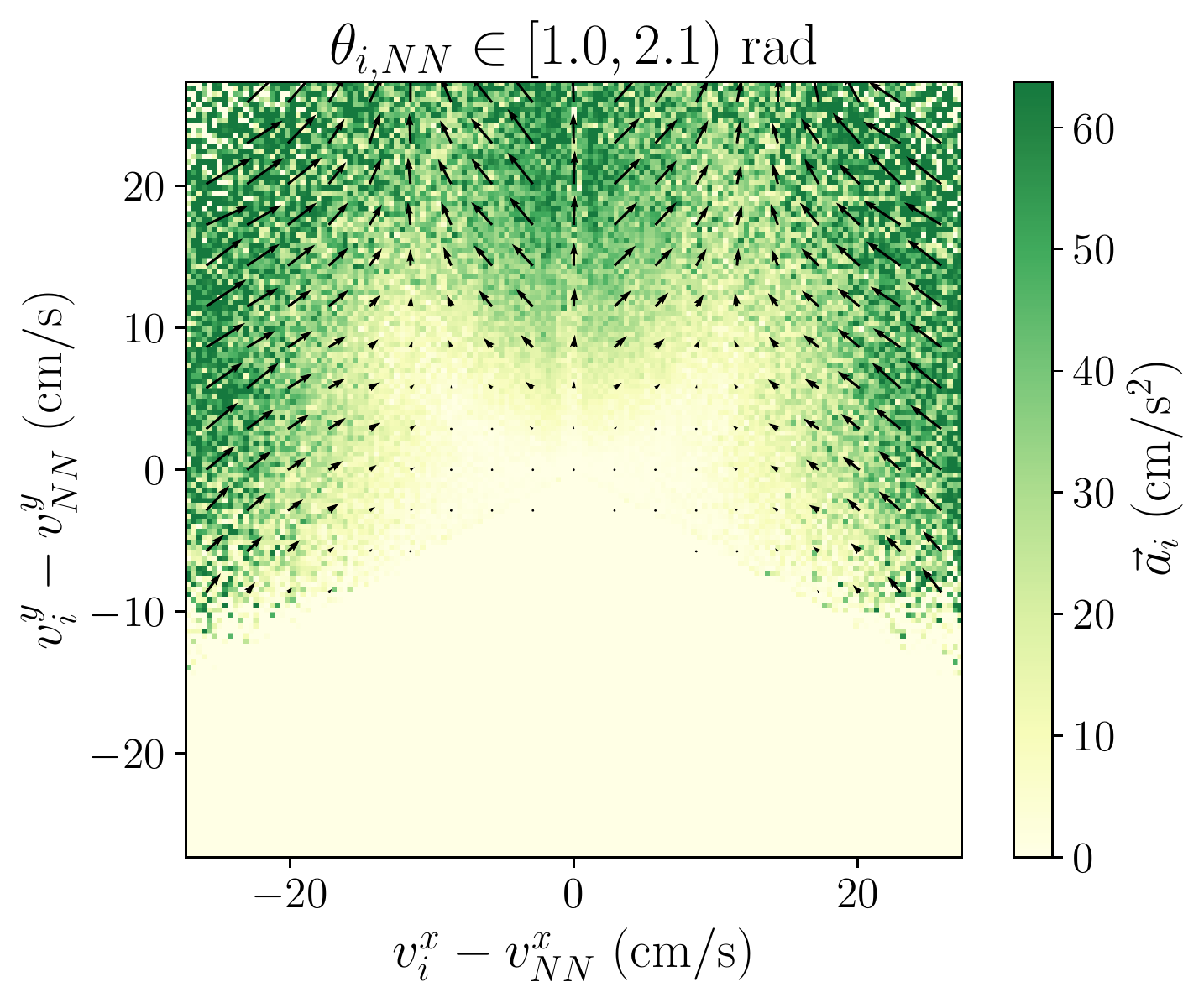}%
}

\subfloat[]{%
  \includegraphics[width=0.4\textwidth]{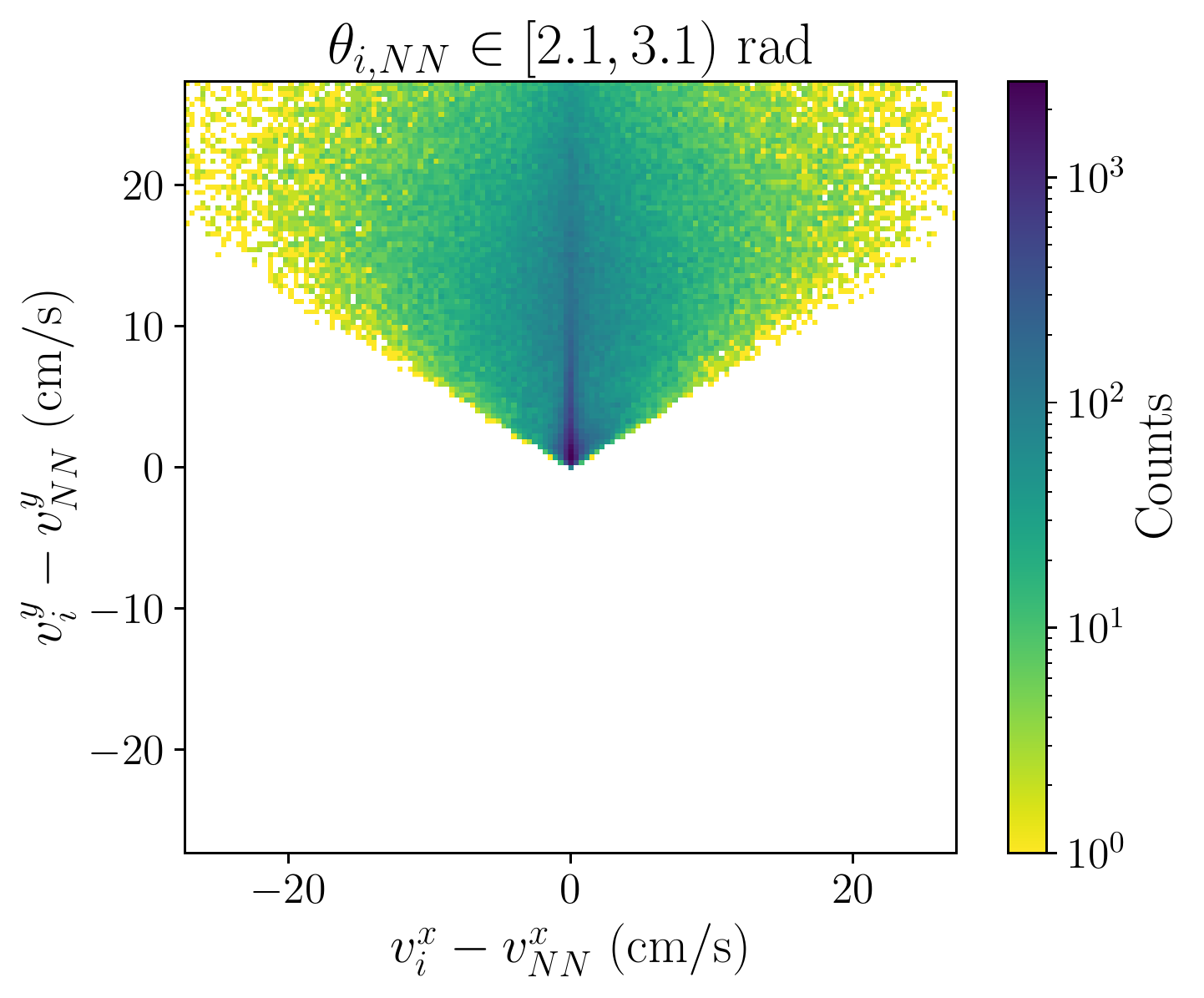}%
}
\hspace{0.01\textwidth}
\subfloat[]{%
  \includegraphics[width=0.4\textwidth]{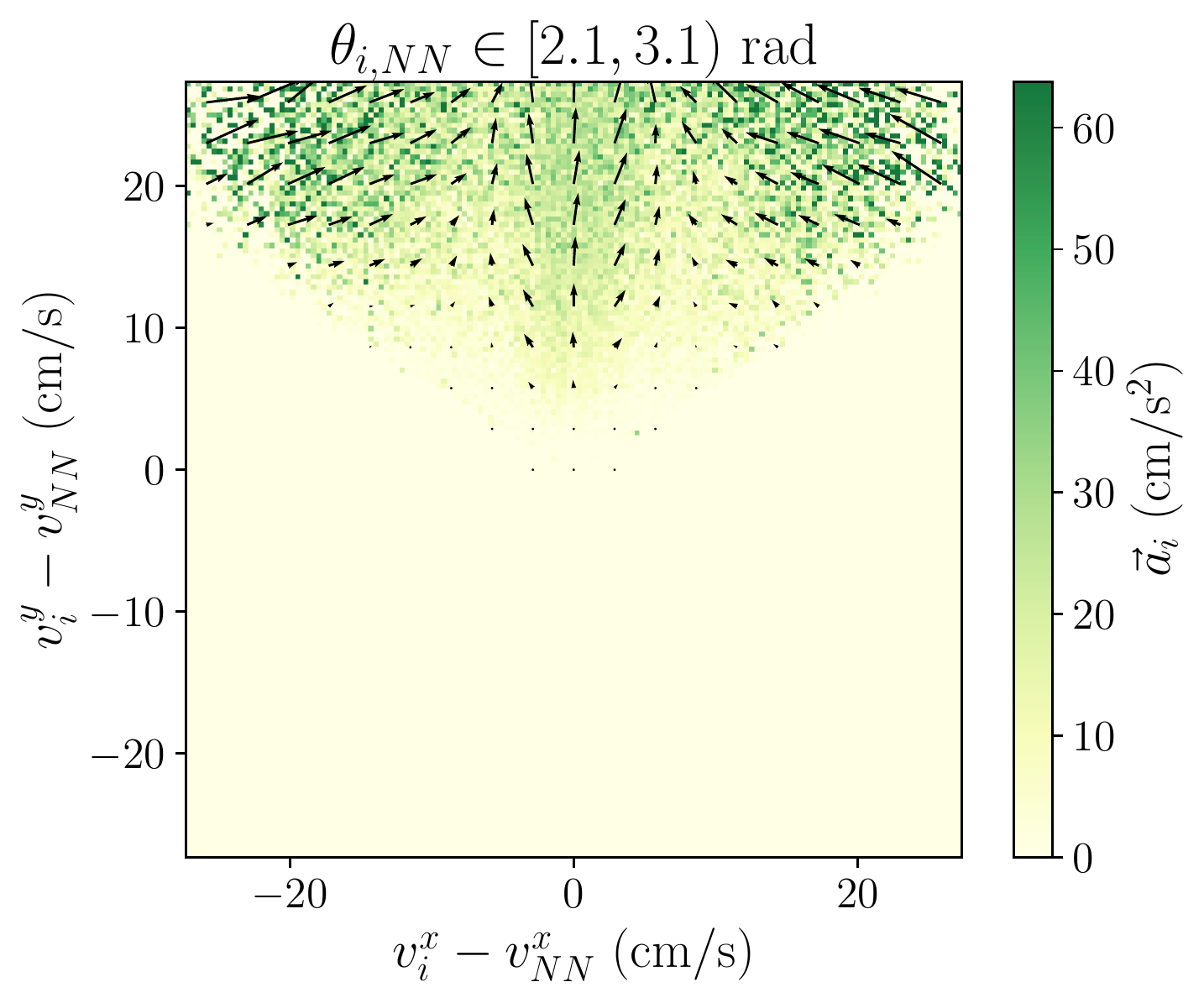}%
}
\caption{Alignment force maps for different heading orientations $\theta_{i,NN}$ between an individual $i$ and its nearest neighbour $NN$ for the experimental data.}\label{supp:fig:alig_headingOrientations}
\end{figure*}

\begin{figure*}[t!p]
\subfloat[]{%
  \includegraphics[width=0.4\textwidth]{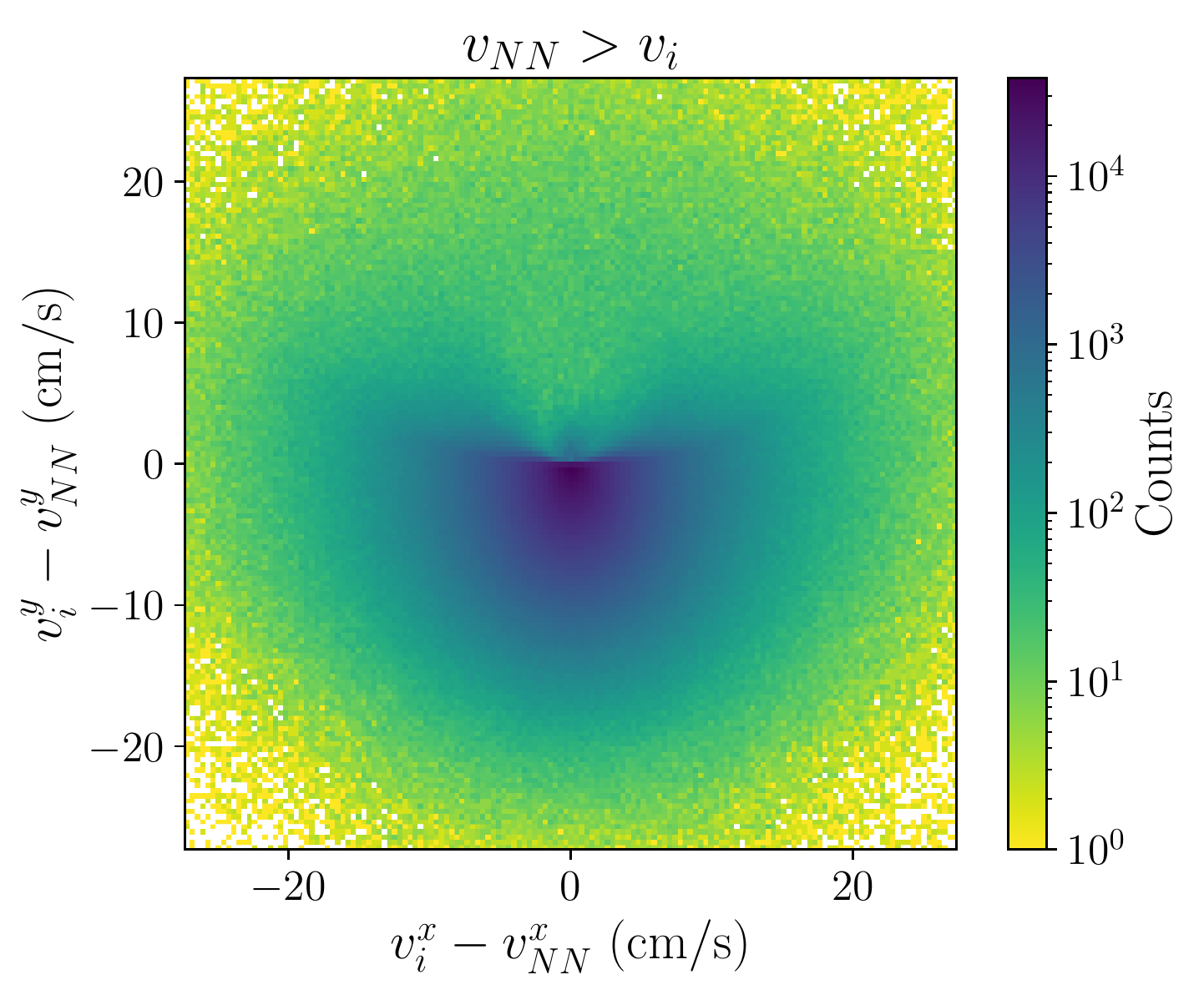}%
}
\hspace{0.01\textwidth}
\subfloat[]{%
  \includegraphics[width=0.4\textwidth]{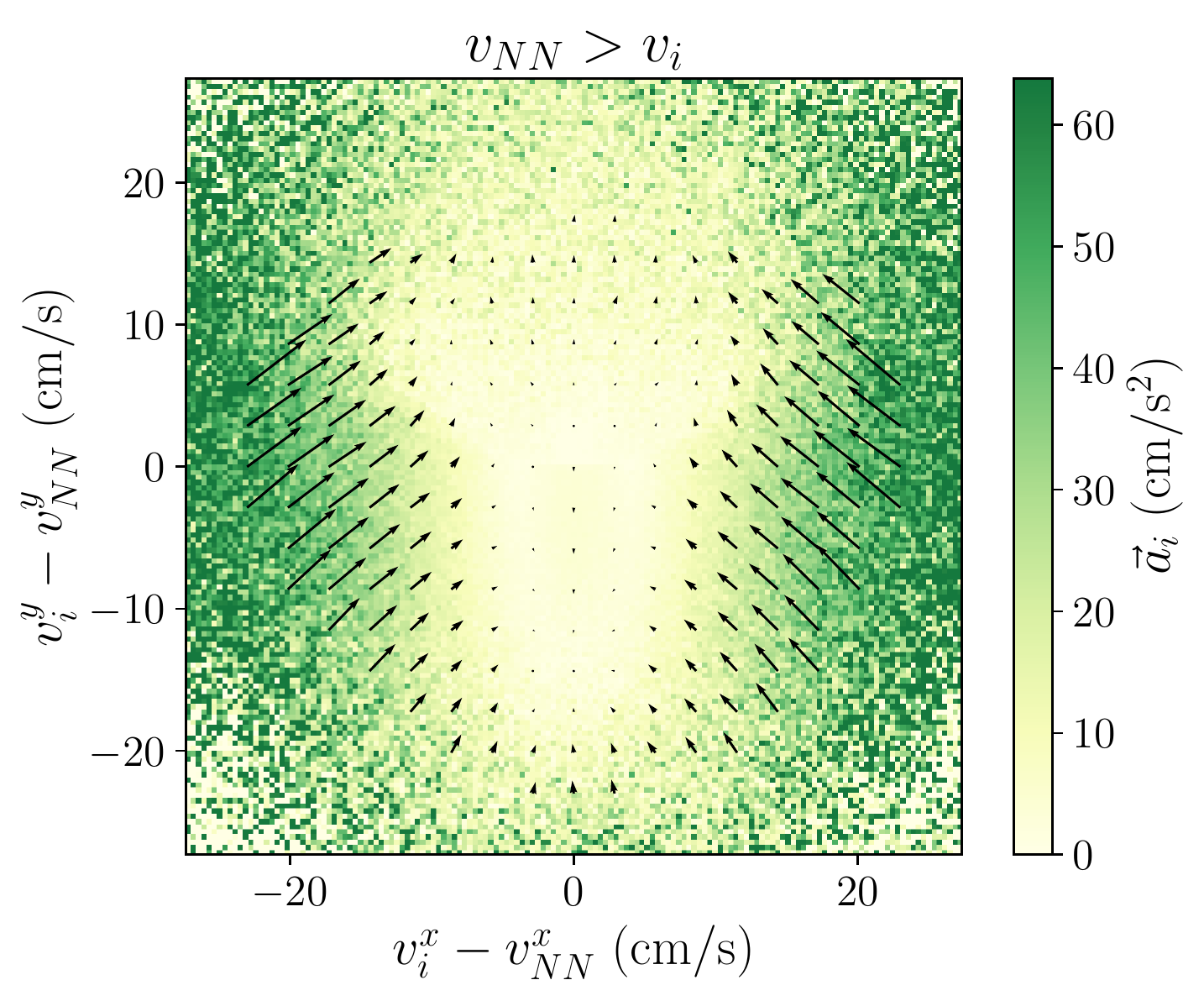}%
}

\subfloat[]{%
  \includegraphics[width=0.4\textwidth]{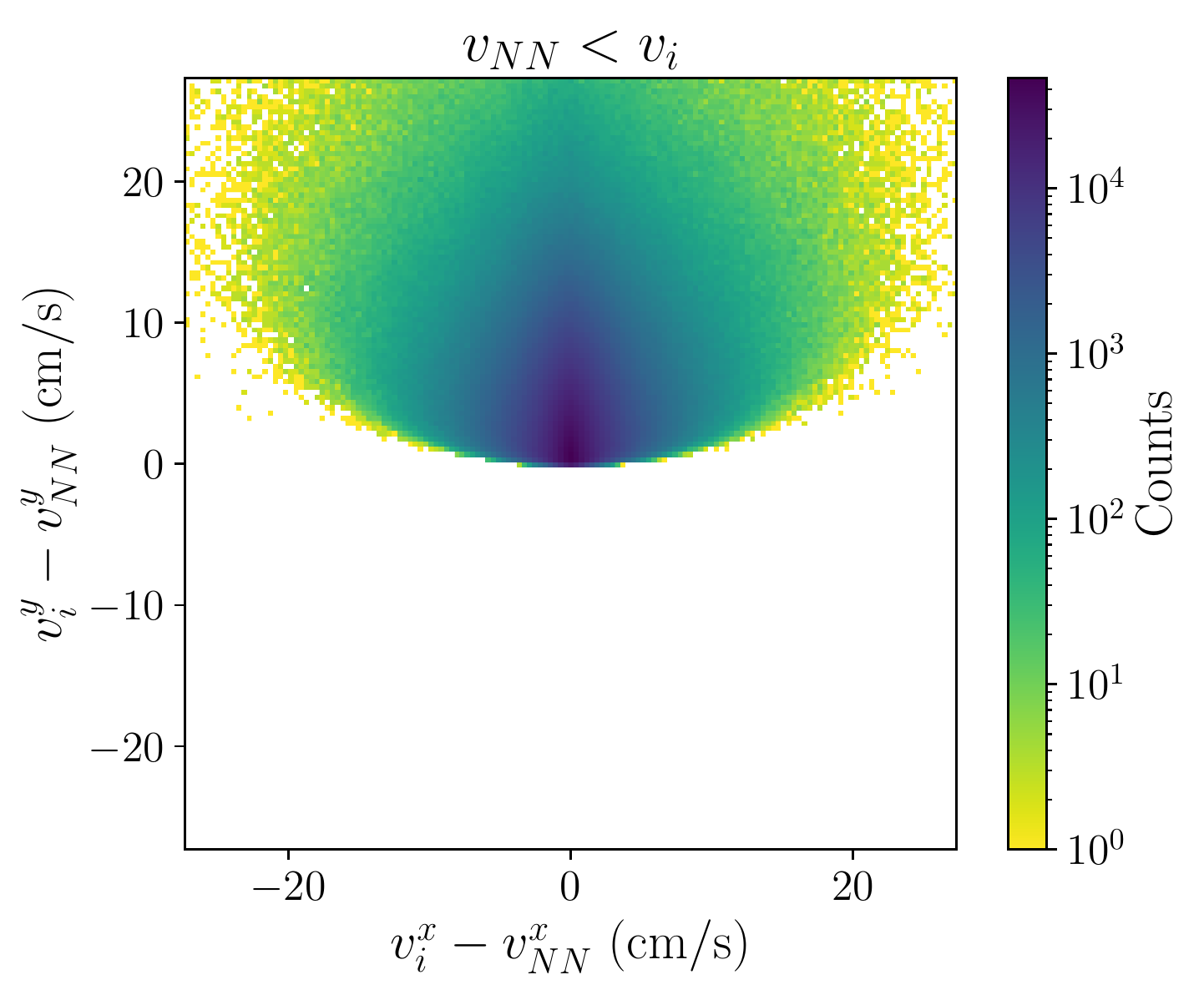}%
}
\hspace{0.01\textwidth}
\subfloat[]{%
  \includegraphics[width=0.4\textwidth]{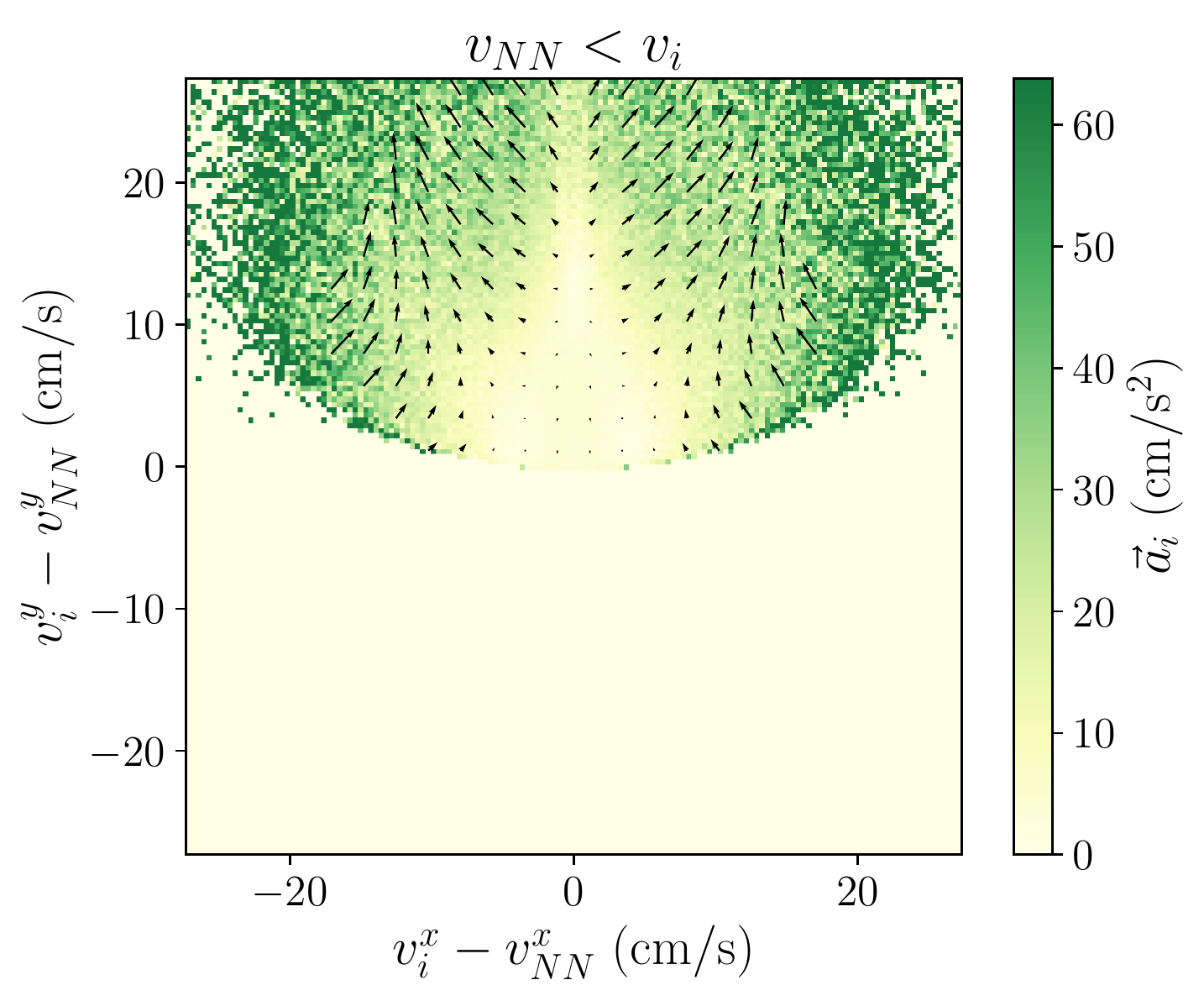}%
}
\caption{Alignment force maps for different relative velocities between the neighbour and the individual for the experimental data. They essentially occupy excluded regions in the force map.} \label{supp:fig:alig_modVelNn_i}
\end{figure*}

\begin{figure*}[t!p]
\subfloat[]{%
  \includegraphics[width=0.4\textwidth]{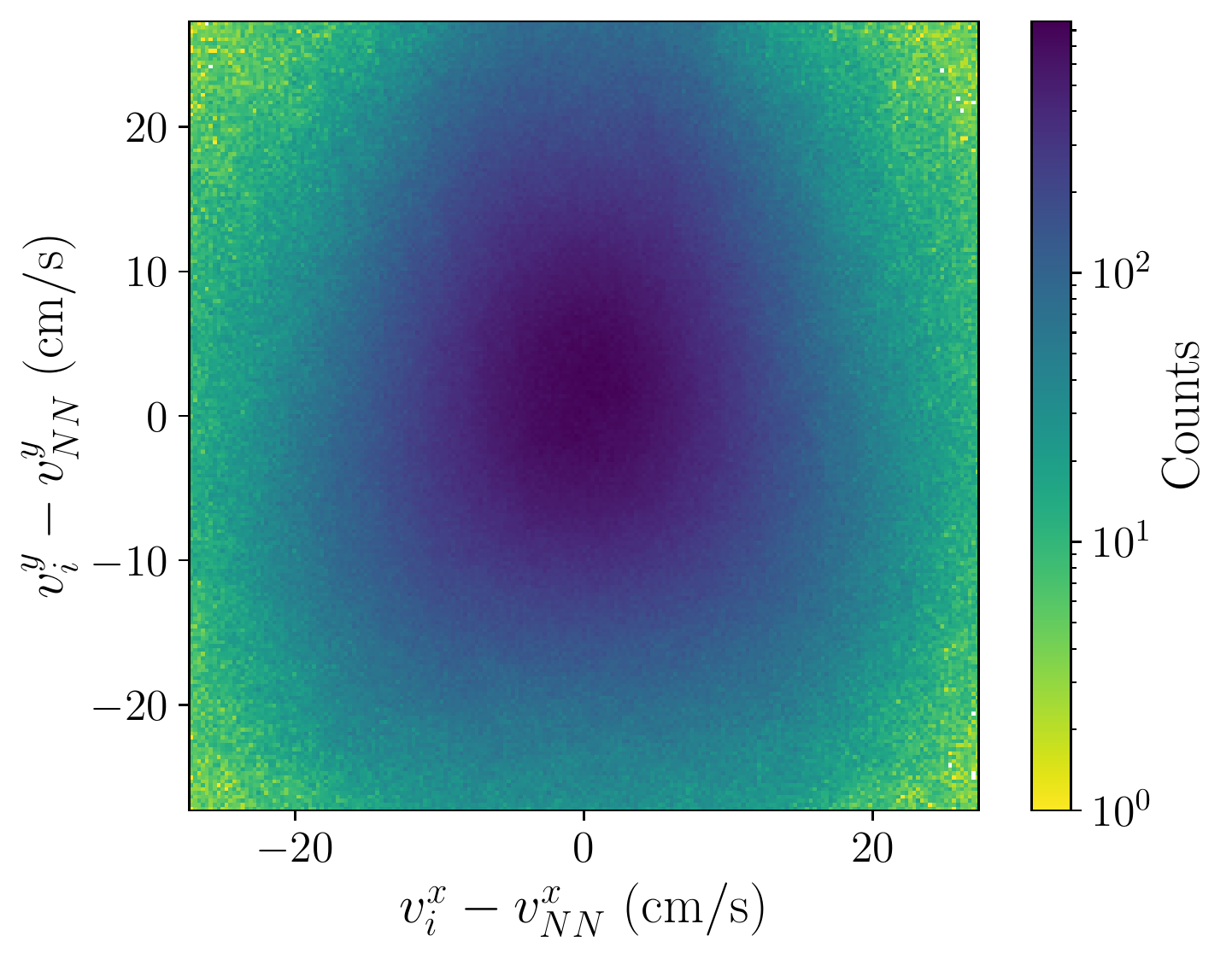}%
}
\hspace{0.01\textwidth}
\subfloat[]{%
  \includegraphics[width=0.4\textwidth]{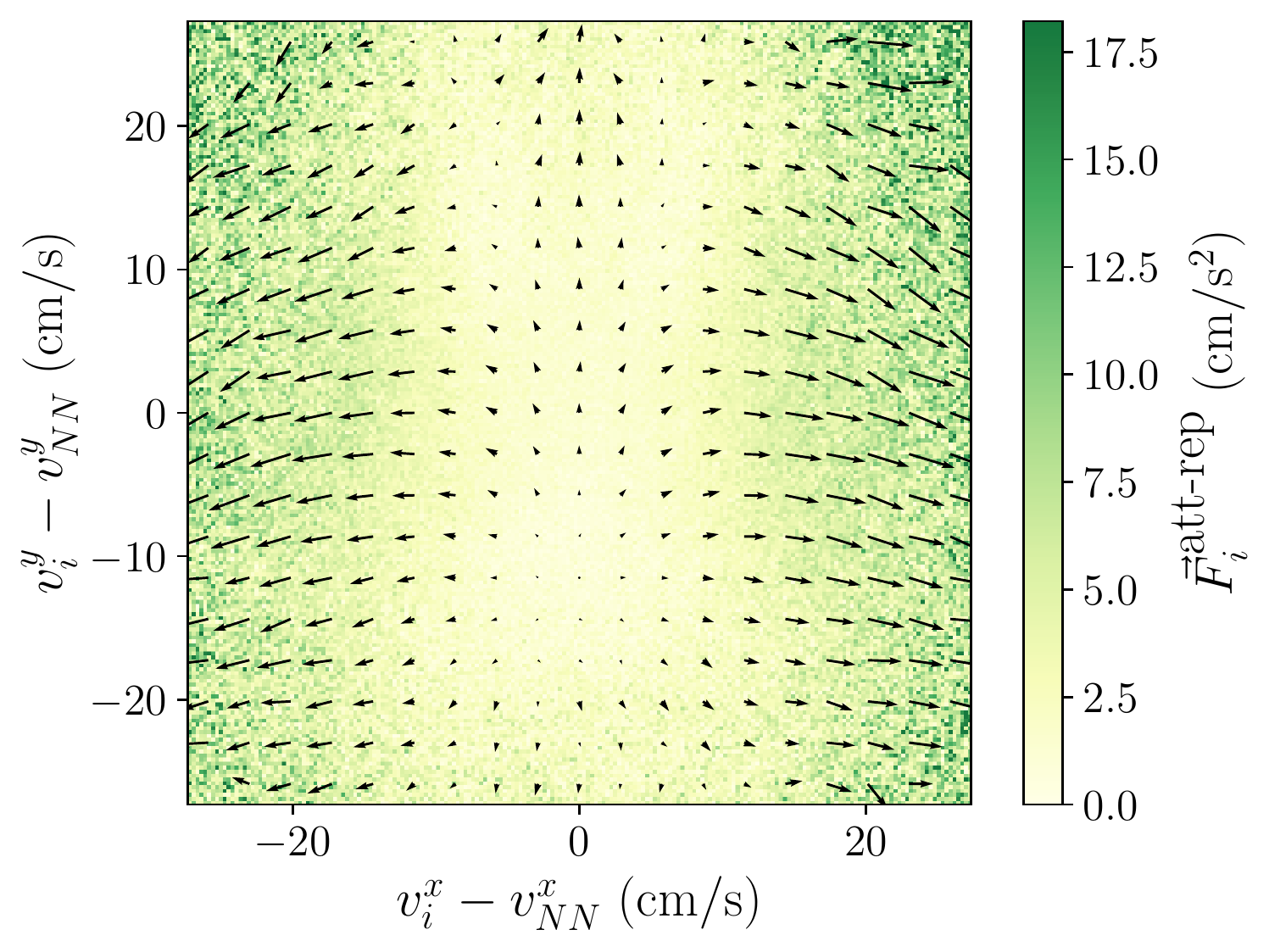}%
}

\subfloat[]{%
  \includegraphics[width=0.4\textwidth]{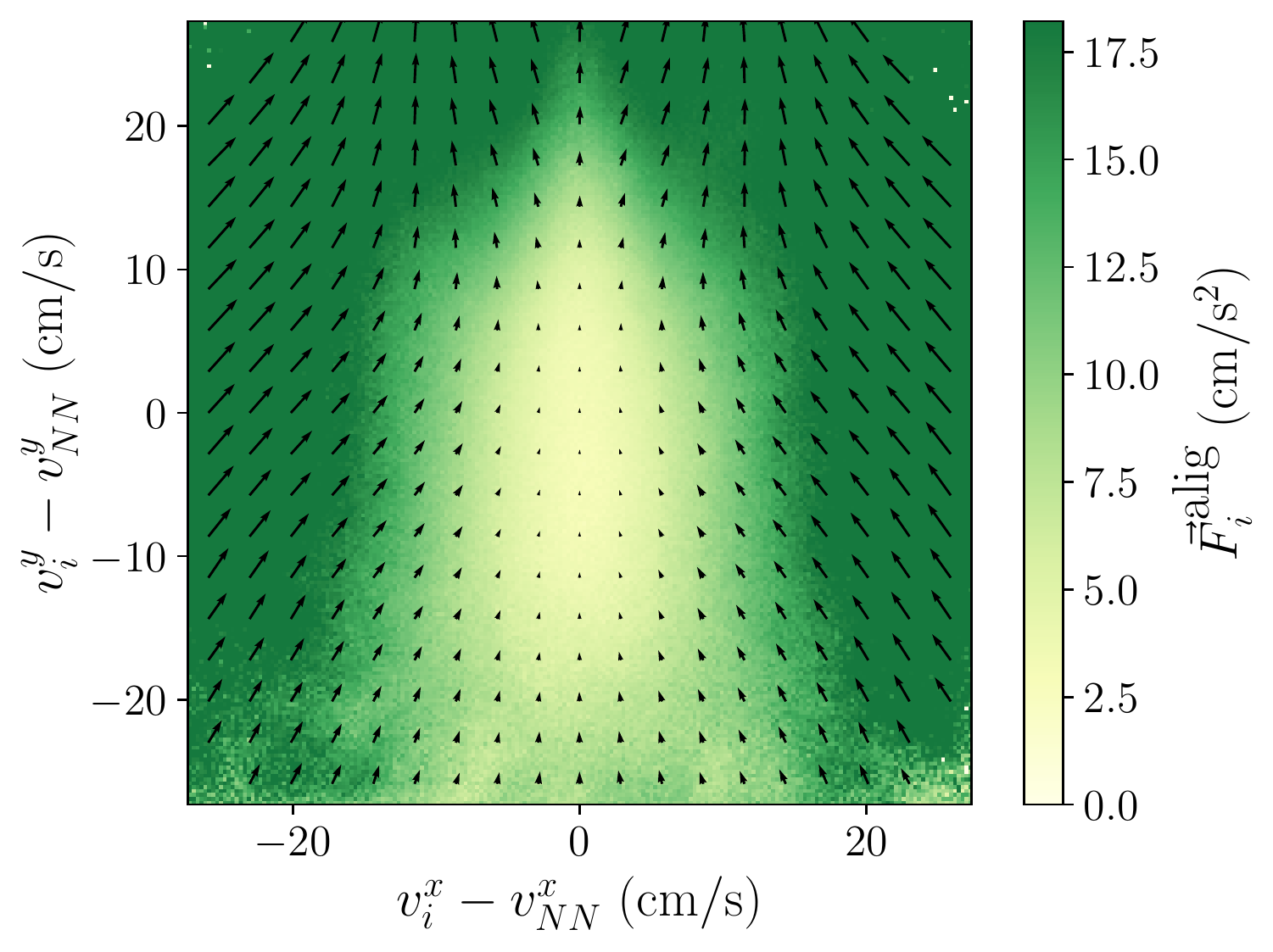}%
}
\hspace{0.01\textwidth}
\subfloat[]{%
  \includegraphics[width=0.4\textwidth]{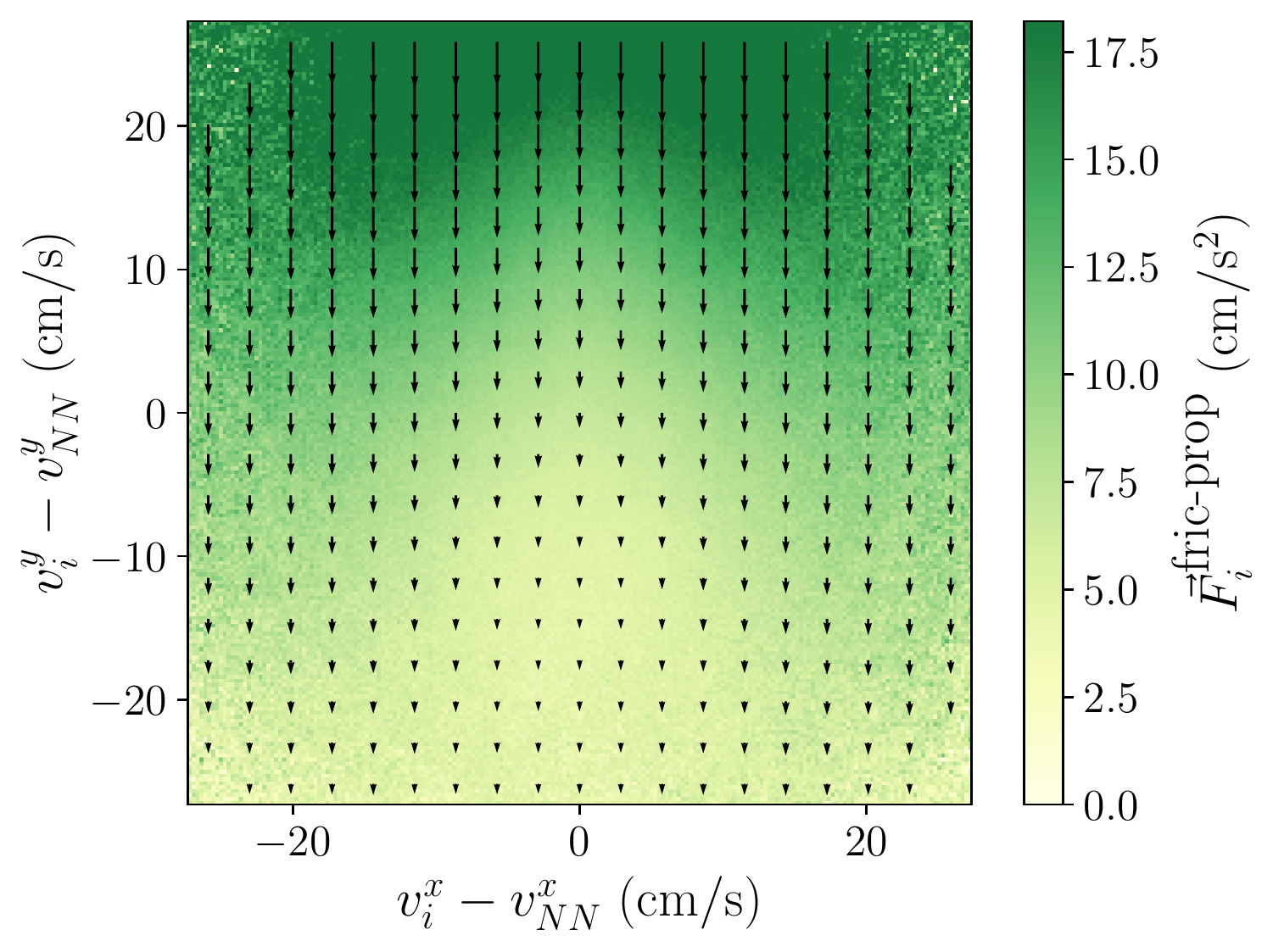}%
}
\caption{(a) Counts, (b) average attraction-repulsion, (c) average alignment and (d) average friction-propulsion force acting on an individual $i$ depending on its relative velocity with the nearest neighbour $NN$ for the explicit anti-alignment model.} \label{supp:fig:expAligmentModel_align}
\end{figure*}

\begin{figure*}[t!p]
\subfloat[]{%
  \includegraphics[width=0.4\textwidth]{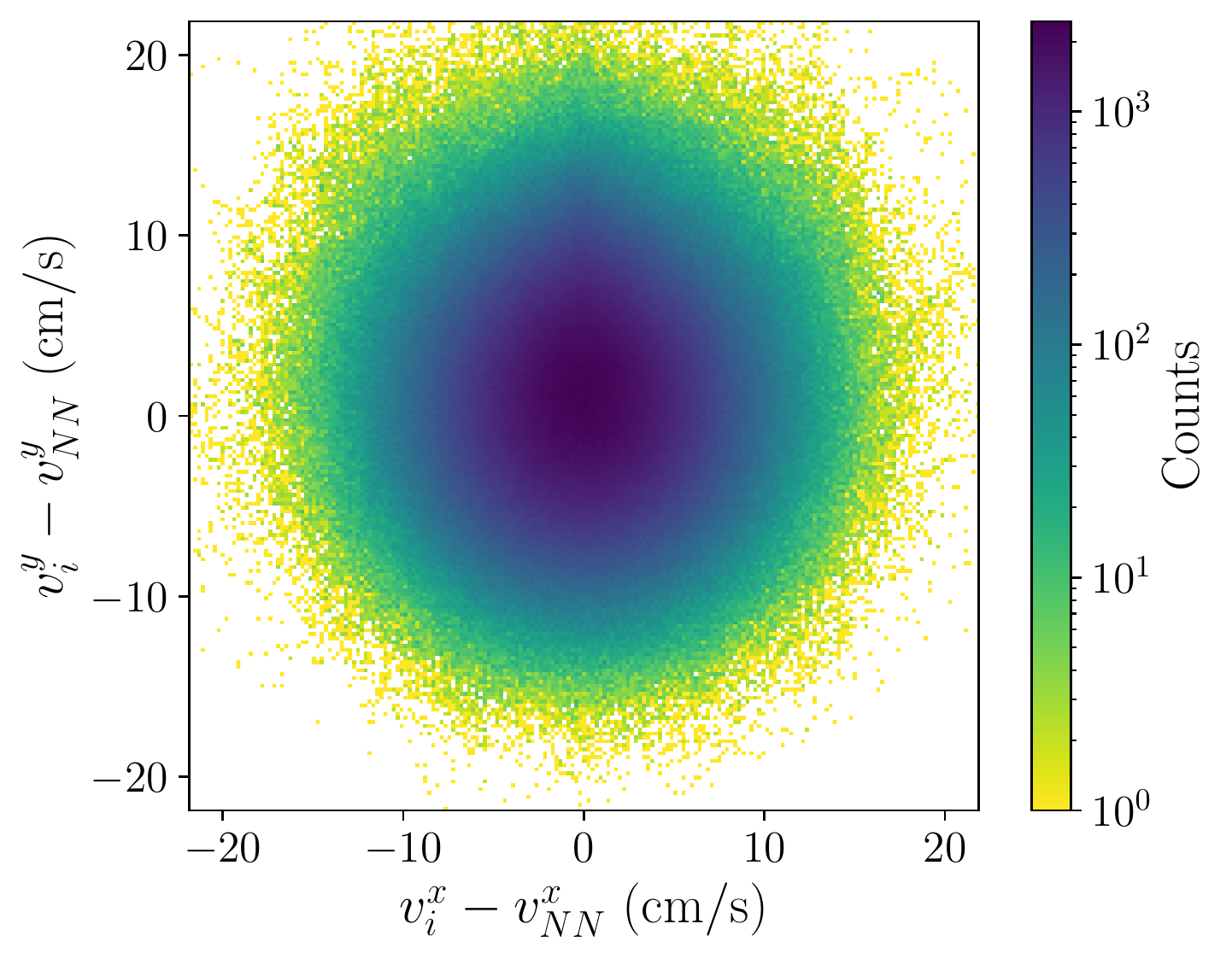}%
}
\hspace{0.01\textwidth}
\subfloat[\label{supp:fig:persistentRandomForce_alig:acc}]{%
  \includegraphics[width=0.4\textwidth]{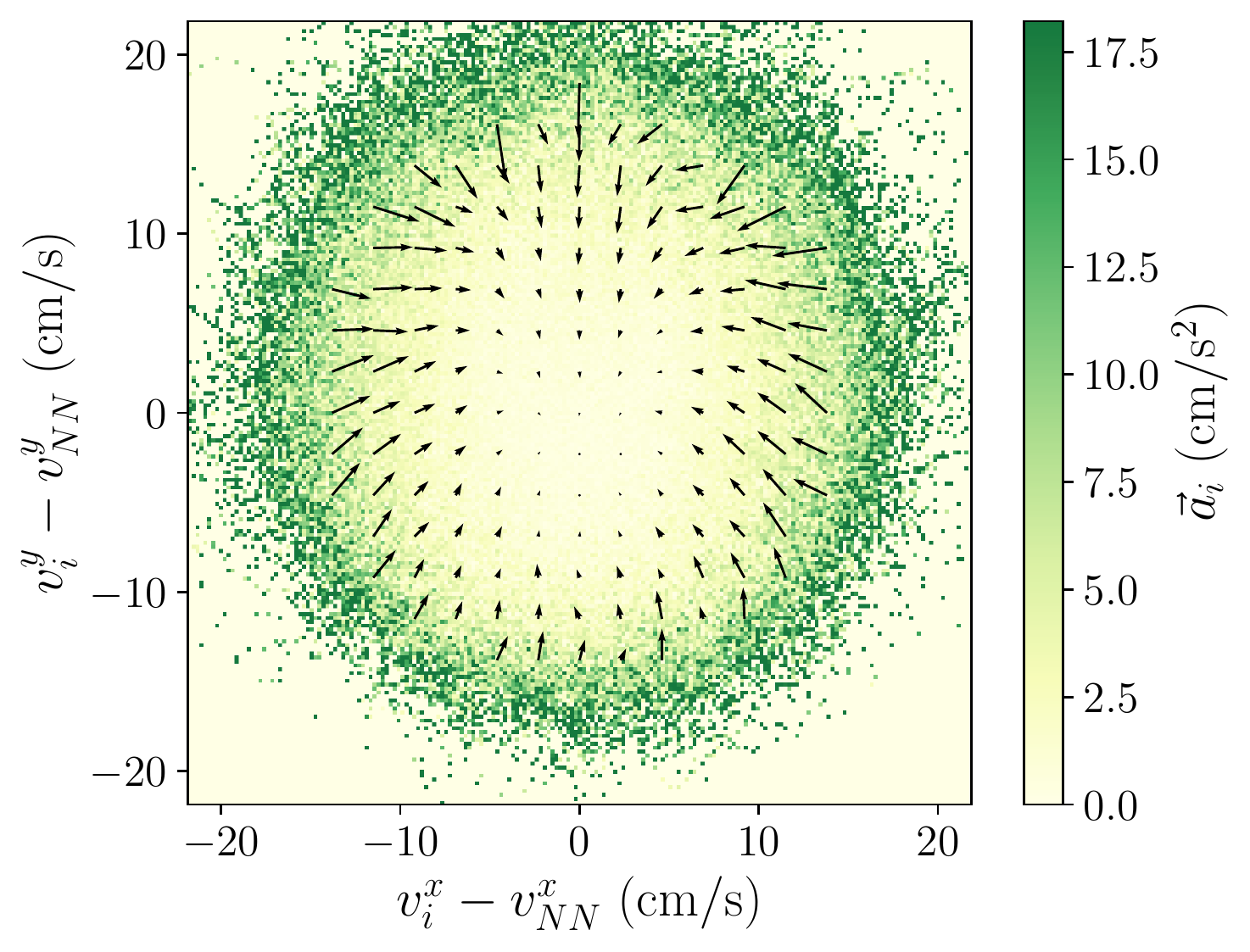}%
}

\subfloat[]{%
  \includegraphics[width=0.4\textwidth]{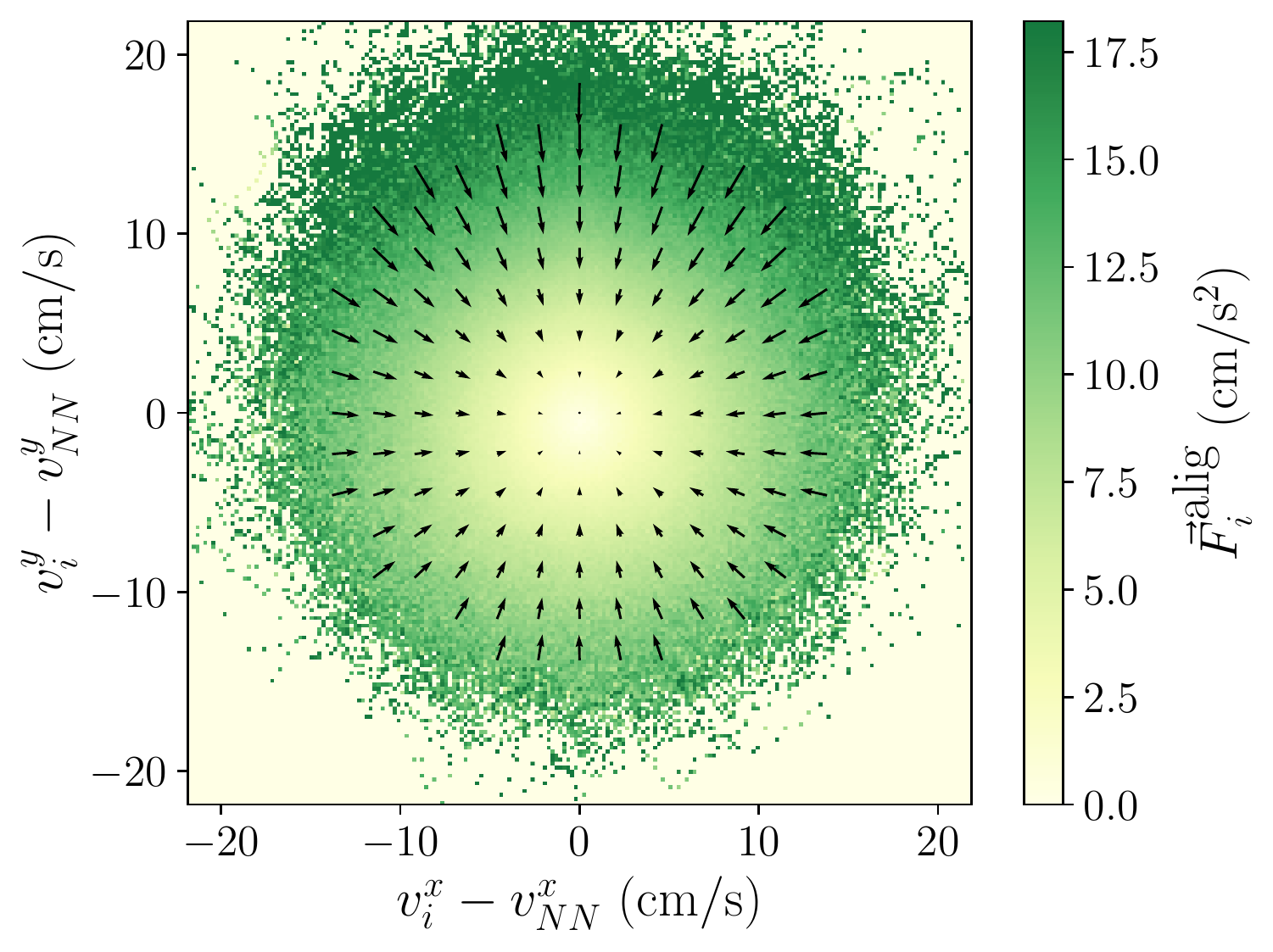}%
}
\hspace{0.01\textwidth}
\subfloat[\label{supp:fig:persistentRandomForce_alig:randomForce}]{%
  \includegraphics[width=0.4\textwidth]{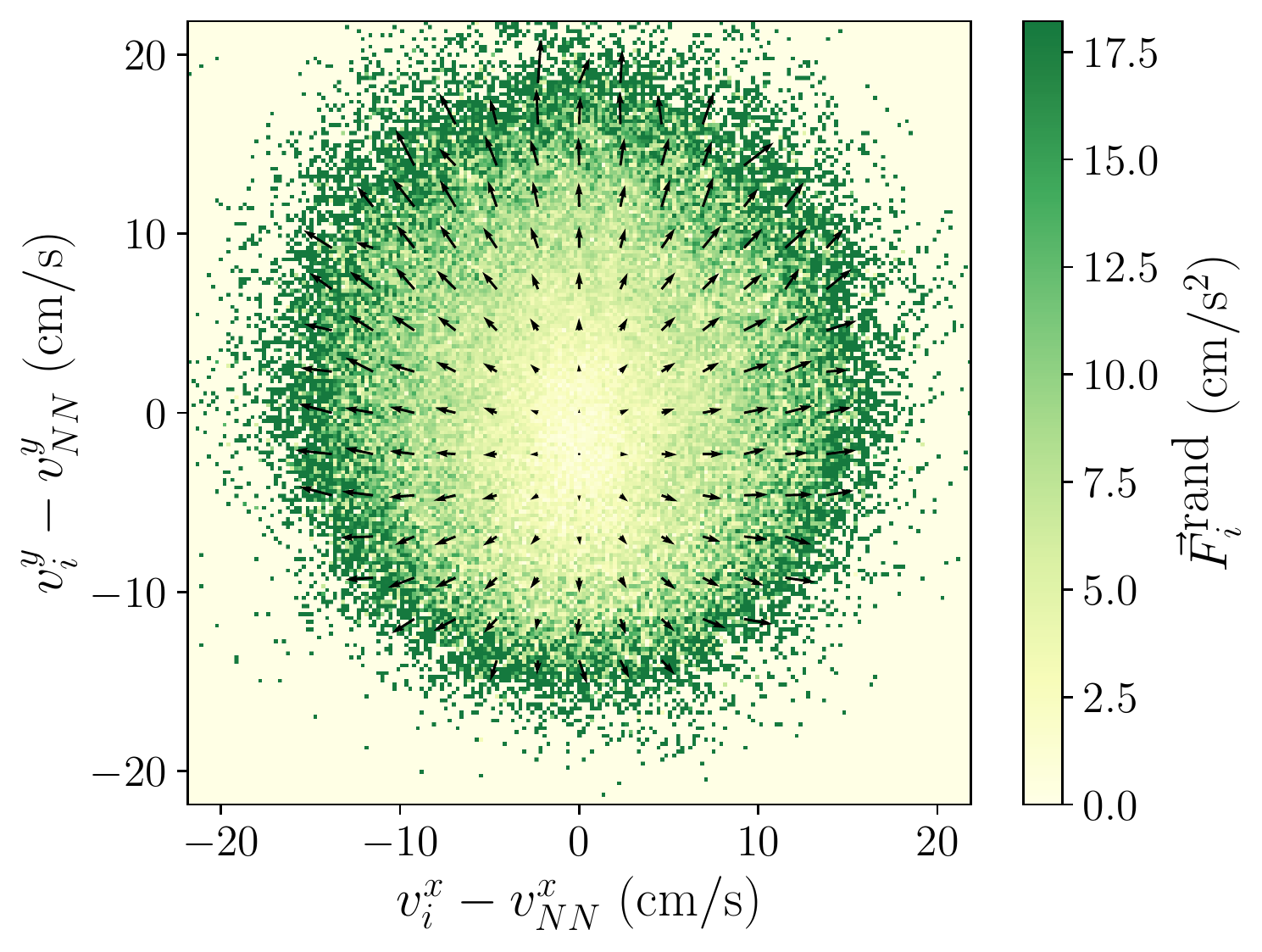}%
}
\caption{(a) Counts, (b) average acceleration (alignment force map), (c) average alignment force and (d) average persistent random force acting on an individual $i$ depending on its relative velocity with the nearest neighbour $NN$ for the standard model with a persistent random force. Here we use an individual has a probability rate $r = 1/20$ to switch off the social forces and experience a persistent random force of amplitude $A = 1.5$ for a time $\tau = 5$. The persistent random force appears as an effective anti-alignment force with arrows pointing outwards. However, in this regime the alignment force dominates and the acceleration is displayed with alignment with arrows pointing inwards.} \label{supp:fig:persistentRandomForce_alig}
\end{figure*}

\begin{figure*}[t!p]
\subfloat[]{%
  \includegraphics[width=0.4\textwidth]{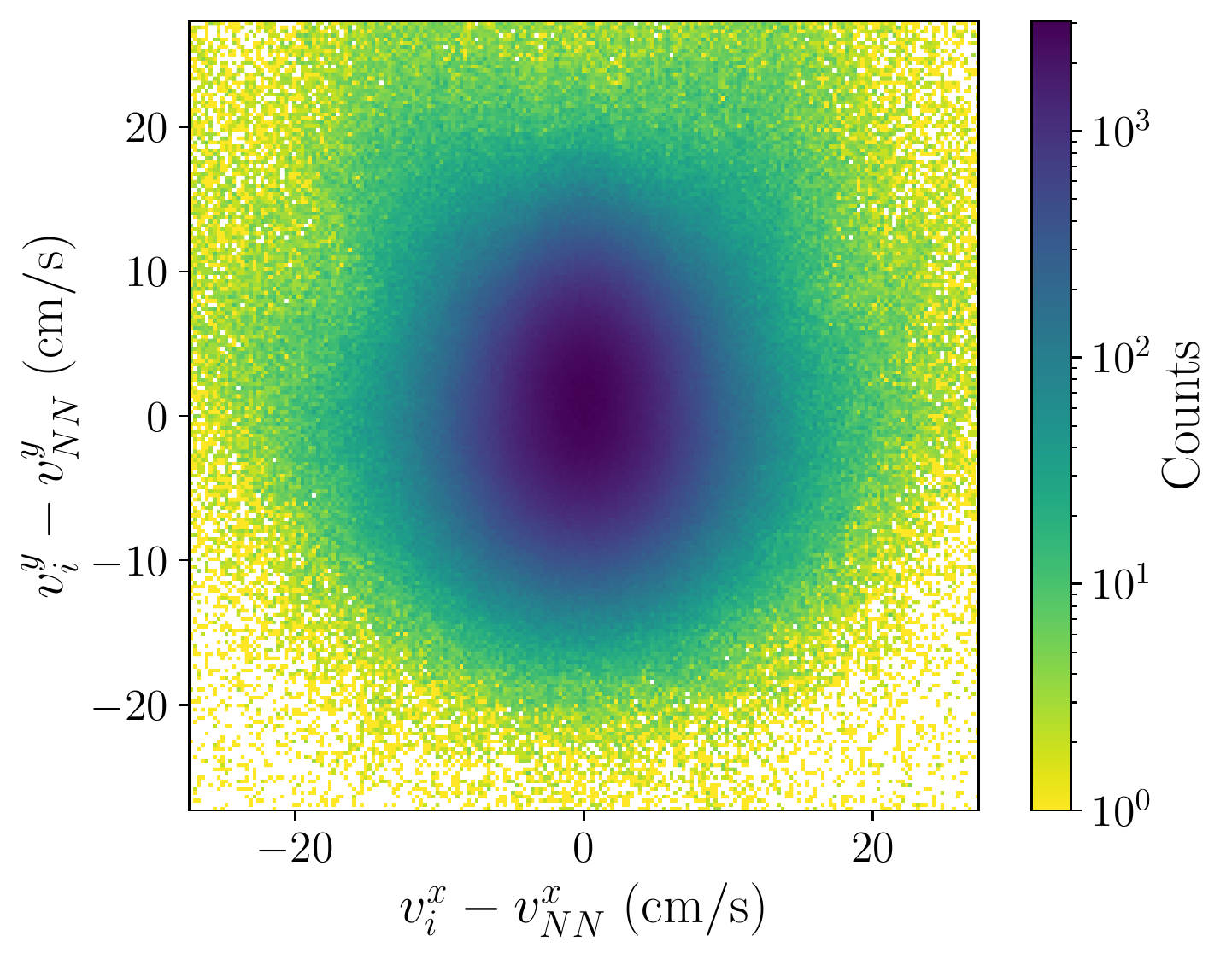}%
}
\hspace{0.01\textwidth}
\subfloat[]{%
  \includegraphics[width=0.4\textwidth]{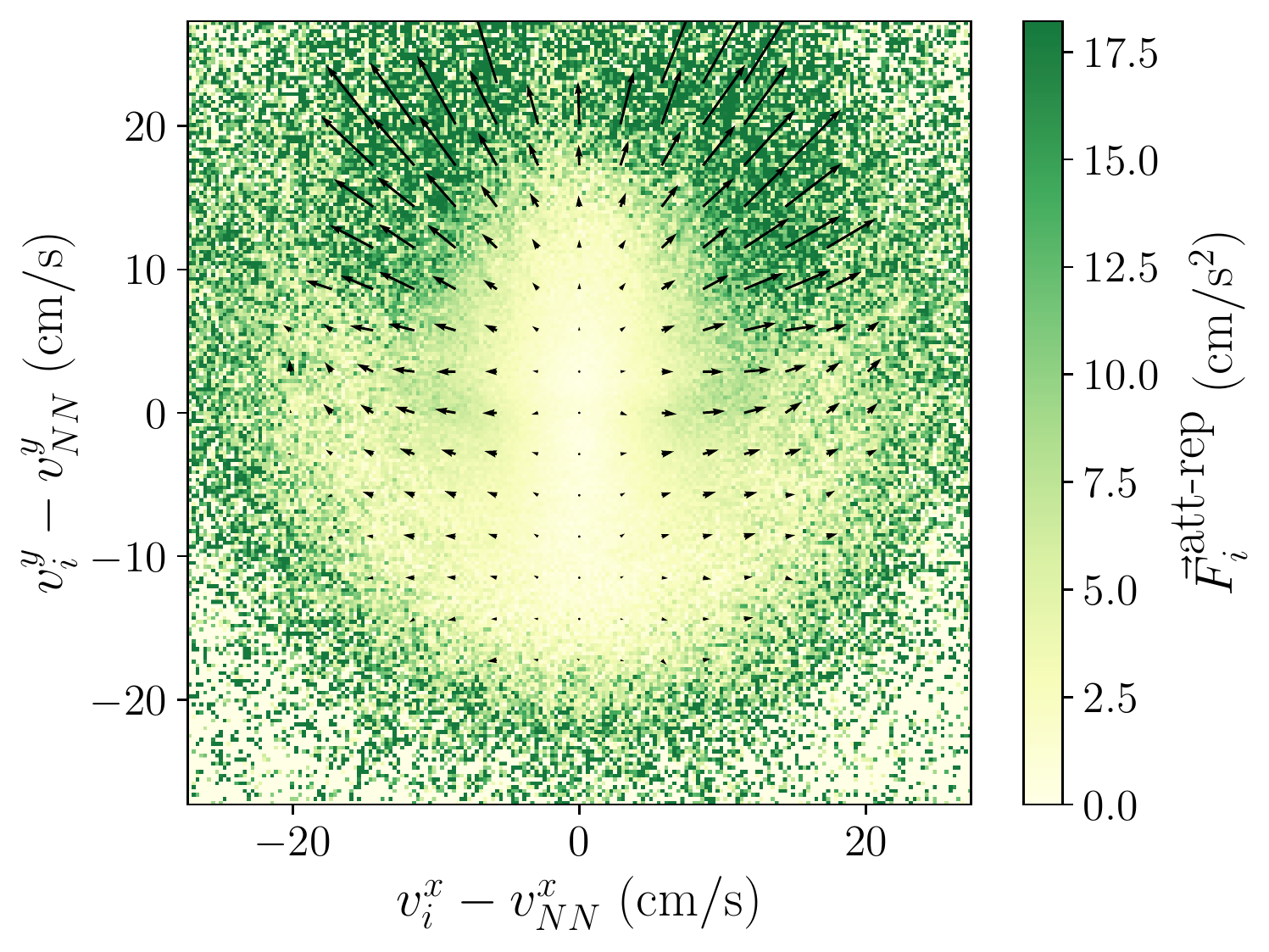}%
}

\subfloat[]{%
  \includegraphics[width=0.4\textwidth]{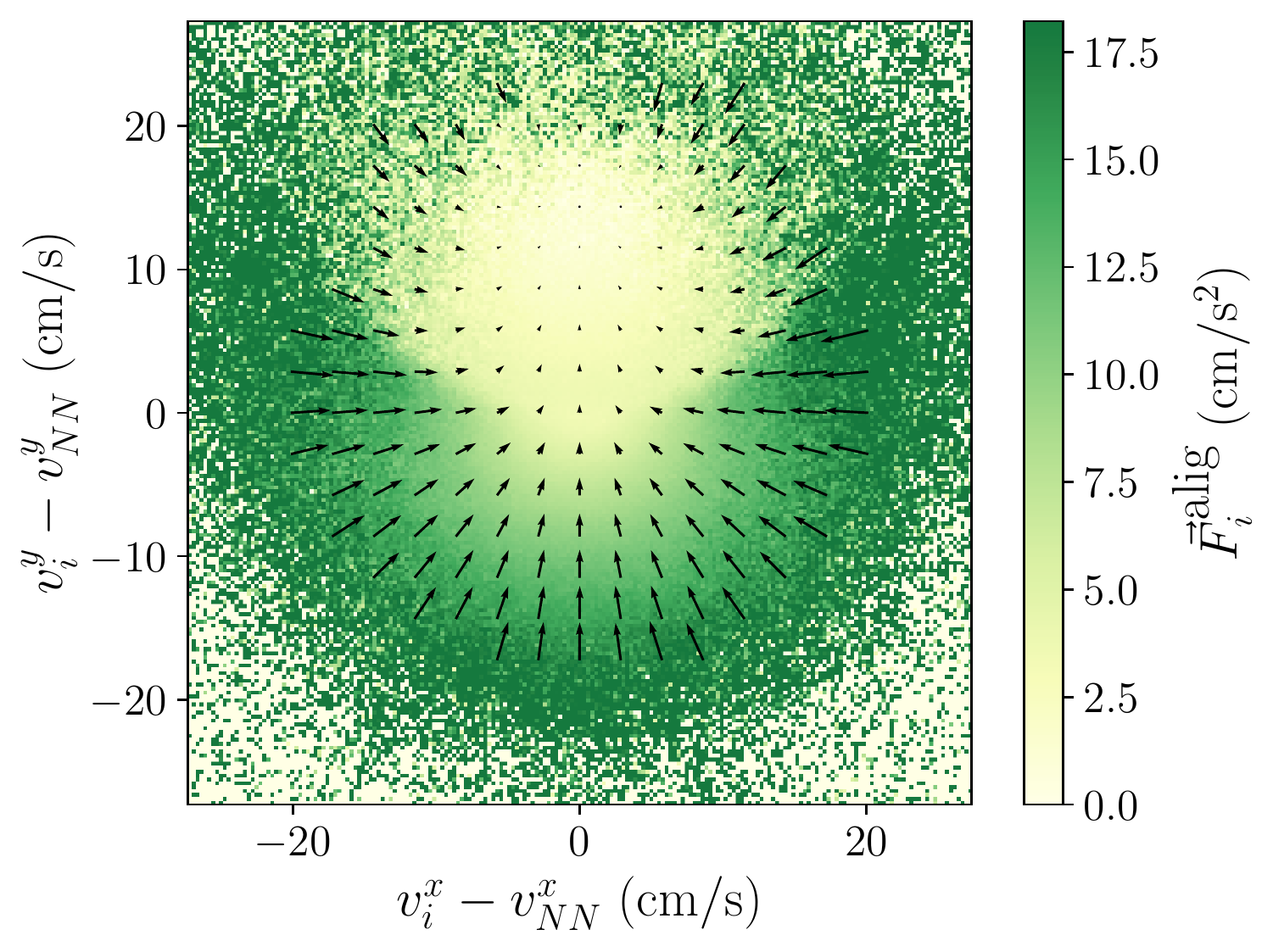}%
}
\hspace{0.01\textwidth}
\subfloat[]{%
  \includegraphics[width=0.4\textwidth]{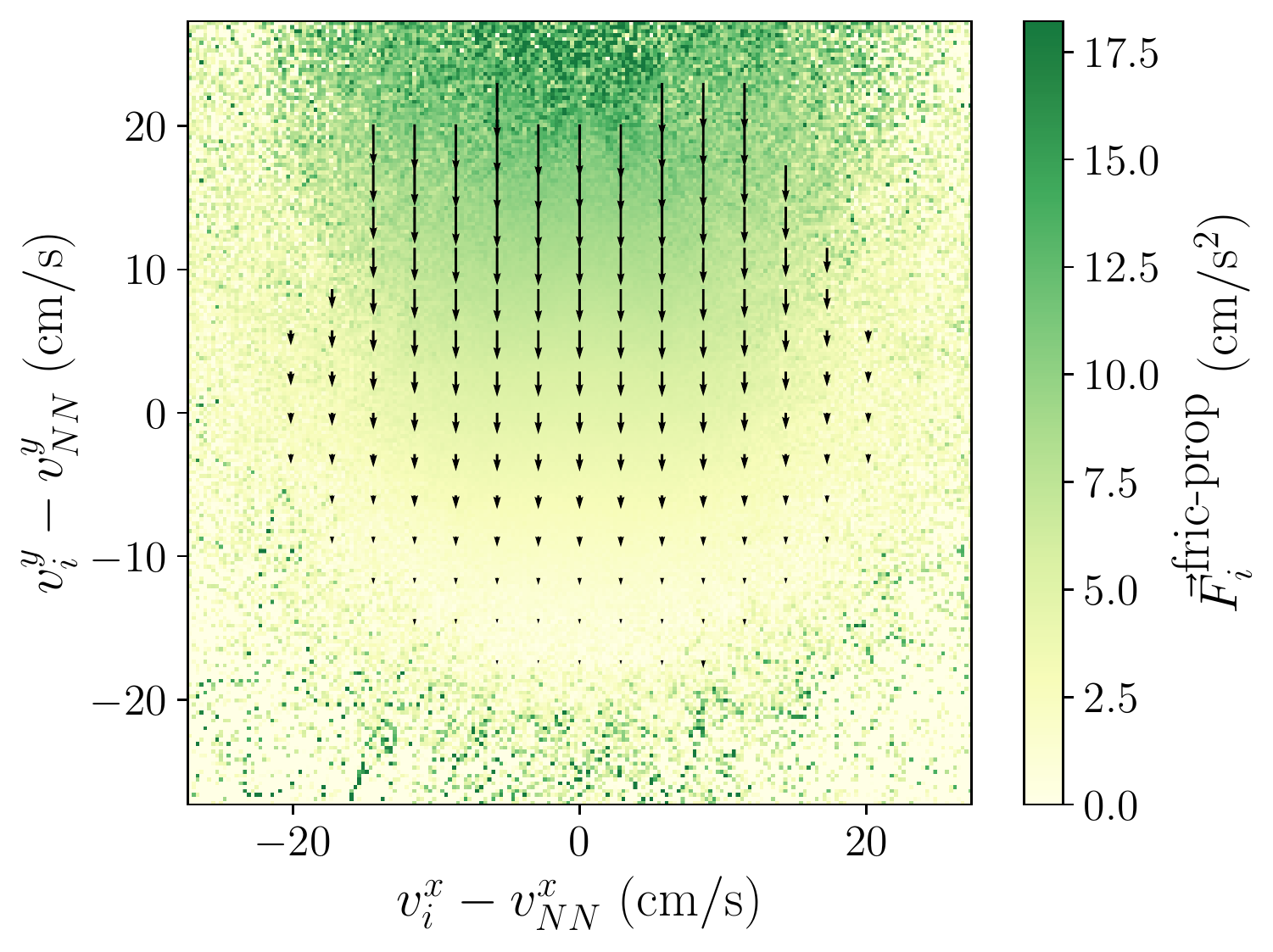}%
}
\caption{(a) Counts, (b) average attraction-repulsion, (c) average alignment and (d) average friction-propulsion force acting on an individual $i$ depending on its relative velocity with the nearest neighbour $NN$ for the selective interactions model. Here the attraction-repulsion force for slower nearest neighbours acts as an effective anti-alignment force with arrows pointing outwards.} \label{supp:fig:selInteractionsModel_align}
\end{figure*}

\begin{figure*}[t!p]
\subfloat[\label{supp:fig:pdfs_comparison:omega}]{%
  \includegraphics[width=0.4\textwidth]{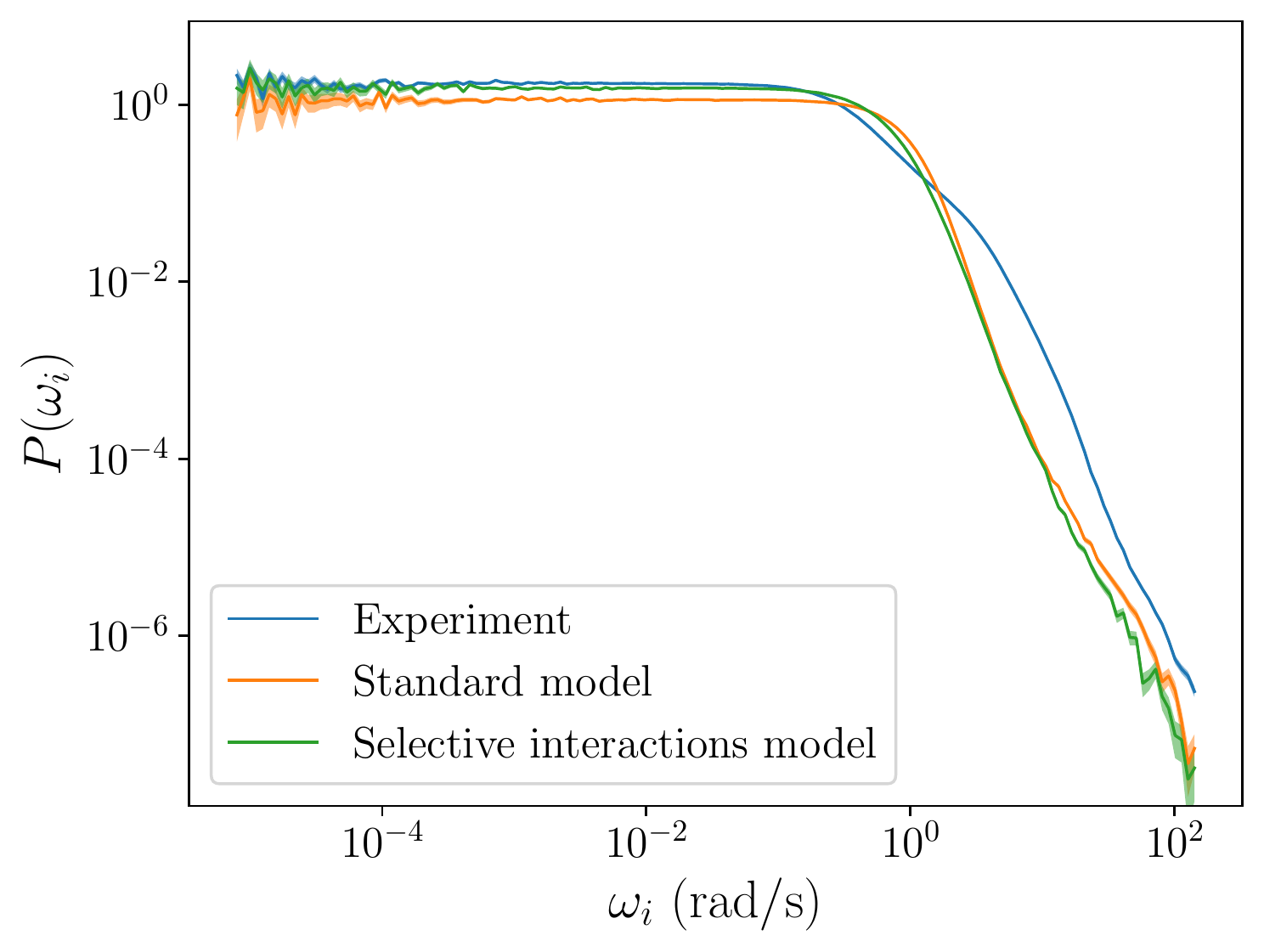}%
}
\hspace{0.01\textwidth}
\subfloat[\label{supp:fig:pdfs_comparison:speeds}]{%
  \includegraphics[width=0.4\textwidth]{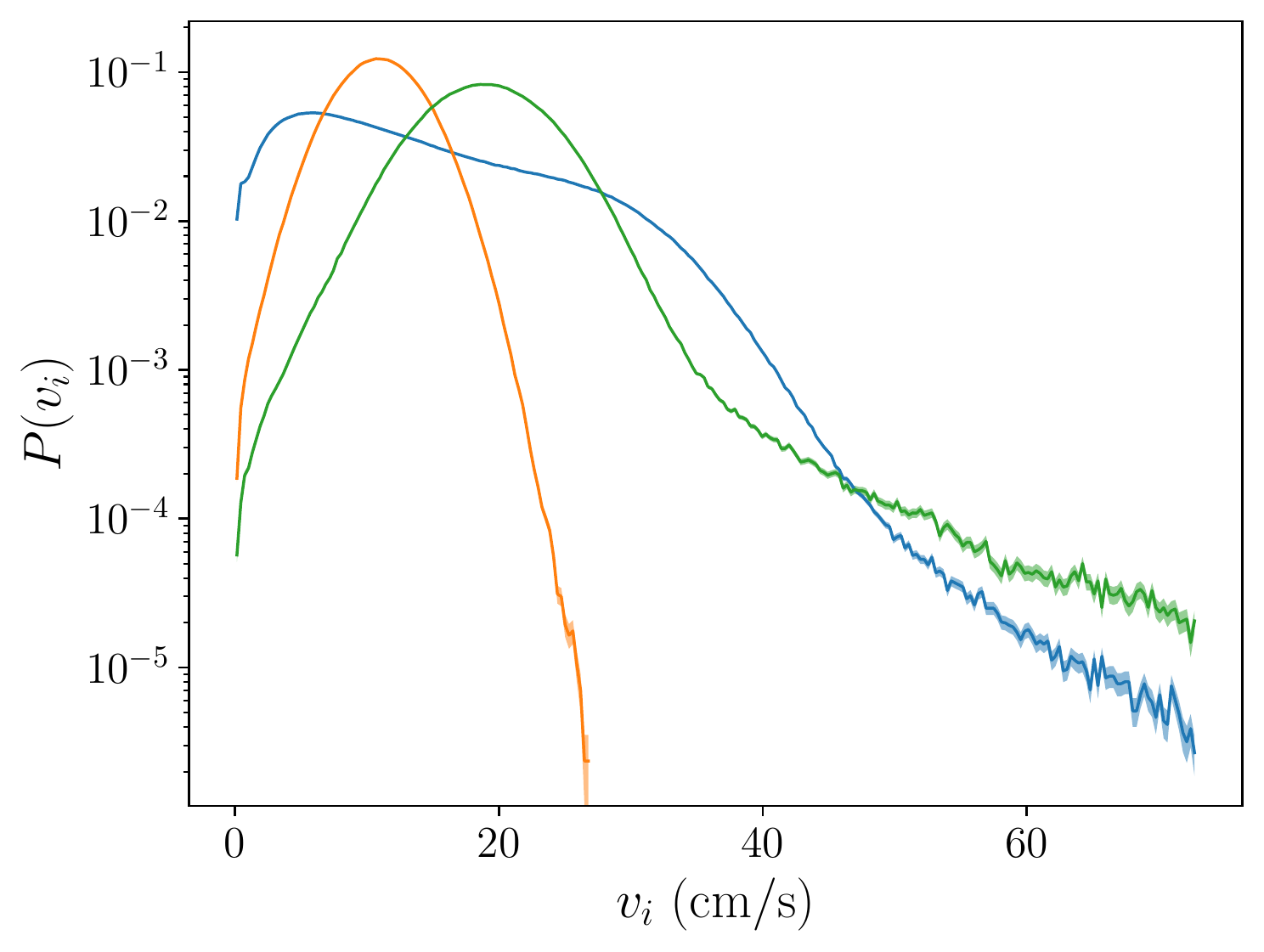}%
}

\subfloat[\label{supp:fig:pdfs_comparison:speed_temporal_evo}]{%
  \includegraphics[width=0.4\textwidth]{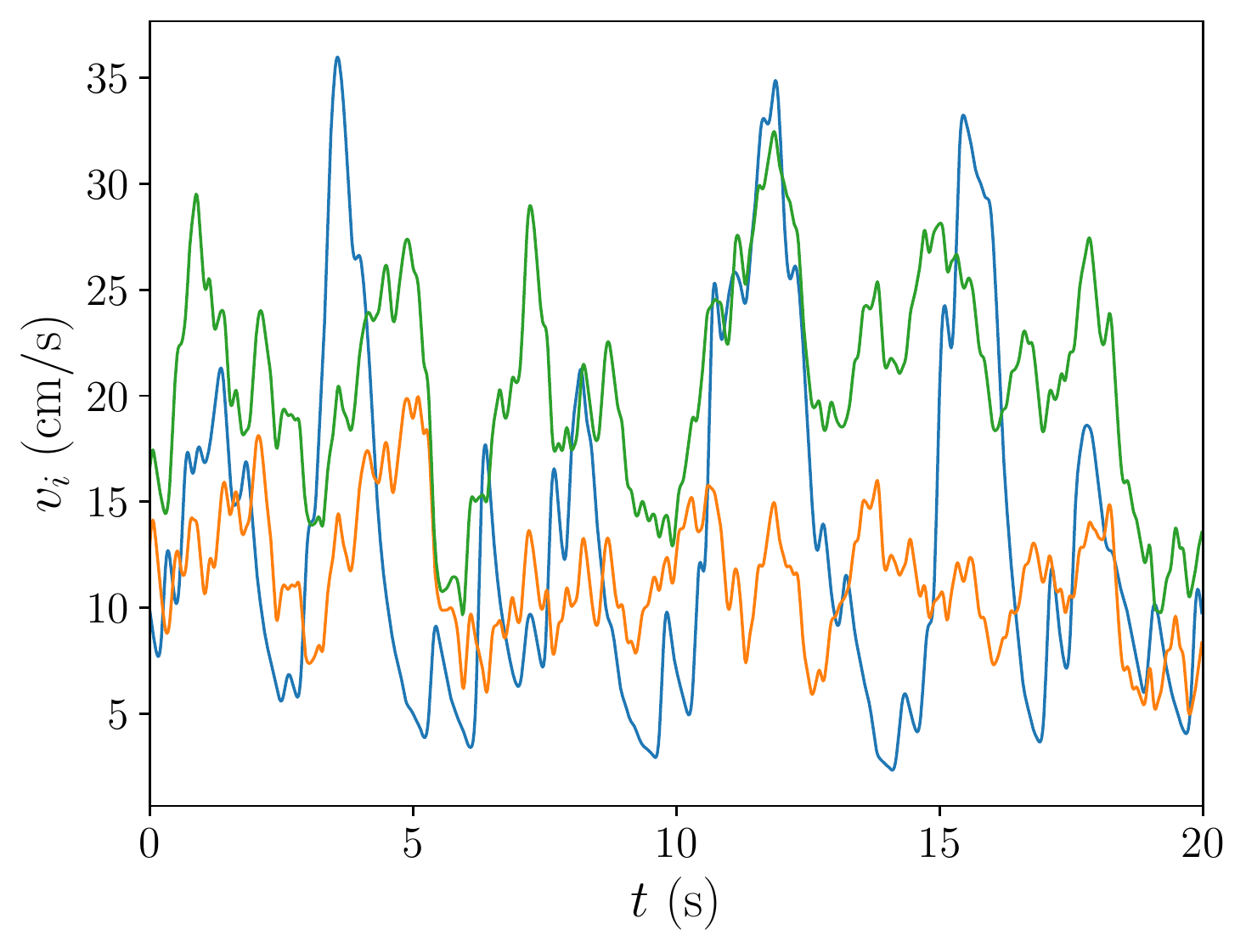}%
}
\caption{(a) PDF of the turning rate $\omega_i$, (b) PDF of the speed $v_i$ and (c) sample of the temporal evolution of an individual speed for the experimental data, the standard model and the selective interactions model.} \label{supp:fig:pdfs_comparison}
\end{figure*}

\begin{figure*}[t!p]
\subfloat[]{%
  \includegraphics[width=0.4\textwidth]{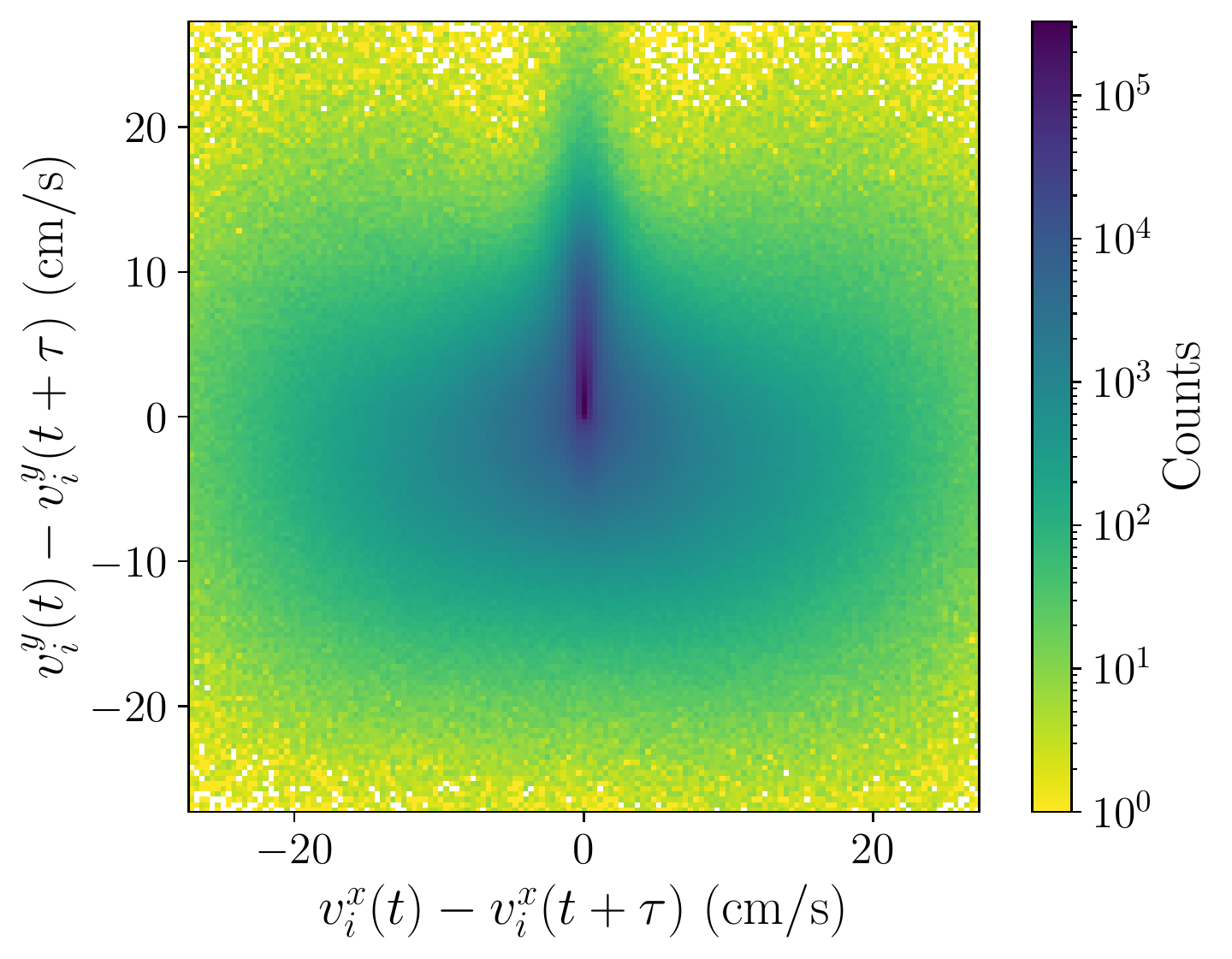}%
}
\hspace{0.01\textwidth}
\subfloat[]{%
  \includegraphics[width=0.4\textwidth]{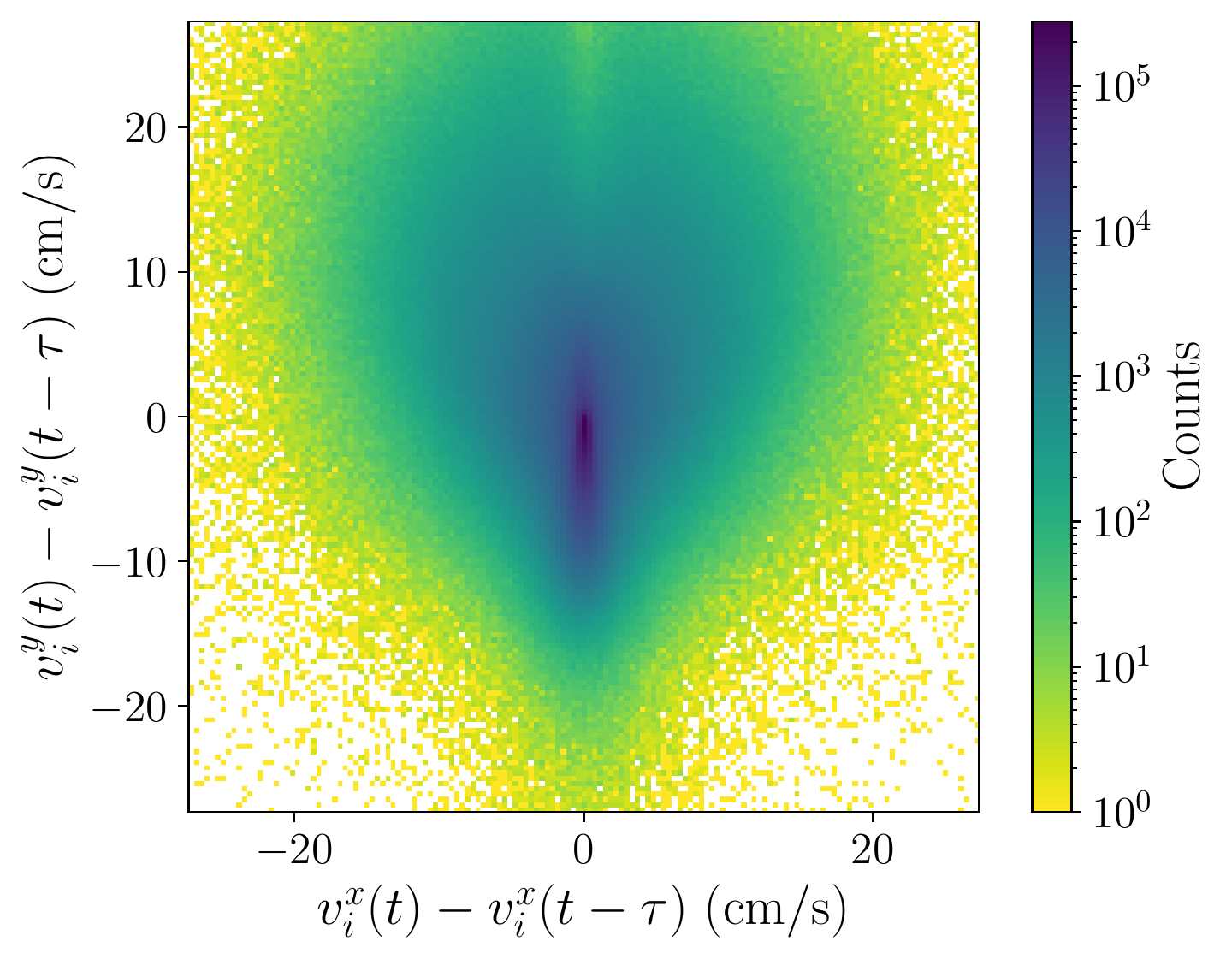}%
}
\caption{Counts for the alignment force maps for the experimental data comparing the individual at present with itself at a (a) positive and (b) negative delay with $\tau = 0.2$~s.} \label{supp:fig:counts_delayedAlignmentForceMap}
\end{figure*}

\begin{figure*}[t!p]
\subfloat[\label{supp:fig:force_maps_Voronoi:Vor_countsatrcRep}]{%
  \includegraphics[width=0.4\textwidth]{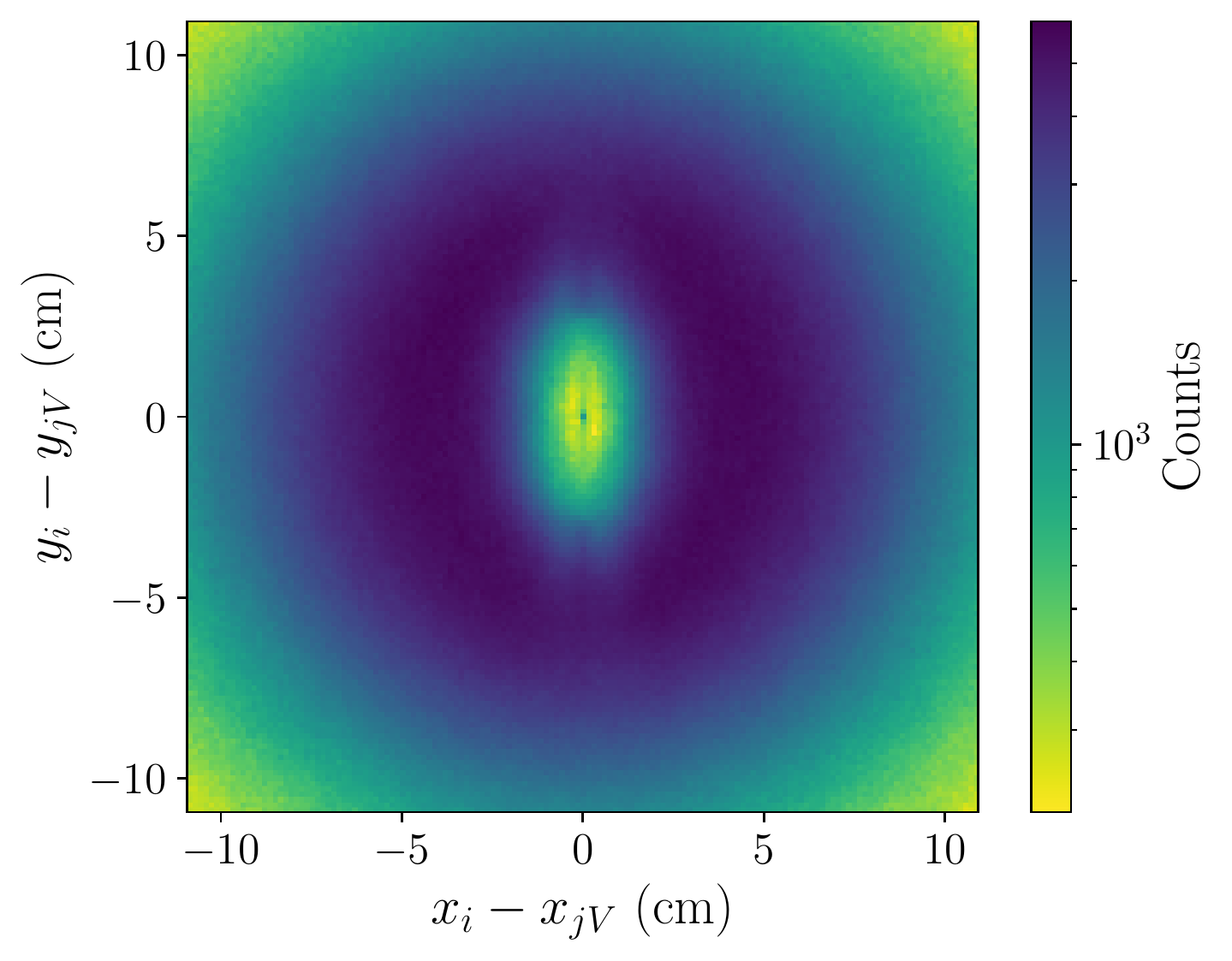}%
}
\hspace{0.01\textwidth}
\subfloat[\label{supp:fig:force_maps_Voronoi:Vor_exp_atracRepModel}]{%
  \includegraphics[width=0.4\textwidth]{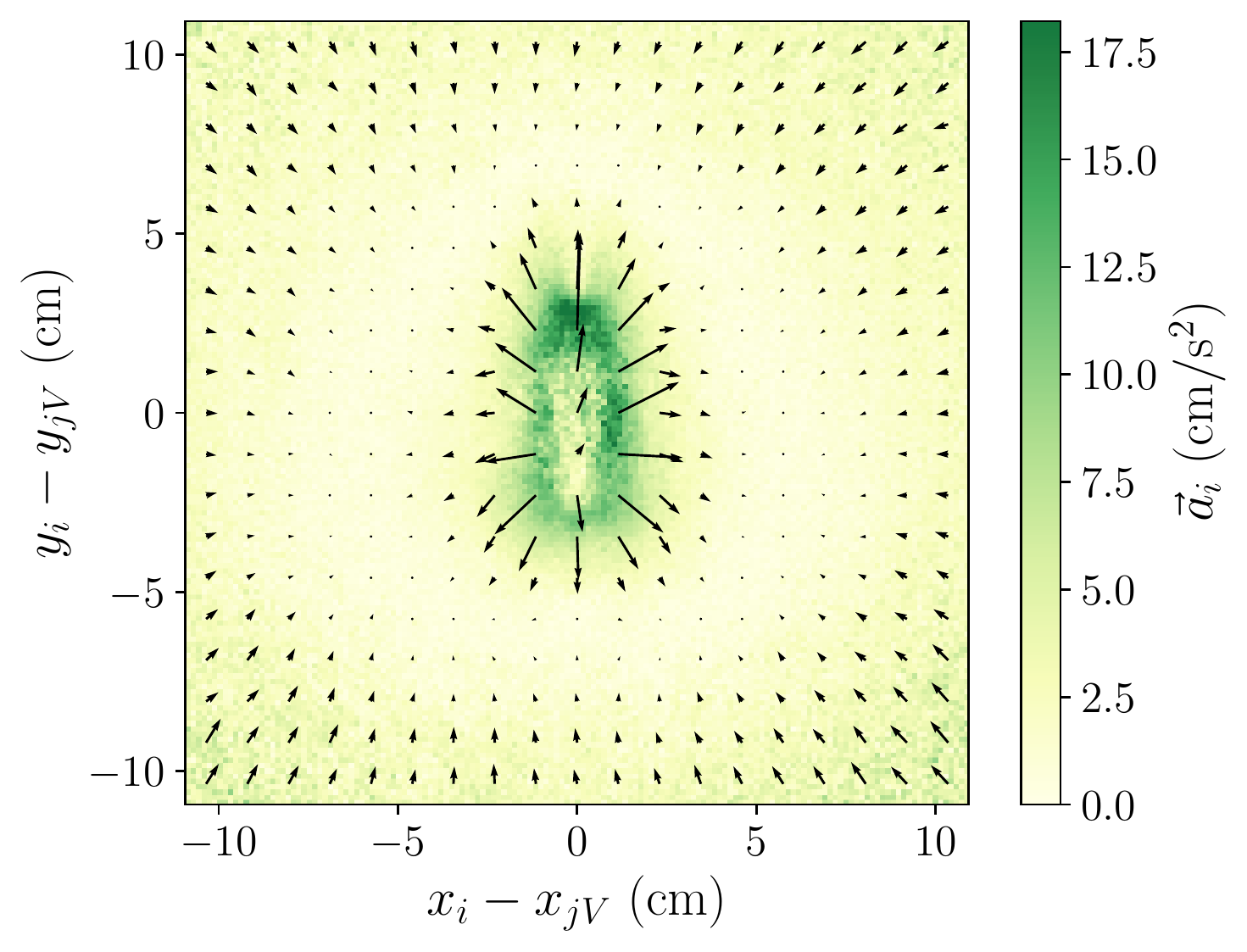}%
}

\subfloat[\label{supp:fig:force_maps_Voronoi:Vor_countsAlig}]{%
  \includegraphics[width=0.4\textwidth]{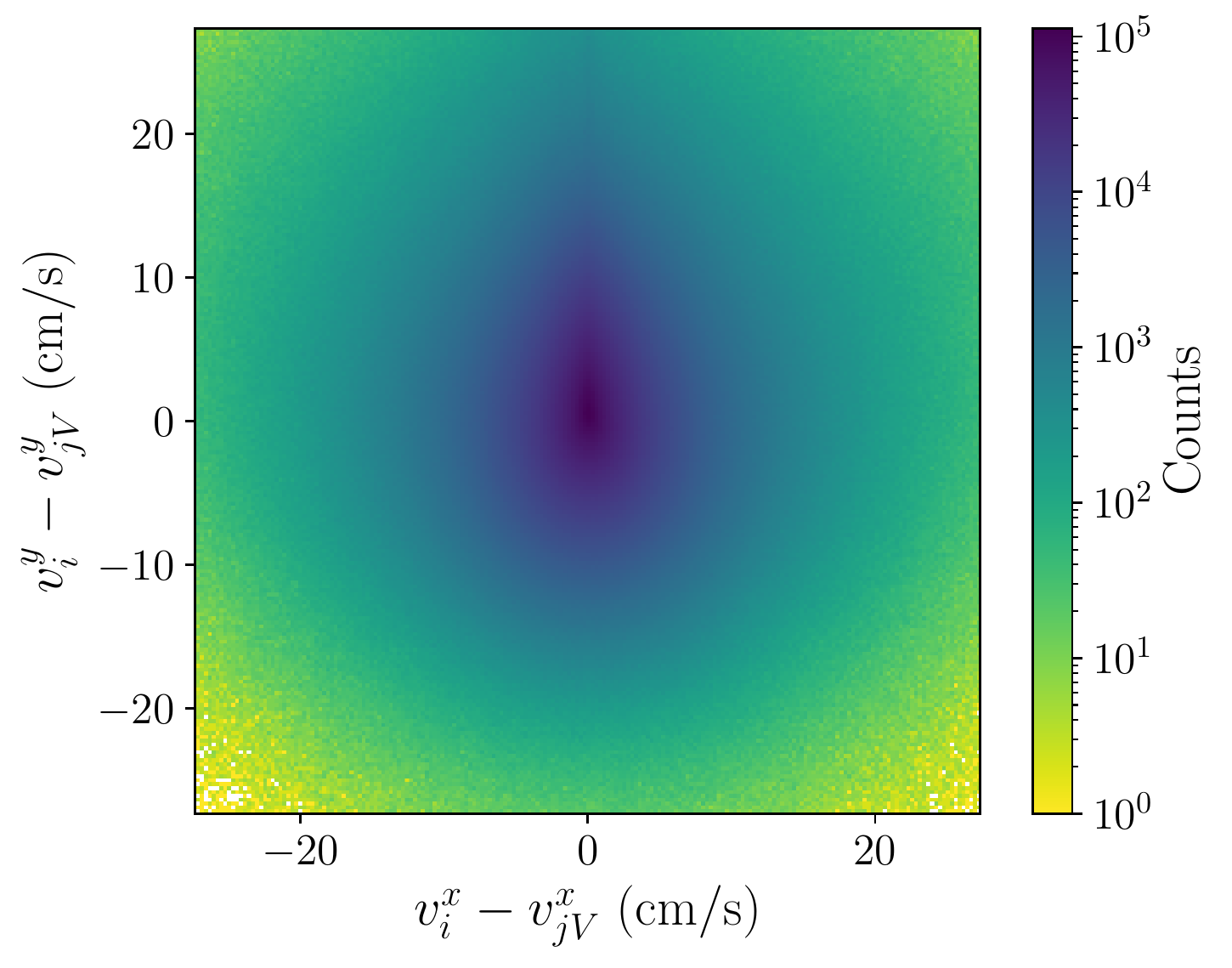}%
}
\hspace{0.01\textwidth}
\subfloat[\label{supp:fig:force_maps_Voronoi:Vor_aligExp}]{%
  \includegraphics[width=0.4\textwidth]{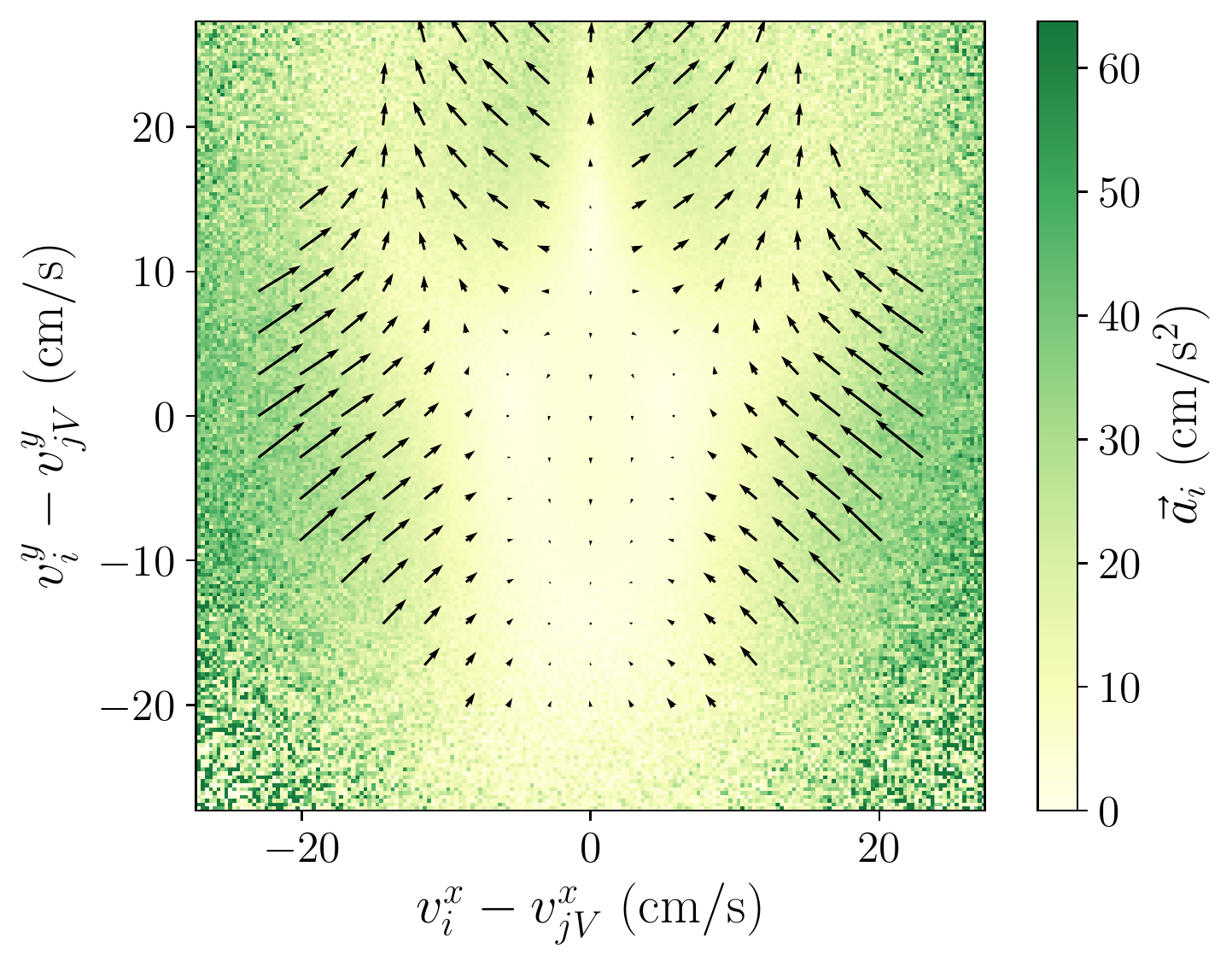}%
}
\caption{For Voronoi neighbours: (a) counts of the relative positions between neighbours, (b) attraction-repulsion force map, (c) counts of the relative velocities between neighbours and (d) alignment force map for the experimental data.} \label{supp:fig:force_maps_Voronoi}
\end{figure*}

\begin{figure*}[t!p]
\subfloat[\label{supp:fig:force_maps_Voronoi:Vor_countsatrcRep}]{%
  \includegraphics[width=0.4\textwidth]{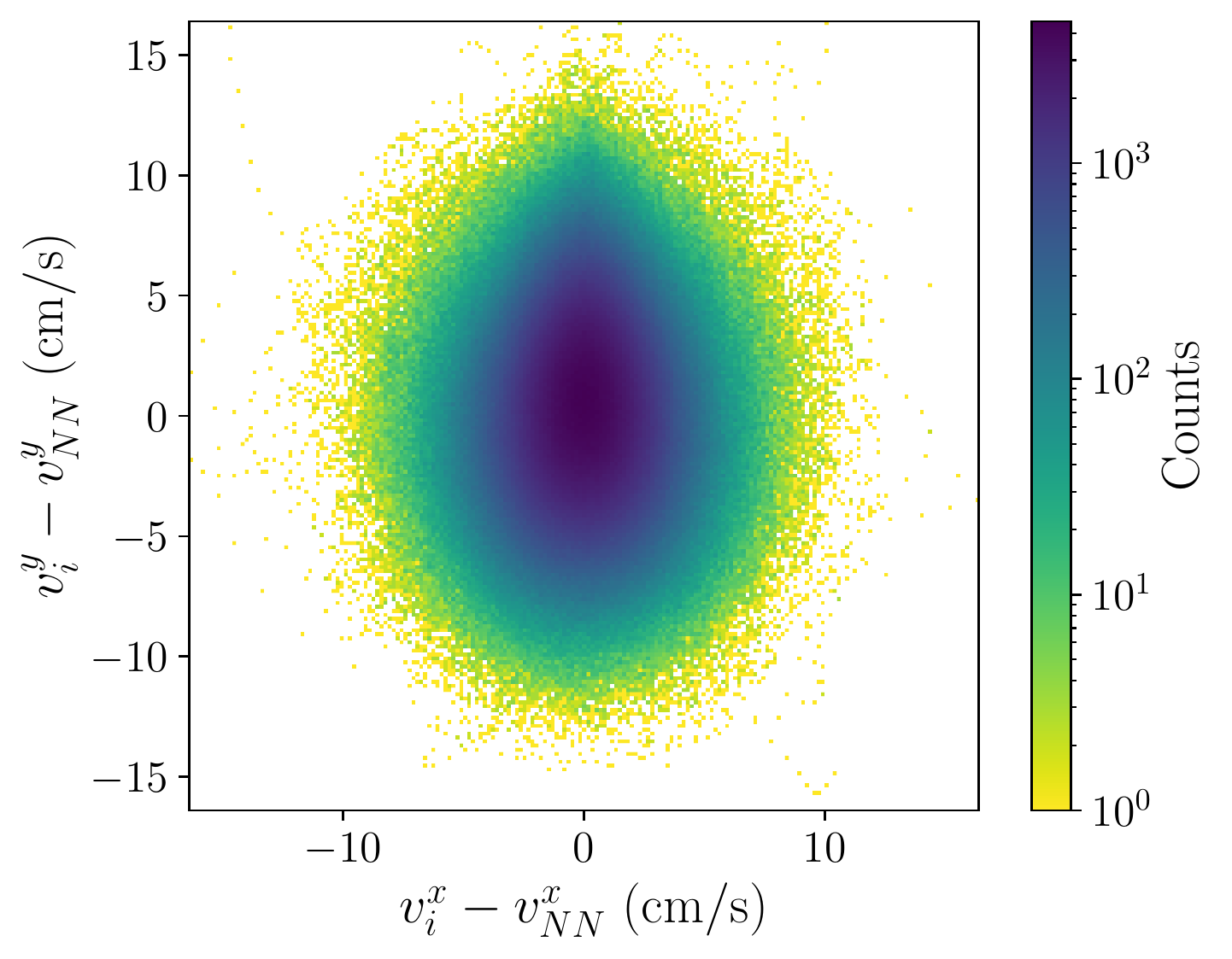}%
}
\hspace{0.01\textwidth}
\subfloat[\label{supp:fig:force_maps_Voronoi:Vor_exp_atracRepModel}]{%
  \includegraphics[width=0.4\textwidth]{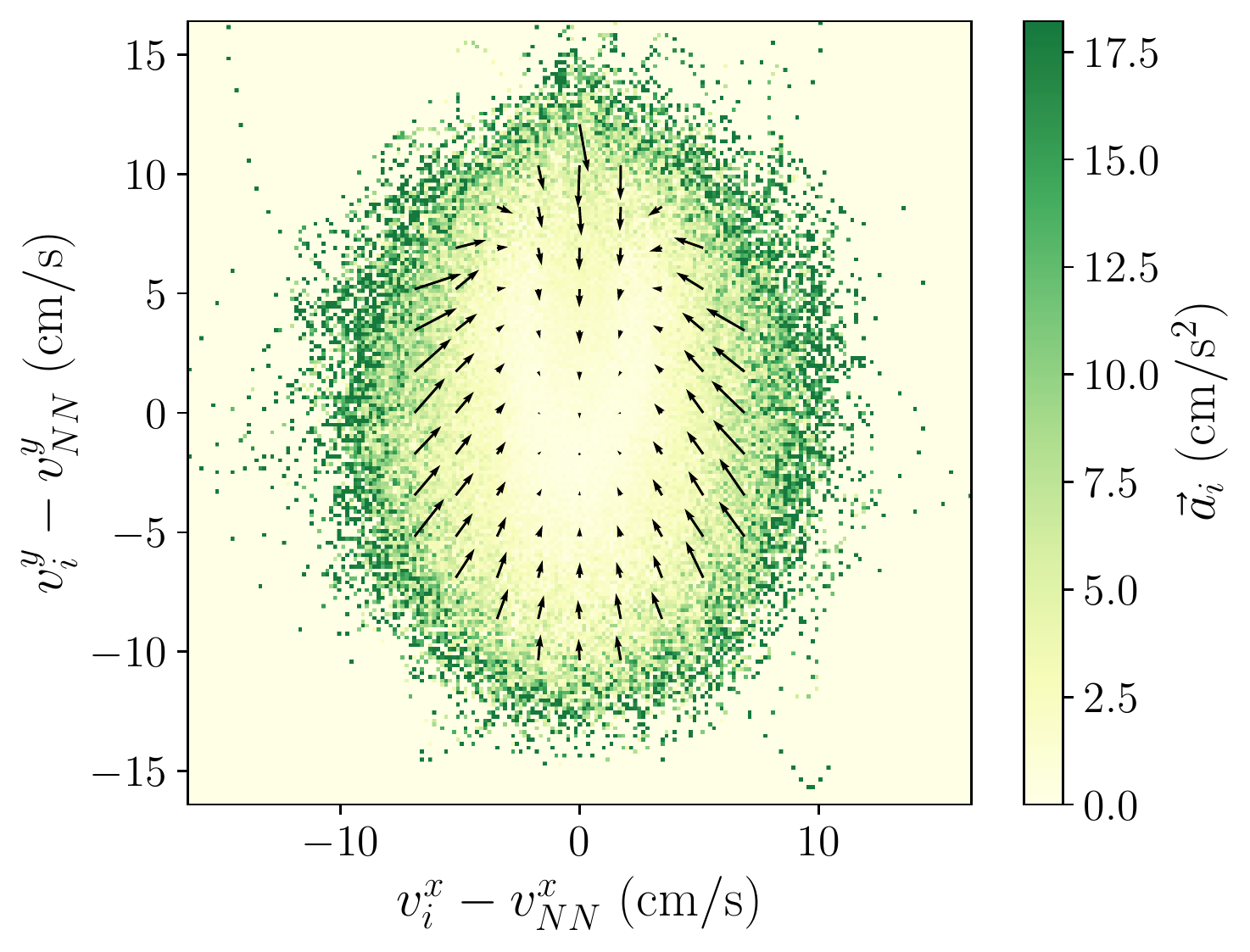}%
}

\subfloat[\label{supp:fig:force_maps_Voronoi:Vor_countsatrcRep}]{%
  \includegraphics[width=0.4\textwidth]{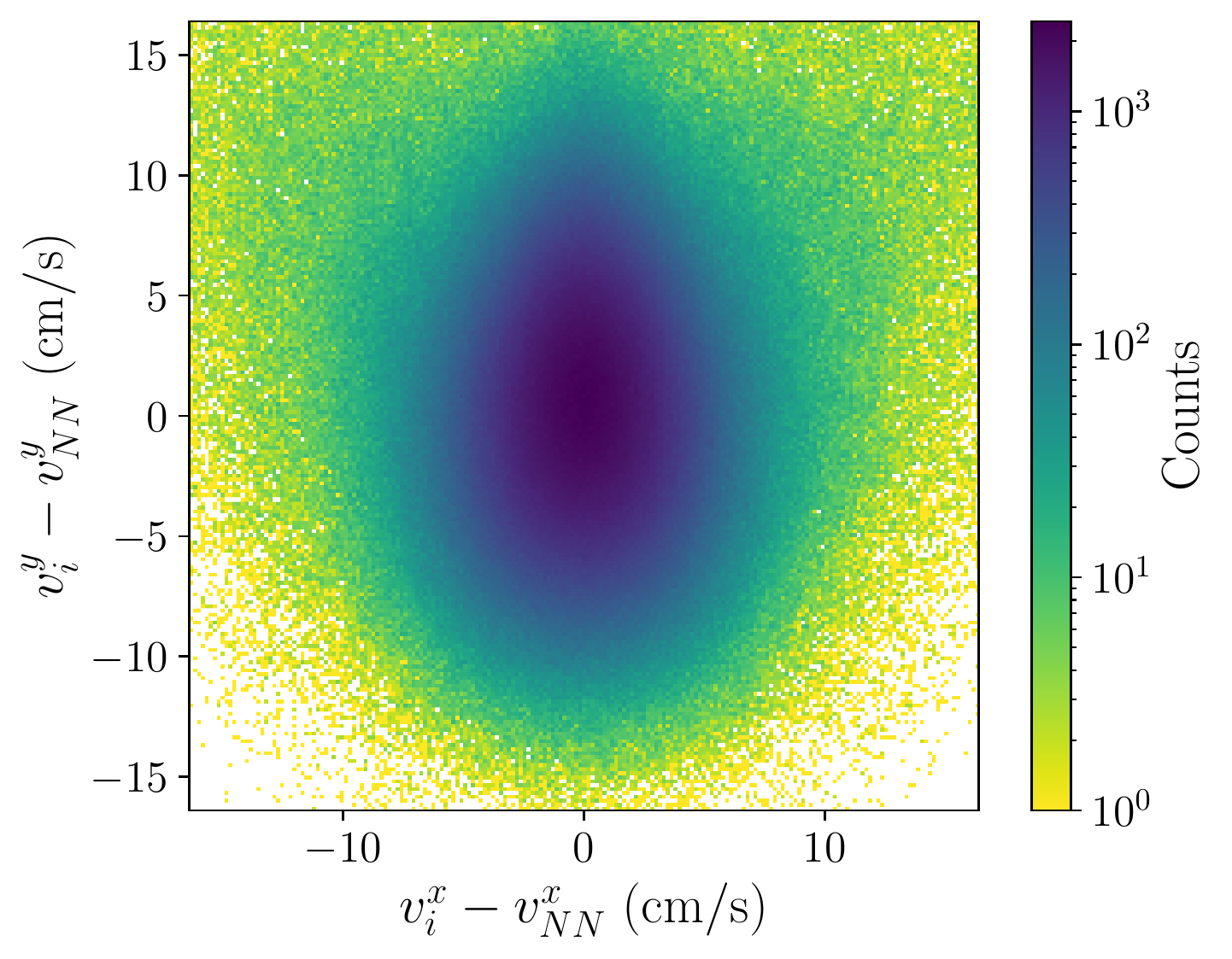}%
}
\hspace{0.01\textwidth}
\subfloat[\label{supp:fig:force_maps_Voronoi:Vor_exp_atracRepModel}]{%
  \includegraphics[width=0.4\textwidth]{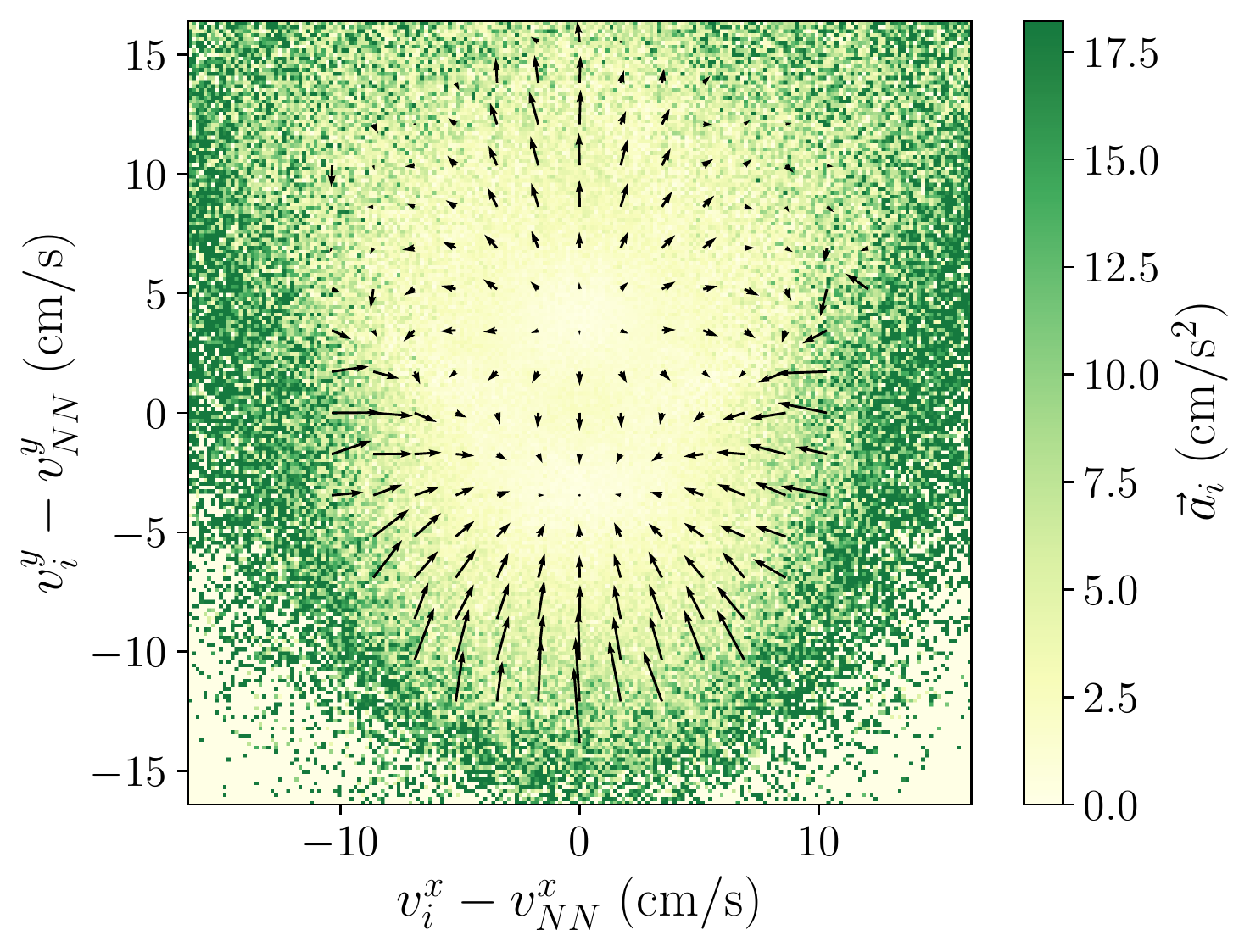}%
}
\caption{Simulations where individuals interact socially only with the nearest neighbour: (a) counts and (b) alignment force map for the standard model, (c) counts and (d) alignment force map for the selective interactions model.} \label{supp:fig:force_maps_simulation_Nn}
\end{figure*}

\clearpage

\end{document}